\documentclass[prx,twocolumn,superscriptaddress,citeautoscript,showpacs,amsart]{revtex4}

\usepackage{pdfpages}
\usepackage{graphicx}
\usepackage{multirow}
\usepackage{color}
\usepackage{bm}
\usepackage{times}
\usepackage{amsmath,bm,amsfonts}
\usepackage{dcolumn}
\usepackage{graphicx}
\usepackage{latexsym}

\usepackage{ulem} 
\usepackage{braket}

\usepackage{cancel}

\def\ket#1{|#1\rangle }
\def\bra#1{\langle #1 |}

\def\bb{\mathbb}

\def\d{\partial}

\begin{document}
\title{Band Topology and Linking Structure of Nodal Line Semimetals with $Z_2$ Monopole Charges}

\author{Junyeong \surname{Ahn}}
\affiliation{Department of Physics and Astronomy, Seoul National University, Seoul 08826, Korea}

\affiliation{Center for Correlated Electron Systems, Institute for Basic Science (IBS), Seoul 08826, Korea}

\affiliation{Center for Theoretical Physics (CTP), Seoul National University, Seoul 08826, Korea}

\author{Dongwook \surname{Kim}}
\affiliation{Department of Physics, Sungkyunkwan University, Suwon 16419, Korea}

\author{Youngkuk \surname{Kim}}
\affiliation{Department of Physics, Sungkyunkwan University, Suwon 16419, Korea}

\author{Bohm-Jung \surname{Yang}}
\email{bjyang@snu.ac.kr}
\affiliation{Department of Physics and Astronomy, Seoul National University, Seoul 08826, Korea}

\affiliation{Center for Correlated Electron Systems, Institute for Basic Science (IBS), Seoul 08826, Korea}

\affiliation{Center for Theoretical Physics (CTP), Seoul National University, Seoul 08826, Korea}

\date{\today}

\begin{abstract}
We study the band topology and the associated linking structure of topological semimetals with nodal lines carrying $Z_{2}$ monopole charges, which can be realized in three-dimensional systems invariant under the combination of inversion $P$ and time reversal $T$ when spin-orbit coupling is negligible. 
In contrast to the well-known $PT$-symmetric nodal lines protected only by $\pi$ Berry phase in which a single nodal line can exist, the nodal lines with $Z_{2}$ monopole charges should always exist in pairs. We show that a pair of nodal lines with $Z_{2}$ monopole charges is created by a {\it double band inversion} (DBI) process, and that the resulting nodal lines are always {\it linked by another nodal line} formed between the two topmost occupied bands.
It is shown that both the linking structure and the $Z_{2}$ monopole charge are the manifestation of the nontrivial band topology characterized by the {\it second Stiefel-Whitney class}, which can be read off from the Wilson loop spectrum. We show that the second Stiefel-Whitney class can serve as a well-defined topological invariant of a $PT$-invariant two-dimensional (2D) insulator in the absence of Berry phase.
Based on this, we propose that pair creation and annihilation of nodal lines with $Z_{2}$ monopole charges can mediate a topological phase transition between a  normal insulator and a three-dimensional weak Stiefel-Whitney insulator (3D weak SWI). 
Moreover, using first-principles calculations, we predict ABC-stacked graphdiyne as a nodal line semimetal (NLSM) with $Z_{2}$ monopole charges having the linking structure.
Finally, we develop a formula for computing the second Stiefel-Whitney class based on parity eigenvalues at inversion invariant momenta, which is used to prove the quantized bulk magnetoelectric response of NLSMs with $Z_2$ monopole charges under a $T$-breaking perturbation.
\end{abstract}

\maketitle

{\it Introduction.|}
Topological semimetals~\cite{
DiracWeyl_review,classification_review,line_review,Herring,3DDirac,Dirac_charge,Dirac_discovery_Cd3As2_1,Dirac_discovery_Cd3As2_2,Dirac_discovery_Na3Bi,unconventional,double_Dirac,Young-Kane,filling-enforced,nonsymmorphic_Zhao-Schnyder,off-centered,hourglass,spinless_hourglass,Murakami,Murakami_classification,nodal,Weyl_pyrochlore1,Weyl_pyrochlore2,Weyl_multilayer,Weyl_discovery1,Weyl_discovery2,Ahn,Park,unified_PT-CP,Z2WeylDirac,realDirac,Z2line,Mikitik,Chiu,Kim,PT_NL_ZrSiS_ncomms,PT_NL_ZrSiS_prb,Fang_inversion,Bzdusek}
are novel states of matter whose band structure features gap-closing points or lines.
Such gapless nodal points or lines are protected by either crystalline symmetry~\cite{Herring,3DDirac,Dirac_charge,Dirac_discovery_Cd3As2_1,Dirac_discovery_Cd3As2_2,Dirac_discovery_Na3Bi,unconventional,double_Dirac,Young-Kane,filling-enforced,nonsymmorphic_Zhao-Schnyder,off-centered,hourglass,spinless_hourglass} or topological invariants~\cite{Murakami,Murakami_classification,nodal,Weyl_pyrochlore1,Weyl_pyrochlore2,Weyl_multilayer,Weyl_discovery1,Weyl_discovery2,Ahn,Park,unified_PT-CP,Z2WeylDirac,realDirac,Z2line,Mikitik,Chiu,Kim,PT_NL_ZrSiS_ncomms,PT_NL_ZrSiS_prb,Fang_inversion,Bzdusek}.
The nodal point (Weyl point) in a Weyl semimetal~\cite{Murakami,nodal,Murakami_classification,Weyl_pyrochlore1,Weyl_pyrochlore2,Weyl_multilayer,Weyl_discovery1,Weyl_discovery2} is a representative example of the latter case.
Due to the quantized monopole charge, Weyl points always exist in pairs~\cite{nodal,Weyl_pyrochlore1,Weyl_pyrochlore2,Weyl_multilayer}.
Moreover, pair creation and annihilation of Weyl points can mediate topological phase transitions between a normal insulator (NI) and a topological insulator in three dimensions (3D)~\cite{Murakami,Murakami_classification,Weyl_pyrochlore1,Weyl_pyrochlore2,Weyl_multilayer,inversion_response,Ramamurthy}. 
Since the origin of the monopole charge is the Berry curvature of complex electronic states, breaking either time reversal $T$~\cite{Weyl_pyrochlore1,Weyl_pyrochlore2,Weyl_multilayer} or inversion $P$~\cite{Murakami,Murakami_classification,Weyl_discovery1,Weyl_discovery2} is a precondition to host a Weyl point~\cite{nodal}. 

However, recent theoretical studies have found that, in the presence of $P$ and $T$ symmetries, a nontrivial monopole charge can exist, carried by a nodal line (NL), when spin-orbit coupling is negligibly weak~\cite{Z2WeylDirac,unified_PT-CP,realDirac,Z2line,Bzdusek,Fang_inversion}.
Here the monopole charge is a $Z_{2}$ number originating from the topology of real electronic states~\cite{Z2WeylDirac,unified_PT-CP,realDirac,Z2line}, which is clearly distinct from the integer monopole charge of Weyl points originating from complex electronic states.
In fact, recently, spinless fermions in $PT$-symmetric systems have received great attention due to the discovery of semimetals with NLs protected by $\pi$ Berry phase~\cite{Mikitik,Chiu,Kim,Z2line}, appearing in various forms including rings~\cite{Ca3P2_1,Ca3P2_2,CaP3family,NCA,BP,AEM1,AEM2,AEM3,CaTe}, crossings~\cite{GN,Cu3PdN,REM,crossing-line}, chains~\cite{chain,chain_experiment,chain_field}, links~\cite{chain_link_Hasan,link,link_SCZ,double-helix,Vanderbilt_link,multi-link,Ezawa,2pi}, knots~\cite{Ezawa,2pi,knot}, nexus~\cite{Volovik,Hyart,triple_point}, and nets~\cite{four-band,net,various}.
However, all the NL belonging to this class do not carry a $Z_{2}$ monopole charge. Because of this, such a NL can exist alone in the Brillouin zone (BZ), which can disappear after shrinking to a point~\cite{Z2line}.
No candidate material has been predicted to host $Z_{2}$-nontrivial NLs ($Z_{2}$NLs) yet.
Although there are preceding theoretical studies on $Z_2$NLs~\cite{Z2line,Bzdusek,Fang_inversion}, the generic feature of the associated band structure topology, which is useful to facilitate material discovery, has not been thoroughly studied.

In this work, we study topological characteristics unique to a nodal line semimetal (NLSM) with $Z_{2}$ monopole charges and propose the first candidate material, ABC-stacked graphdiyne.
In particular, we describe the mechanism for creating $Z_{2}$NLs and the linking structure between them, which originates from the underlying global topological characteristics of real electronic states represented by {\it the second Stiefel-Whitney (SW) class}. 
The linking structure exists between a $Z_{2}$NL near the Fermi energy $E_{F}$ and another NL below $E_{F}$, similar to the linking structure predicted in 5D Weyl semimetal recently~\cite{5DWeyl}.
This demonstrates that, in contrast to the common belief, the topological property of NLSM is determined not only by the local band structure near crossing points at $E_F$ but also by the global topological structure of all occupied bands below $E_{F}$.

\begin{figure}[t!]
\includegraphics[width=8.5cm]{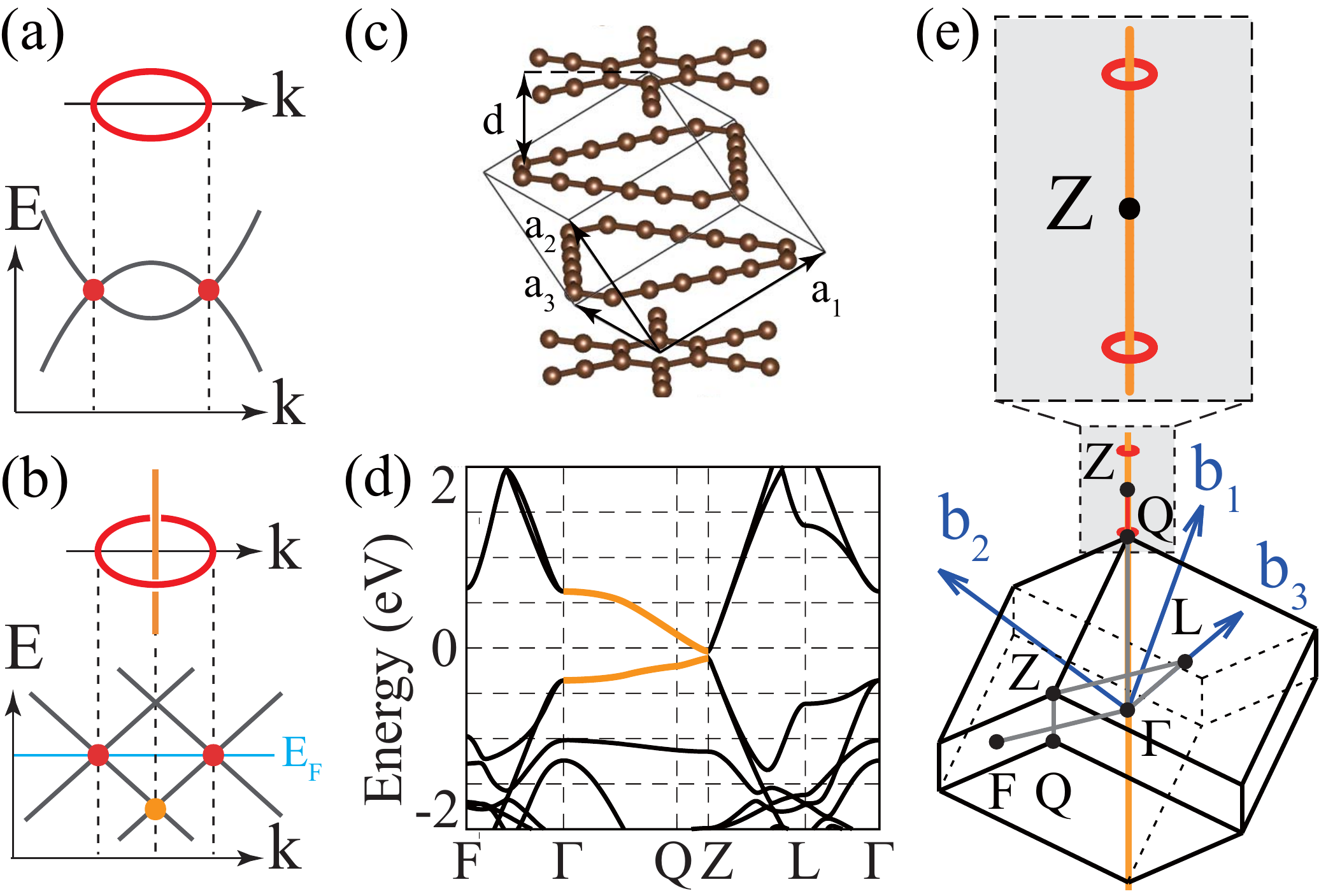}
\caption{
(a) Band structure near a nodal line (NL) with zero $Z_{2}$ monopole charge.
(b) Band structure near a NL carrying a unit $Z_{2}$ monopole charge ($Z_2$NL) linked with another nodal line (NL$^*$) below the Fermi level ($E_{F}$).
(c) Atomic structure of ABC-stacked graphdiyne.
(d) Band structure of ABC-stacked graphdiyne where thick orange lines indicate degenerate NLs above and below $E_{F}$.
(e) The shape of two $Z_2$NLs (red loops) at $E_{F}$ ($E=0$) linked with a NL$^*$ below $E_{F}$ (yellow line) in ABC-stacked graphdiyne.
Here, for clarity, only the NL$^*$ linked with Z2NLs is shown whereas other unlinked NL$^*$s are not plotted.
}
\label{material}
\end{figure}

{\it Band crossing in $PT$-invariant spinless fermion systems.|}
$Z_2$-trivial NLs can be described as follows~\cite{Kim,Z2line}.
Since $(PT)^{2}=+1$ in the absence of spin-orbit coupling, $PT$ operator can be represented by $PT=K$ where $K$ denotes the complex conjugation.
In this basis, the $PT$ invariance of the Hamiltonian, $PTH({\bf k})(PT)^{-1}=H({\bf k})$, requires $H({\bf k})$ to be real. 
Then the effective two-band Hamiltonian near a band crossing point can be written as
$H({\bf k})=f_0({\bf k})+f_1({\bf k})\sigma_x+f_3({\bf k})\sigma_z,$
where $\sigma_{x,y,z}$ are the Pauli matrices for the two crossing bands and $f_{0,1,3}(\bf{k})$ are real functions of momentum $\bf{k}$=$(k_{x},k_{y},k_{z})$.
Because closing the band gap requires only two conditions $f_{1,3}({\bf k})=0$ to be satisfied whereas there are three independent variables $k_{x,y,z}$, the generic shape of band crossing points is a line.

On the other hand, to describe $Z_{2}$NLs, one needs to consider a four-band Hamiltonian as first proposed in~\cite{Z2line}. 
When the reality condition is imposed, $H({\bf k})$ can include three $4\times 4$ anticommuting matrices, which indicates that a 3D massless Dirac fermion can exist.
The Dirac point is stable against the gap opening because the mass terms, which are imaginary, are forbidden. 
However,  there are other allowed real matrix terms that can deform the Dirac point into a NL.
For instance, let us consider the following Hamiltonian introduced in~\cite{Z2line},
\begin{align}
\label{nlsm}
H({\bf k})
&=k_x\sigma_x+k_y\tau_y\sigma_y+k_z\sigma_z+m\tau_z\sigma_z,
\end{align}
where $\tau_{x,y,z}$ and $\sigma_{x,y,z}$ are Pauli matrices.
The energy eigenvalues are $E=\pm\sqrt{k_x^2+\left(\rho\pm |m|\right)^2}$ where $\rho=\sqrt{k_y^2+k_z^2}$.
One can see that the conduction and valence bands touch along the closed loop (a $Z_{2}$NL) satisfying $k_x=0$ and $\rho=|m|$.
Moreover, two occupied bands cross along another line along $\rho=0$ (NL$^*$), which is linked with the $Z_{2}$NL.
Because of this linking, the $Z_{2}$NL is stable and distinct from trivial NLs.
As $m\rightarrow 0$, the linking requires that the $Z_{2}$NL shrinks to a Dirac point.
As $m$ becomes finite after sign reversal, the size of the $Z_{2}$NL increases again. It can never disappear by itself.
Because a single $Z_{2}$NL is stable, only a pair of $Z_{2}$NLs can be created by band inversions.

{\it $Z_{2}$NLs in ABC-stacked graphdiyne.|}
Our first-principles calculations predict that ABC-stacked graphdiyne realizes $Z_2$NLs with the linking structure. ABC-stacked graphdiyne is an ABC stack of 2D graphdiyne layers composed of a sp$^2$-sp carbon network of benzene rings connected by ethynyl chains. [See Fig.~\ref{material}(c).] Recently, Nomura \textit{et. al.}~\cite{graphdiyne} theoretically proposed ABC-stacked graphdiyne as a NLSM. Here we show that the NLs in this material are $Z_2$NLs.
Consistent with~\cite{graphdiyne}, we find NLs occurring off the high-symmetry $Z$ point of the BZ. While the electronic band structure displays band gap along the high-symmetry lines as shown in Fig.~\ref{material}(d), the valence and conduction bands cross off the high-symmetry $\bf k$ points along a pair of closed NLs colored in red in Fig.~\ref{material}(e). Additionally, we find that two topmost occupied bands form another NL [the orange line in Fig.~\ref{material}(e)], which pierces the red NLs, manifesting the proposed linking structure.
Interestingly, the effective four-band Hamiltonian describing ABC-stacked graphdiyne near $E_{F}$ is identical to Eq.(\ref{nlsm})~\cite{graphdiyne}, indicating the generality of our theory.

{\it Double band inversion (DBI).|}
\begin{figure}[t!]
\includegraphics[width=8.5cm]{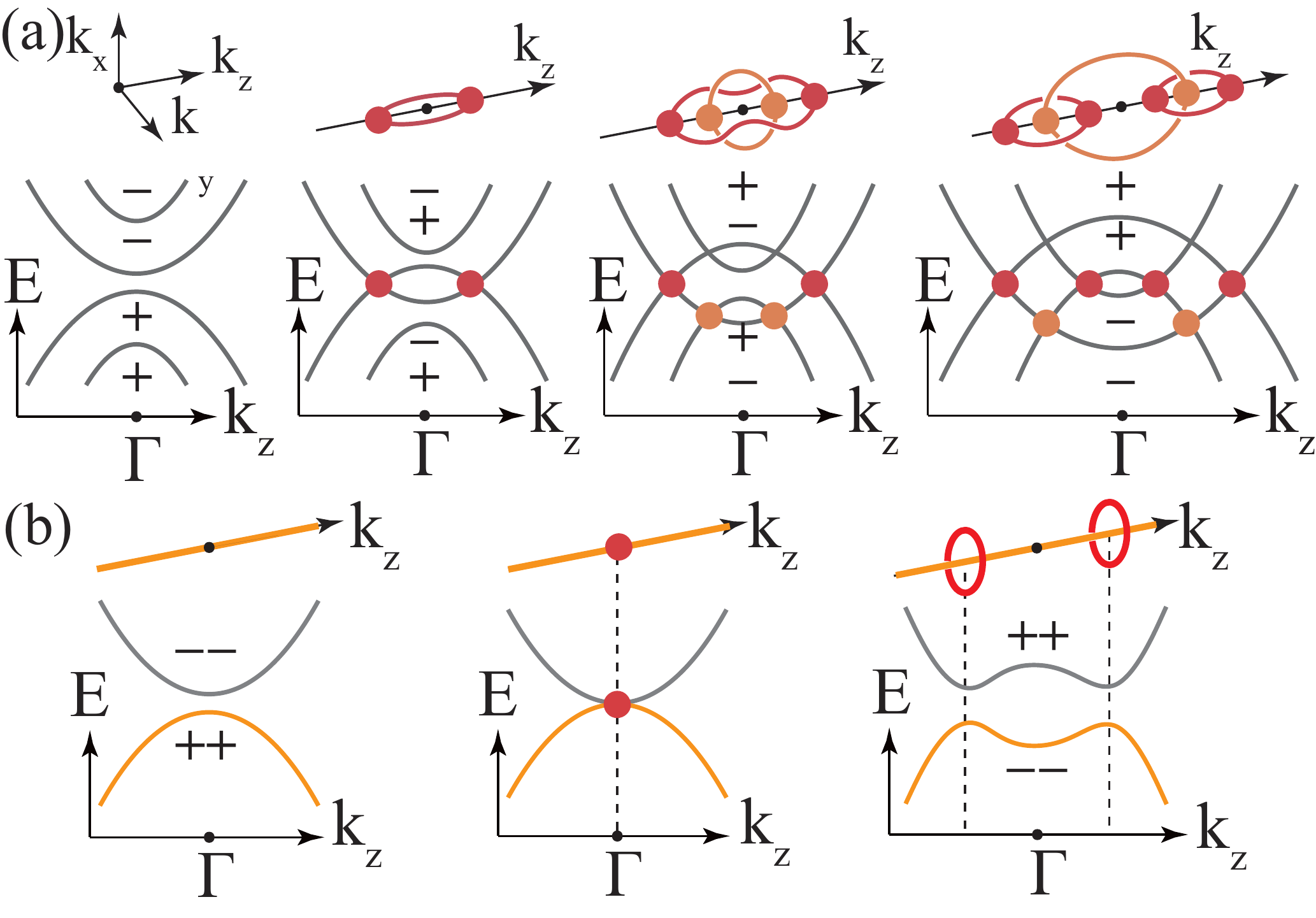}
\caption{
(a) Evolution of band structure during a double band inversion (DBI).
Red (Orange) points and lines indicate the crossing between the conduction and valence bands (two occupied bands). $\pm$ indicate the inversion eigenvalues at the $\Gamma$ point.
(b) A variant of DBI process realized in graphdiyne.
Due to the three-fold rotation symmetry, both the valence and conduction bands are degenerate along the $k_{z}$ axis, thus NL$^*$ exists before $Z_2$NLs are created.
}
\label{DBI}
\end{figure}
Let us illustrate a generic mechanism for a pair creation of $Z_{2}$NLs in inversion symmetric systems, which is comprised of consecutive band inversions, dubbed {\it a double band inversion} (DBI).
For concreteness, we describe a DBI by using the Hamiltonian in Eq.~(\ref{nlsm}) after the replacement $k_z\rightarrow|{\bf k}|^2-M$.
The evolution of the band structure during the DBI is illustrated in Fig.~\ref{DBI}(a) as a function of the parameter $M$.
As we increase $M$ from $M<-|m|$, the first band inversion occurs at $M=-|m|$ between the top valence and bottom conduction bands, creating a trivial NL.
Then, the inversion at $M=0$ between two occupied (unoccupied) bands generates another NL below (above) $E_F$, which we call NL$^*$.
The last band inversion at $M=|m|$ between two inverted bands near $E_{F}$ splits the trivial NL into two $Z_{2}$NLs linked by the NL$^*$ below $E_{F}$ [Fig.~\ref{DBI}(a)].
During the DBI, each occupied (unoccupied) band crosses both of two unoccupied (occupied) bands, which explains why the minimal number of bands required to create a $Z_{2}$NL is four.
In ABC-stacked graphdiyne, both valence and conduction bands are doubly degenerate along the high-symmetry $k_{z}$ axis due to three-fold rotation symmetry, thus NL$^*$ exists from the beginning. In such a system, a single band crossing immediately inverts two occupied and two unoccupied bands having opposite parities, generating a $Z_{2}$NL pair as shown in Fig.~\ref{DBI}(b). In noncentrosymmetric systems, $Z_{2}$NL pair creation occurs in a similar manner, that is, by splitting a trivial NL into a $Z_{2}$NL pair which are linked with another NL below $E_{F}$ as described in~\cite{supp}.

{\it $Z_2$ monopole charge, linking number, and the second Stiefel-Whitney class.|}
Here we give a formal proof for the equivalence between the $Z_{2}$ monopole charge and the linking number, based on the correspondence between the $Z_2$ monopole charge and the second Stiefel-Whitney (SW) class implied by K-theory~\cite{Gomi}.

\begin{figure}[t!]
\includegraphics[width=8.5cm]{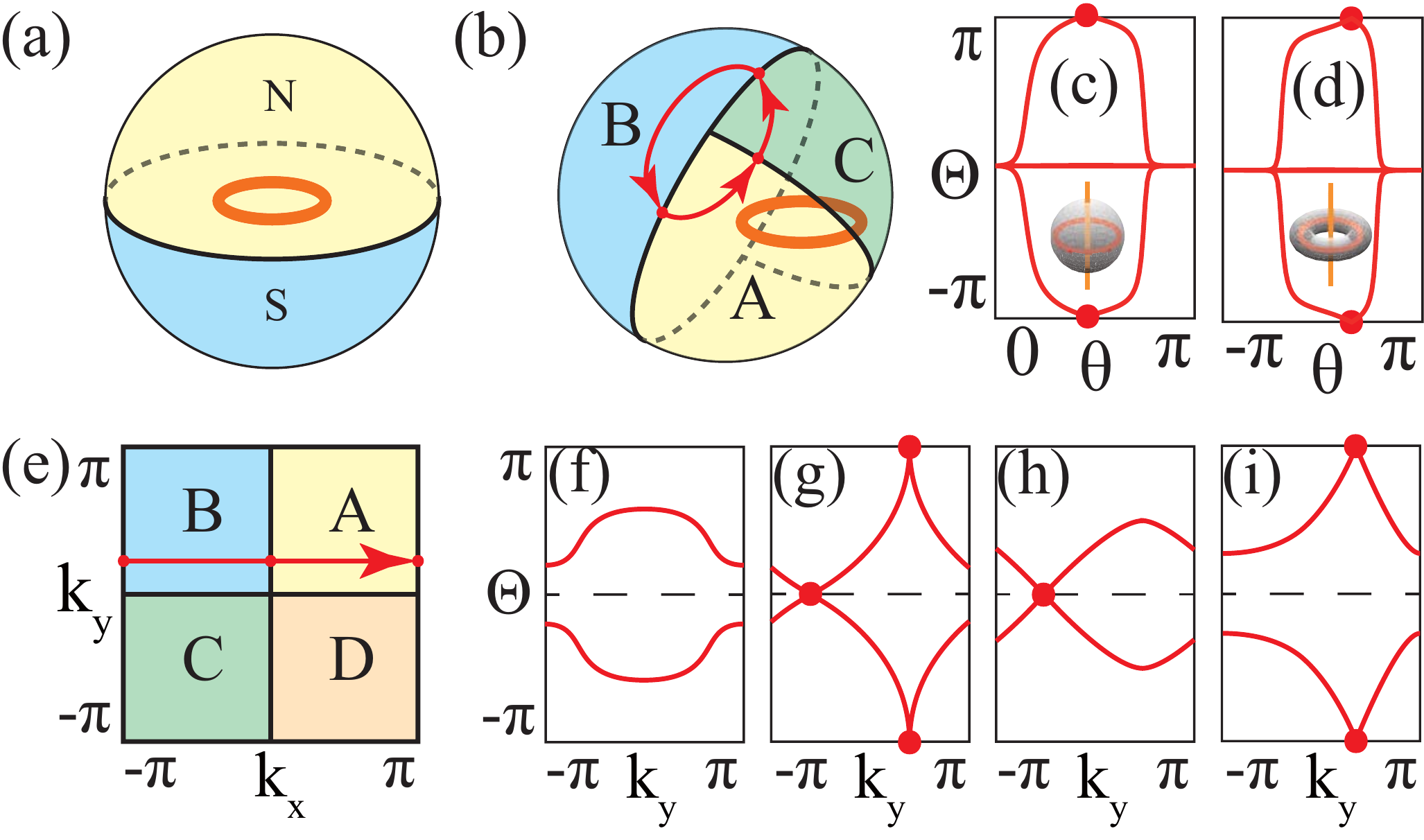}
\caption{
(a, b) The wrapping sphere covered by two or three patches.
(c, d) Wilson loop spectrum for ABC-stacked graphdiyne computed on a sphere or a torus 
wrapping a $Z_{2}$NL. 
(e) A torus covered by four patches.
(f) The Wilson loop spectrum on a torus with $(w_{1,y},w_{2})=(0,0)$ when $N_{\text{occ}}=2$. Similar spectra with $(w_{1,y},w_{2})=(0,1)$, $(1,0)$, $(1,1)$ are shown in (g), (h), and (i), respectively.
}
\label{wilsonloop}
\end{figure}

The $Z_{2}$ invariant was originally defined in~\cite{Z2line} as follows.
First, we take real occupied states by imposing $PT\ket{u_{n\bf k}}=\ket{u_{n\bf k}}$.
Then we consider a sphere surrounding a NL, which is divided into two patches (the northern and southern hemispheres) overlapping along the equator as shown in Fig.~\ref{wilsonloop}(a). One can find smooth real states $\ket{u^{N}_{n\bf{k}}}$ ($\ket{u^{S}_{n\bf{k}}}$) on the northen (southern) hemisphere. On the overlapping circle, $\ket{u^{N,S}_{n\bf{k}}}$ are connected by a smooth transition function $t^{NS}({\bf k})\in \text{SO}(N_{\text{occ}})$ in a way that $\ket{u^{S}_{n\bf{k}}}=t^{NS}_{mn}({\bf k})\ket{u^{S}_{m\bf{k}}}$, where $N_{\text{occ}}$ denotes the number of occupied bands. 
Let us note that, since the real occupied states are orientable on a sphere, transition functions can be restricted to $\text{SO}(N_{\text{occ}})$~\cite{supp}.
The homotopy group $\pi_{1}(\text{SO}(N_{\text{occ}}>2))=Z_{2}$ indicates that there is a $Z_{2}$-type obstruction for defining real smooth state on the sphere, which is nothing but the $Z_{2}$ monopole charge of NLs. 
Because $\pi_{1}(\text{SO}(2))=Z$, the winding number of $t^{NS}({\bf k})$ is an integer invariant when $N_{\text{occ}}=2$.
In this case, the $Z_2$ monopole charge is defined by the parity of the winding number.

Now we make a connection between the $Z_{2}$ monopole charge and the second SW class $w_2$.
$w_{2}$ characterizes the obstruction to lifting transition functions of real occupied states to their double covering group~\cite{pin_structure,DeWitt-Morette_v2,Nakahara}.
When $w_{2}=0$ ($w_{2}=1$), the lifting is allowed (forbidden).
For simplicity, let us first consider the case with $N_{\text{occ}}=2$ so that the transition function $t^{NS}({\bf k})=\exp(i\theta({\bf k})\sigma_{y})$, where $\sigma_{x,y,z}$ are the Pauli matrices for two occupied bands.
When the $Z_{2}$ monopole charge on the sphere is 0 (1), the angle $\theta({\bf k})$ evolves from 0 to $4n\pi$ ($(4n+2)\pi$) with an interger $n$, because $t^{NS}$ is periodic along the equator and has an even (odd) winding number. Now let us ask whether it is possible to take a lift $t^{NS}\rightarrow \tilde{t}^{NS}$ from SO$(2)$ to its double covering group U$(1)$ while the periodicity of $\tilde{t}^{NS}$ is kept. To answer this, one defines a two-to-one mapping $\pi:\text{U}(1)\rightarrow \text{SO}(2)$ by using $\tilde{t}^{NS}(\theta)=\exp(i\frac{\theta}{2})$ and $t^{NS}(\theta)=\exp(i\theta\sigma_{y})$.
Let us note that when $t^{NS}(\theta)$ has an even (odd) winding number with $\theta\in[0,4n\pi]$ ($\theta\in[0,(4n+2)\pi ]$), $\tilde{t}^{NS}(\theta)$ is periodic (non-periodic), thus the lifting from $t^{NS}$ to $\tilde{t}^{NS}$ is well-defined (ill-defined). 
The same argument applies to the case with $N_{\rm occ}>2$~\cite{pin_structure}.
The $Z_{2}$ monopole charge is thus identified with $w_2$.

{To derive the equivalence between $w_{2}$ and the linking number, let us continuously deform the sphere wrapping a NL $\gamma$, by gluing the north and south poles at the center, into a thin torus completely enclosing $\gamma$. As long as the band gap remains finite during the deformation, $w_{2}$ is invariant since the gluing of the north and south poles does not creat a monopole, which is further confirmed numerically as shown in Fig.~\ref{wilsonloop}(c,d).} We assume that the torus is thin enough so that all occupied bands on it are non-degenerate. In this limit, according to the Whitney sum formula~\cite{Hatcher,Hatcher_AT}, $w_{2}$ safisfies the following relations modulo two~\cite{supp}
\begin{align}
w_{2}
=\sum_{n<m}\left[w_{1,\phi}({\cal B}_n)w_{1,\theta}({\cal B}_m)-w_{1,\phi}({\cal B}_m)w_{1,\theta}({\cal B}_n)\right]
\end{align}
where $w_{1,\phi/\theta}({\cal B}_n)$ is the first SW class of the $n$th occupied band ${\cal B}_n$ along the toroidal/poloidal cycle on the torus wrapping $\gamma$.
As shown in~\cite{supp}, the first SW class $w_{1,\phi/\theta}({\cal B}_n)$ corresponds to the Berry phase $\Phi_{n,\phi/\theta}$ of the $n$th band along $\phi/\theta$ calculated in a smooth complex gauge, and it characterizes the orientability of the occupied states.
Through a direct calculation of the Berry phase in a Coulomb gauge, we find that~\cite{supp}
\begin{align}
w_{2}
=\sum_{\tilde{\gamma}_{j}}{\rm Lk}(\gamma,\tilde{\gamma}_{j}),
\end{align}
where ${\rm Lk}(\gamma,\tilde{\gamma}_{j})=\frac{1}{4\pi}\oint_{\gamma}d{\bf k}\times\oint_{\tilde{\gamma}_{j}} d{\bf p}\cdot \frac{{\bf k}-{\bf p}}{|{\bf k}-{\bf p}|^3}$ is the linking number~\cite{Ricca-Nipoti} between $\gamma$ and another NL $\tilde{\gamma}_{j}$ formed by the occupied band degeneracy.
Let us notice that NLs formed between unoccupied bands do not contribute to the linking number because the monopole charge is defined by occupied bands.
For the model in Eq.~(\ref{nlsm}) with $N_{\text{occ}}=2$, $\Phi_{1,\phi}=\pi$, $\Phi_{1,\theta}=\pi$, $\Phi_{2,\phi}=\pi$, and $\Phi_{2,\theta}=0$, so ${\rm Lk}=1$ as expected.

{\it Wilson loop method for computing $w_{2}$.|}
$w_{2}$ can be computed efficiently by using the Wilson loop technique~\cite{Z2line,Bzdusek,Wilson_loop,inversion_Wilson_loop,group_cohomology}.
The relation between the Wilson loop spectrum and the $Z_2$ monopole charge can be proved by using the definition of $w_2$~\cite{Nakahara,DeWitt-Morette_v2} as explicitly shown in~\cite{supp}.
In general, on a 2D closed manifold with coordinates $(\phi,\theta)$, the Wilson loop operator along $\phi$ at a fixed $\theta$ is defined by~\cite{Wilson_loop,inversion_Wilson_loop,group_cohomology}
 $W_{(\phi_0+2\pi,\theta)\leftarrow (\phi_0,\theta)}
=\lim_{N\rightarrow \infty}F_{N-1}F_{N-2}...F_{1}F_{0}$
where $F_{j}$ is the overlap matrix at $\phi_j=\phi_0+2\pi j/N$ with matrix elements $[F_{j}]_{mn}=\braket{u_{m}(\phi_{j+1},\theta)|u_{n}(\phi_{j},\theta)}$, and $\phi_{N}=\phi_0$.
On the wrapping sphere covered by three patches shown in Fig.~\ref{wilsonloop}(b),
the Wilson loop operator $W_{0}(\theta)\equiv W_{(2\pi,\theta)\leftarrow (0,\theta)}$ becomes
$W_{0}(\theta)=
t^{AB}W_{(2\pi,\theta)\leftarrow (\pi,\theta)}
t^{BC}W_{(\pi,\theta)\leftarrow (\pi/2,\theta)}
t^{CA}W_{(\pi/2,\theta)\leftarrow (0,\theta)}$,
where $t^{AB}_{mn}=\braket{u^A_m(0,\theta)|u^B_n(2\pi,\theta)}$, $t^{BC}_{mn}=\braket{u^B_m(\pi,\theta)|u^C_n(\pi,\theta)}$, and $t^{CA}_{mn}=\braket{u^C_m(\pi/2,\theta)|u^A_n(\pi/2,\theta)}$. 
Let us take a parallel-transport gauge defined by
$\ket{u^\alpha_{p;n}(\phi,\theta)}
=[W^{\alpha}_{(\phi,\theta)\leftarrow (\phi^{\alpha}_{0},\theta)}]_{mn}\ket{u^\alpha_m(\phi,\theta)}$, 
where $\phi^{\alpha}_{0}=0,\pi,\pi/2$ for $\alpha=A,B,C$, respectively, and $W^{\alpha}$ is defined with smooth states within the patch $\alpha$.
Then the Wilson loop operator becomes
\begin{align}
W_{0}(\theta)
=W_{p,0}(\theta)
=t^{AB}_p(\theta)t^{BC}_p(\theta)t^{CA}_p(\theta),
\end{align}
where $W_p$ and $t_p$ are the Wilson loop operator and the transition function in the parallel-transport gauge. 
Let ue note that, in this gauge, $W_{0}(\theta)$ is simply given by the product of transition functions along the $\phi$ cycle. 
Since $W_{0}(0,\pi)=1$ due to the consistency condition at triple overlaps~\cite{supp}, the image of the map $W_{0}(\theta)$ for $\theta\in[0,\pi]$ forms a closed loop. Then $w_{2}$ is given by the parity of the winding number of $W_{0}(\theta)$~\cite{supp}, which can be obtained gauge-invariantly from its eigenvalue $\Theta(\theta)$~\cite{Wilson_loop,Bzdusek}.
We apply the Wilson loop technique to ABC-stacked graphdiyne, and find that the Z2NLs carry nontrivial monopole charges. Figure \ref{wilsonloop}(c) shows the first-principles calculations of the Wilson loop spectrum computed on a sphere wrapping a $Z_{2}$NL. The single crossing on the $\Theta=\pi$ line indicates the odd winding number, leading to $w_{2}=1$. Fig.~\ref{wilsonloop}(d) shows that the Wilson loop spectrum computed on a torus is also nontrivial.
These first-principles results confirm the NLSM phase that we proposed here hosted in ABC-stacked graphdiyne.

{\it 2D SW insulator (SWI).|}
Using $w_{2}$ computed on a 2D BZ torus, we can define a new $PT$-invariant 2D topological insulator characterized by $w_2$ when $w_1=0$ (i.e., $w_{1,\phi}=w_{1,\theta}=0$).
To prove this, we consider a 2D BZ torus with coordinates $(\phi,\theta)=(k_{x},k_{y})$ [Fig.~\ref{wilsonloop}(e)].
Then $w_{2}$ is again given by the spectral degeneracy of the Wilson loop $W_{0}(\theta)$ on the torus, as shown in~\cite{supp}.

Let us first consider $N_{\text{occ}}=2$ case.
We calculate $W_0$ along an orientable cycle, because otherwise the Wilson loop spectrum has no stable crossing points such that it does not show the topological property.
One can always choose such an orientable cycle~\cite{supp}.
Then, there are four $Z_2$ homotopy classes of Wilson loop spectra shown in Fig.~\ref{wilsonloop}(f-i).
They are classified by the parity of the number of linear crossing points on $\Theta=0$ and $\Theta=\pi$.
A spectrum corresponds to $w_2=0$ ($w_2=1$) when it has an even (odd) linear crossing points on $\Theta=\pi$.
Fig.~\ref{wilsonloop}(f,g) and \ref{wilsonloop}(h,i) are distinguished by the total number of linear crossing points, which is even (odd) since $w_{1,\theta}=0$ ($w_{1,\theta}=1$)~\cite{supp}.

Notice that the topology of the spectrum in Fig.~\ref{wilsonloop}(h) and (i) differs only by an overall shift of the eigenvalues by $\pi$, whereas those in Fig.~\ref{wilsonloop}(f,g) are invariant under the shift.
This indicates that $w_2$ is independent (dependent) of the unit cell choice when $w_{1,\theta}=0$ $(w_{1,\theta}=1)$, because the Wilson loop eigenvalues correspond to the Wannier centers for insulators~\cite{Wilson_loop}.
Indeed, the same unit cell dependence exists for any even $N_{\text{occ}}$ whereas $w_2$ is independent of the unit cell choice for any odd $N_{\text{occ}}$~\cite{supp}.
Therefore, $w_{2}$ is a well-defined topological invariant when $w_1=0$.
We may call the insulator characterized by $w_2=1$ as a 2D SW insulator (SWI).
This is a new kind of fragile topological phase~\cite{fragile,Disconnected,BuildingBlock}  since it can be trivialized when bands with $(w_{1},w_{2})=(1,0)$ are added.

{\it Topological phase transition.|}
As a sphere wrapping a $Z_{2}$NL can be continuously deformed to two parallel 2D BZs, one with $w_{2}=1$ and the other with $w_{2}=0$, a $Z_2$NL can be considered as a critical state separating a 2D NI and a 2D SWI.
 Accordingly, the pair creation and annihilation of $Z_{2}$NLs can mediate a topological phase transition between a 3D NI and a 3D weak SWI, a vertical stacking of 2D SWIs.
The presence of two NL$^*$s formed between occupied bands clearly distinguishes a 3D weak SWI from a NI.
Interestingly, first-principles calculations show that ABC-stacked graphdiyne turns into a 3D weak SWI after pair annihilation of $Z_{2}$NLs under about 3 $\%$ of a uniaxial tensile strain applied along the $z$ direction. [See~\cite{supp}.]

{\it Discussion.|}
Let us discuss about measurable properties of NLSM with $Z_2$NLs.
Unfortunately, its surface states are generally not robust due to $P$ breaking on the surface~\cite{Z2line}.
Nevertheless, our study suggests that observing the linking structure using angle-resolved photoemission spectroscopy~\cite{bulkARPES} can provide strong evidence for $Z_2$NLs.
Moreover, the bulk magnetoelectric response under magnetic field can provide another evidence.
When $P$ and $T$ are individually symmetries of the system, the number of pairs of $Z_{2}$NLs ($N_{mp}$) can be determined from the inversion eigenvalues of the occupied bands at inversion-invariant momenta (IIM).
Since a DBI changes two inversion eigenvalues at an IIM, $N_{mp}$ is given by the sum of the number of negative eigenvalue pairs over all IIM~\cite{Fang_inversion,supp}.
Let us note that, in $P$-invariant insulators with broken $T$, two times magnetoelectric polarizability $2P_{3}$ is determined by inversion eigenvalues in the same way as $N_{mp}$ is~\cite{inversion}.
This implies that a NLSM with an odd number of $Z_{2}$NL pairs turns into an axion insulator, which can host chiral hinge modes along the domain wall~\cite{RotationAnomaly,Khalaf,Khalaf-Po-Vishwanath-Watanabe}, when the band gap is opened due to a $T$-breaking perturbation such as magnetic field~\cite{supp}.
We believe that the theoretical prediction given in the present work can be experimentally tested in ABC-stacked graphdiyne in near future.

{\it Acknowledgment.|}
J.A. was supported by IBS-R009-D1.
B.-J.Y. was supported by the Institute for Basic Science in Korea (Grant No. IBS-R009-D1) and Basic Science Research Program through the National Research Foundation of Korea (NRF) (Grant No. 0426-20170012, No.0426-20180011), the POSCO Science Fellowship of POSCO TJ Park Foundation (No.0426-20180002), and the U.S. Army Research Office under Grant Number W911NF-18-1-0137.
Y.K. was supported by Institute for Basic Science (IBS-R011-D1) and NRF grant funded by the Korea government (MSIP) (No. S-2017-0661-000). 
D. K. was supported by Samsung Science and Technology Foundation under Project Number SSTF-BA1701-07 and Basic Science Research Program through NRF funded by the Ministry of Education (NRF-2018R1A6A3A11044335). The computational calculations were performed using the resource of Korea institute of Science and technology information (KISTI).
We appreciate the helpful discussions with Yoonseok Hwang, Sungjoon Park, Eunwoo Lee, Ken Shiozaki, Haruki Watanabe, and Akira Furusaki.

{\it Note Added.|}
Recently, fragile topology in $Z_2$-nontrivial NLSMs was also explored in \cite{moTe2}; the results of that work is consistent with our conclusions.

\setcounter{section}{0}
\setcounter{figure}{0}
\setcounter{equation}{0}

\renewcommand{\thefigure}{S\arabic{figure}}
\renewcommand{\theequation}{S\arabic{equation}}
\renewcommand{\thesection}{SI \arabic{section}}

\tableofcontents

\section{Outline}

In this Supplemental Material, we provide detailed derivations of the results shown in the main text.
In Sec.~\ref{sec.1stSW} and \ref{sec.2ndSW}, we connect the known definition of the 1D and 2D topological invariants (i.e., the Berry phase and the $Z_2$ monopole charge) to the first and second Stiefel-Whitney classes.
For a rigorous treatment of the second Stiefel-Whitney class, we introduce the concept of the \v{C}ech cohomology in  Sec.~\ref{sec.2ndSW}, which can be used to check the Whitney sum formula. 
This formula is crucial for proving the equivalence between the linking number of nodal lines and the second Stiefel-Whitney class and for deriving the Wilson loop method in Sec.~\ref{sec.Wilson}.
In Sec.~\ref{sec.Wilson}, we provide a simple formula for calculating the second Stiefel-Whitney class from inversion eigenvalues. The derivation of this formula is based on the flux integral formula proposed in Ref.~\onlinecite{realDirac}, which we review in Sec.~\ref{sec.Euler}.
In Sec.~\ref{sec.Wilson}, we demonstrate various types of pair creations of $Z_2$ monopoles and the corresponding creation of the linking structure. We also explain the double band inversion process in more detail.
Finally, in Sec.~\ref{sec.material}, we provide the detailed information of the first principles calculations of the ABC-stacked graphdiyne that was not presented in the main text.
In particular, we show the Wilson loop spectrum which verifies that a tensile strain can induce the topological phase transition in ABC-stacked graphdiyne from a $Z_2$-nontrivial nodal line semimetal to a 3D weak Stiefel-Whitney insulator.

\section{The First Stiefel-Whitney Class}
\label{sec.1stSW}

The topological phases of $PT$-symmetric one-dimensional (1D) systems are classified by the Berry phase: there are two distinct classes with 0 and $\pi$ Berry phase modulo $2\pi$.
In 3D systems, the Berry phase serves as a topological invariant protecting nodal lines.
As a nodal line generates $\pi$ Berry phase over any closed loop encircling a line segment, it is topologically protected when the system is $PT$-symmetric.
Here, we show how the Berry phase is related to the global structure of quantum states.
While quantum states $\ket{u_{n\bf k}}$ can always be smoothly defined over a 1D submanifold in the Brillouin zone, they may not be smoothly defined after we require the reality condition $PT\ket{u_{n\bf k}}=\ket{u_{n\bf k}}$ because of the topological obstruction characterized by the topological invariant so-called the first Stiefel-Whitney class.
In this section, we show that the Berry phase in (complex) smooth gauges is identical to the first Stiefel-Whitney class in real gauges.
This result will be useful in the next section for deriving the linking number from the 2D topological invariant.

\subsection{Quantization of Berry phase due to $PT$ symmetry}

Let us first review the quantization of Berry phase in the presence of $PT$ symmetry.
The Berry connection $A_{mn}({\bf k})=\braket{u_{m\bf k}|i\nabla_{\bf k}|u_{n\bf k}}$ satisfies the following $PT$ symmetry constraint
\begin{align}
A({\bf k})
&=-(G^{\dagger}({\bf k})A({\bf k})G({\bf k})+G^{\dagger}({\bf k})i\nabla_{\bf k}G({\bf k}))^*,
\end{align}
which follows from $PT\ket{u_{n\bf k}}=G_{mn}({\bf k})\ket{u_{m\bf k}}$.

Suppose a set ${\cal B}_i$ of bands is isolated from the other bands by the band gap.
By taking a trace over ${\cal B}_i$, we have an abelian Berry connection
\begin{align}
{\rm Tr}A({\bf k})
&=-\frac{i}{2}\nabla_{\bf k}\log \det G({\bf k}),
\end{align}
where the trace and determinant are defined over ${\cal B}_i$.
By integrating the abelian Berry connection, we have a quantized phase
\begin{align}
\oint_C{\rm Tr}A({\bf k})=-\frac{i}{2}\oint_C\nabla_{\bf k}\log \det G({\bf k})=p\pi,
\end{align}
where $p=\oint_C \nabla_{\bf k}\phi$ is the winding number of $\det G=e^{i\phi}$.
Because of this quantization of the Berry phase, the $\pi$ Berry phase of a nodal line cannot be adiabatically changed to zero.

\subsection{Berry phase and the first Stiefel-Whitney class}
\label{subsec.orientation}

Now we review the definition of the orientation of vector spaces and quantum states, and then we show that the Berry phase in a smooth complex gauge characterizes the orientability of quantum states in a real gauge.
This property allows the Berry phase computed with smooth complex states to be interpreted as the first Stiefel-Whitney class of real quantum states.

The orientation of a real vector space refers to the choice of ordered basis.
Any two ordered bases are related to each other by a unique nonsingular linear transformation.
When the determinant of the transformation matrix is positive (negative), we say the bases have the same (different) orientation.
After choosing an ordered reference basis $\{v_1,v_2,...\}$, the orientation of another basis $\{u_1,u_2,...\}$ is specified to be positive (negative) when the basis have the same (different) orientation with respect to the reference basis.

Real quantum states in the Brillouin zone are real unit basis vectors defined at each momentum (In other words, quantum states have the structure of a real vector bundle over the Brillouin zone.).
The basis can be defined smoothly in a local patch, but may not be smooth over a closed submanifold $\cal M$ of our interest.
We say quantum states are orientable over $\cal M$ when local bases can be glued only with transition functions with positive determinant, i.e., all transition functions are orientation-preserving.
When quantum states are orientable, they are classified into two classes with the positive and negative orientation in the same way as the real vector spaces are.

Because the orientability of real quantum states is determined by their the global structure, it is encoded in the form of a topological invariant.
Over a closed 1D manifold, the $Z_2$ topological invariant which measures the orientability of real quantum states is the first Stiefel-Whitney class $w_1$~\cite{Hatcher}.
Real quantum states are orientable (non-orientable) when $w_1=0$ ($w_1=1$).

Although the term {\it the first Stiefel-Whitney class} may be unfamiliar to readers, it is in fact equivalent to the well-known quantized Berry phase defined in a smooth complex basis.
It can be observed by investigating how the $\pi$ Berry phase computed with smooth complex states affects the real states given by a gauge transformation. 
In order to make states real, the $\pi$ Berry phase should be eliminated by a local phase rotation of the states because the diagonal components of the Berry connection are zero when the states are real.
\begin{align}
0=\int^{2\pi}_0dk\;{\rm Tr}\tilde{A}=\int^{2\pi}_0dk({\rm Tr}A+i\nabla_{k}\log \det g),
\end{align}
where $\tilde{A}_{mn}=\braket{\tilde{u}_{m\bf k}|i\nabla_{\bf k}|\tilde{u}_{n\bf k}}$ and $A_{mn}=\braket{u_{m\bf k}|i\nabla_{\bf k}|u_{n\bf k}}$ are the Berry connection given by real states $\ket{\tilde{u}_{n\bf k}}$ and by smooth complex states $\ket{u_{n\bf k}}$, respectively, and $g$ is the gauge transformation matrix defined by $\ket{\tilde{u}_{nk}}=g_{mn}(k)\ket{u_{mk}}$.
Integrating the $\log \det g$ term, we have
\begin{align}
\frac{\det g(2\pi)}{\det g(0)}=\exp \left[-i \int^{2\pi}_0dk\;{\rm Tr}A\right].
\end{align}
Thus, when the total Berry phase is nontrivial, the real states $\ket{\tilde{u}_{nk}}$ require an orientation-reversal to transit from $k=2\pi$ to $k=0$, because it follows from $\ket{u_{n(2\pi)}}=\ket{u_{n(0)}}$ that $\ket{\tilde{u}_{n(2\pi)}}=[g^{-1}(0)g({2\pi})]_{mn}\ket{\tilde{u}_{m(0)}}$.
We conclude that the first Stiefel-Whitney class $w_1$ for a closed curve $C$ in the Brillouin zone is
\begin{align}
w_1|_C=\frac{1}{\pi}\oint_C d{\bf k}\cdot {\rm Tr}{\bf A}({\bf k}),
\end{align}
where $A$ is the Berry connection calculated in a complex smooth gauge.

As an example, let us consider the Su-Schuriffer-Heeger (SSH) model~\cite{SSH} in a real basis
\begin{align}
H_{\rm SSH}=\sin k \sigma_x+(t+\cos k)\sigma_z.
\end{align}
This Hamiltonian is symmetric under $PT=K$.
It is well-known that this system describes an insulator which is topologically trivial when $|t|> 1$ and nontrivial when $|t|<1$.
Let us see how the topology manifests on the occupied state.
The occupied state is
\begin{align}
\label{SSH_VB}
\ket{u_v}=\frac{e^{i\phi_v(k)}}{N(k)}
\begin{pmatrix}
\sin k\\
t+\cos k-\sqrt{(t+\cos k)^2+\sin^2k}
\end{pmatrix},
\end{align}
where $\phi_v(k)$ is an arbitrary overall phase factor and $N(k)$ is a positive normalization factor.

\begin{figure}[t!]
\includegraphics[width=8.5cm]{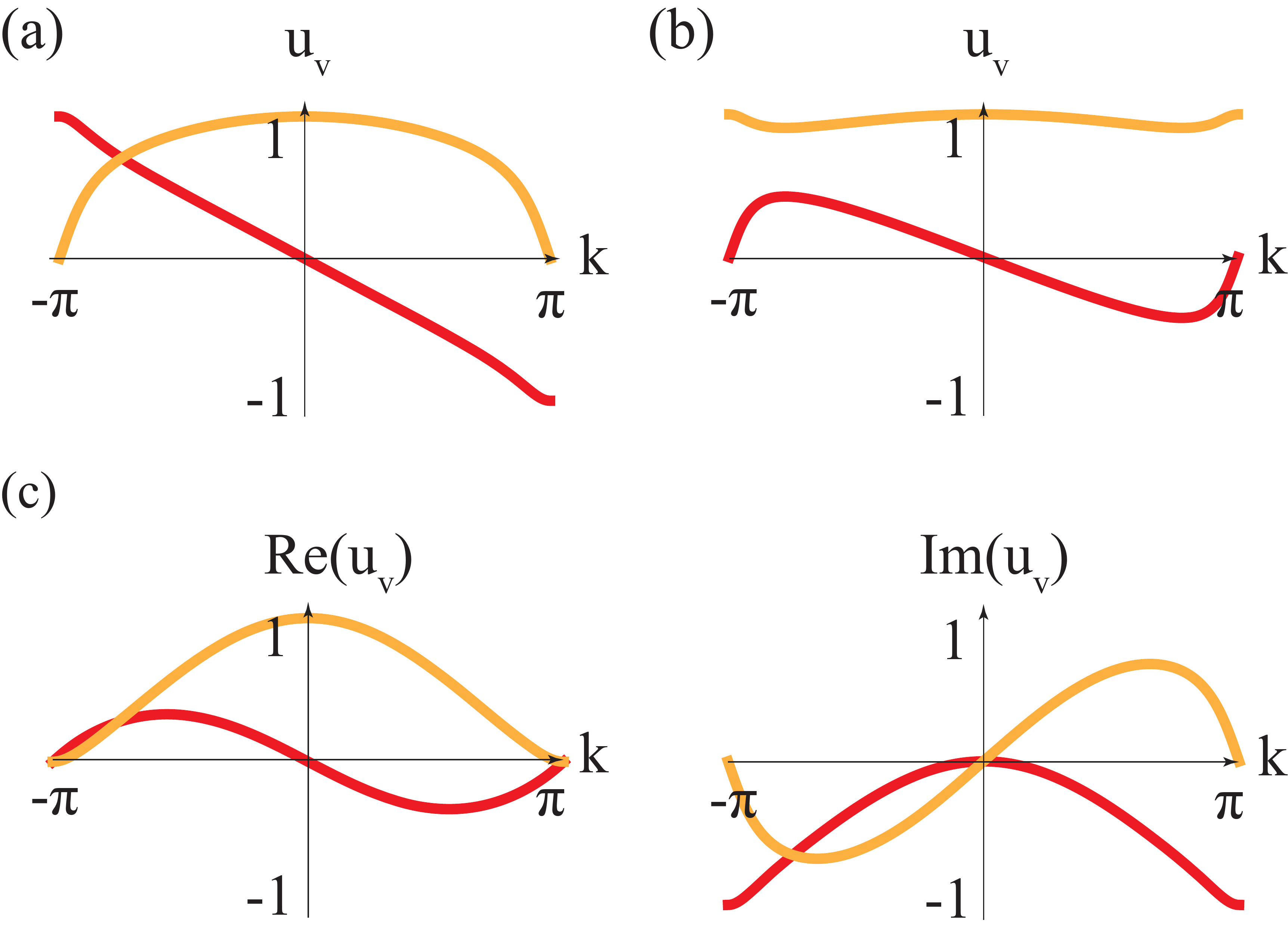}
\caption{Amplitude of the occupied state in the SSH model.
Red and orange lines show the first and second component of the occupied state in Eq.~(\ref{SSH_VB}).
(a,b): Reality condition, $\phi_v(k)=0$ or $\pi$ modulo $2\pi$, is imposed on the occupied state.
The occupied state is anti-periodic for
(a) $t=0.7$
and periodic for
(b) $t=1.3$.
(c): $\phi_v(k)=k/2$ and $t=0.7$. When a complex phase is allowed, the occupied state can be made smooth even for $|t|<1$.
}
\label{orientation}
\end{figure}

First, we impose the reality condition on the occupied state over the whole Brillouin zone, i.e., $e^{i\phi_v(k)}=\pm 1$ at each $k$.
As shown in Fig.~\ref{orientation}(a), when $|t|<1$, the real occupied state can be made smooth over $-\pi<k<\pi$, but the boundaries $k=\pm \pi$ should be glued with an orientation-reversing transition function.
The occupied state is thus non-orientable.
In contrast, it is orientable when $|t|>1$ as shown in Fig.~\ref{orientation}(b).

Next, we relax the reality condition on the occupied state.
By taking $\phi_v(k)=k/2$, we can make the occupied state globally smooth even when $|t|<1$ [See Fig.~\ref{orientation}(c)].
The cost of taking smoothness is to have a nontrivial Berry phase.
In this gauge, we have $A=\braket{u_v|i\nabla_k|u_v}=1/2$  and thus  $\int^{\pi}_{-\pi}dk A=\pi$.

In fact, the first Stiefel-Whitney class determines the orientability of real states even in higher dimensions~\cite{Hatcher}.
From the analysis in 1D, we find
\begin{align}
\frac{\det g({\bf q})}{\det g({\bf p})}=\exp \left[-i \int^{\bf q}_{\bf p}d{\bf k}\cdot {\rm Tr}{\bf A}({\bf k})\right].
\end{align}
$\det g$ is globally smooth when the Berry phase is trivial over every closed cycle.
Otherwise, $\det g$ becomes discontinuous at some points so that real states are non-orientable as in the 1D case.
Thus, real states are orientable over an arbitrary dimensional closed manifold $\cal M$ if and only if the total Berry phase, which is calculated in a smooth complex gauge, is trivial for any 1D closed loop in $\cal M$.

\section{The Second Stiefel-Whitney Class}
\label{sec.2ndSW}

$PT$-symmetric two-dimensional systems are topologically classified (according to the K-theory) by a $Z_2$ invariant called the second Stiefel-Whitney class~\cite{Gomi}.
In three-dimensional systems, the second Stiefel-Whitney class defines the $Z_2$ monopole charge of a gap-closing object when the invariant is defined over the 2D manifold enclosing the object.
Historically, it was first discovered by Ho$\check{\rm r}$ava~\cite{Horava} using K-theory that gap-closing points in three-dimensional systems described by real quantum states are classified by a $Z_2$ invariant.
Subsequently, Dirac points in spinless $PT$-symmetric systems were found to be the gap-closing points carrying the nontrivial $Z_2$ monopole charges by Morimoto and Furusaki~\cite{Z2WeylDirac} (and also by Zhao et al.~\cite{unified_PT-CP,realDirac}).
More recently, however, Fang {\it et al.}~\cite{Z2line} showed that the Dirac point is not stable against small perturbations: it deforms into a nodal line which still carries the $Z_2$ monopole charge.
In this section, we show that the known expression for the $Z_2$ monopole charge corresponds to the definition of the second Stiefel-Whitney class as the obstruction to the spin structure.
We also present a formal definition of the second Stiefel-Whitney class in terms of the \v{C}ech cohomology.
This definition will be useful when we identify the $Z_2$ monopole charge with the topology of the Wilson loop spectrum in the next section.
It is also useful for demonstrating a nontrivial property of the second Stiefel-Whitney class, so-called the Whitney sum formula.
Then, by using the Whitney sum formula, we show that the $Z_2$ monopole charge of a nodal line is identical to its linking number modulo two.
Finally, we comment on the flux integral form of the second Stiefel-Whitney class.

\subsection{Obstruction to Spin and Pin structures}

Here, we begin by recalling how spinors and pinors are defined in arbitrary dimensions.
The spin and pin structures for real quantum states are defined in an analogous way as physical spinors are defined, although they are independent of the physical spin.
Then, we show that the nontrivial $Z_2$ monopole charge of a nodal line induces an obstruction to the existence of the spin structure over a sphere enclosing the nodal line.
This identifies the $Z_2$ monopole charge with the second Stiefel-Whitney class $w_2$, because $w_2$ is the topological invariant characterizing the obstruction to the existence of a spin structure of real orientable quantum states~\cite{DeWitt-Morette_v2}.

\subsubsection{Spin and Pin groups}

Spinors in three spatial dimensions are the objects transforming under the double covering group of the orthogonal group $\text{SO}(3)$.
Let us recall that spinors are the objects that transforms only half under spatial rotations in the following sense.
For an object with azimuthal angular momentum $J_z$, the $\Theta$ rotation around the $z$-axis is represented by
\begin{align}
R_{J,z}(\Theta)=\exp(-iJ_z\Theta).
\end{align}
A spinor has a half-integer $J_z$, while a vector or a tensor has an integer $J_z$.
Accordingly, while the vectors and tensors are invariant under any $2\pi$ rotation, spinors get a phase $(-1)$ when the system is rotated by $2\pi$ and come back to their original state after another $2\pi$ rotation.
Therefore, the group of rotations for spinors are twice larger than that for vectors.
The former and the latter are $\text{SU}(2)$ and $\text{SO}(3)$, and
the two-to-one mapping $\pi:\text{SU}(2)\rightarrow \text{SO}(3)$ is given by 
$
\pi:\exp(-i\Theta_iS_i)\rightarrow \exp(-i \Theta_iL_i),
$
where $S_{i=1,2,3}=\sigma_i/2$, and $(L_{i=1,2,3})_{jk}=-i\epsilon_{ijk}$.
The double covering group of $\text{SO}(3)$, which is $\text{SU}(2)$, is called the spin group $\text{Spin}(3)$, because it is responsible for the transformation of spinors under rotations in real space.



Similary, spinors in arbitrary spatial dimensions $d$ are defined as the objects transforming under $\text{Spin}(d)$, the double covering group of $\text{SO}(d)$.
In general, the two-to-one mapping $\pi:\text{Spin}(d)\rightarrow \text{SO}(d)$ is given by
\begin{align}
\pi:\exp(-i\Theta_{\mu\nu}S_{\mu\nu})\rightarrow \exp(-i\Theta_{\mu\nu}L_{\mu\nu}),
\end{align}
where $\mu,\nu=1,...,d$, $\Theta_{\mu\nu}$ is the angle of rotation in the $x_\mu-x_\nu$ plane,
$
S_{\mu\nu}
=-\frac{i}{4}(\Gamma_{\mu}\Gamma_{\nu}-\Gamma_{\nu}\Gamma_{\mu}),
$
$
(L_{\mu\nu})_{\alpha\beta}
=-i(\delta_{\mu\alpha}\delta_{\nu\beta}-\delta_{\mu\beta}\delta_{\nu\alpha}),
$
and $\Gamma_{\mu}$ is the $2^{[d/2]}\times 2^{[d/2]}$ Gamma matrix satisfying the Clifford algebra $\{\Gamma_{\mu},\Gamma_{\nu}\}=2\delta_{\mu\nu}$, where the bracket is the greatest integer function.

We can also construct a spinor which transforms under the double covering group of $\text{O}(d)$ not only that of $\text{SO}(d)$.
A spinor having this property is called a pinor, and the double covering group of $\text{O}(d)$ is called $\text{Pin}(d)$ 
\footnote{
Notice that $\text{O}(d)$ is written by dropping a $S$ from $\text{SO}(d)$. 
[S]pinor transforms under the double covering group $\text{[S]pin}(d)$ of $\text{[S]O}(d)$.
}.
Because $\text{O}(d)$ can be obtained by adding a mirror operation, e.g., the $x_1$-mirror operation $M_{x_1}:x_1\rightarrow -x_1$ to $\text{SO}(d)$, it is enough for constructing $\text{Pin}(d)$ to consider the lift of $M_{x_1}$ to its double covering group. 
There are two choices of the two-to-one mapping~\cite{Witten_RMP}
\begin{align}
\pi^+: \tilde{M}_{x_1}(=\pm \Gamma_1) &\rightarrow  M_{x_1},\notag\\
\pi^-:\tilde{M}_{x_1}(=\pm i\Gamma_1) &\rightarrow  M_{x_1},
\end{align}
where $(\tilde{M}_{x_1})^2=I$ and $-I$, respectively, for $\pi^+$ and $\pi^-$ (we use $I$ to denote the identity element in the double covering group, while we use 1 for the original group.).
The double covering group of $\text{O}(d)$ is called $\text{Pin}^+(d)$ in the former and $\text{Pin}^-(d)$ in the latter.

\subsubsection{Spin and Pin structures}

Now we explain what are the spin and pin structures of real quantum states in the Brillouin zone (i.e., spin and pin structures on real vector bundles).
Let us recall that the transition function is defined by
\begin{align}
\ket{u^B_{n\bf k}}=t^{AB}_{mn}({\bf k})\ket{u^A_{m\bf k}}
\end{align}
on an overlap $A\cap B$ of two open covers $A$ and $B$.
$t^{AB}$ belongs to $\text{O}(N_{\text{occ}})$ for real occupied states $\ket{u_{m\bf k}}$, where $N_{\text{occ}}$ is the number of occupied bands.
In analogy to defining a pinor, 
we may define pinor states $\ket{\Psi_{\alpha \bf k}}$.
They are smooth over each cover of  $\ket{u_{m\bf k}}$, and their transition function $\tilde{t}^{AB}$ is given by lifting $t^{AB}$ to its double covering group $\text{Pin}^{\pm}(N_{\text{occ}})$.
\begin{align}
\ket{\Psi^B_{\beta\bf k}}=\tilde{t}^{AB}_{\alpha\beta}({\bf k})\ket{\Psi^A_{\alpha\bf k}}
\end{align}
over $A\cap B$, where $\pi(\tilde{t}^{AB})=t^{AB}$ for the two-to-one projection $\pi:\text{Pin}^{\pm}(N_{\text{occ}})\rightarrow \text{O}(N_{\text{occ}})$.
We have two choices for $\tilde{t}^{AB}({\bf k})$ over each overlap $A\cap B$ because we are free to choose the sign of it.
A choice of the signs over all overlaps is called a pin structure~\cite{Witten_RMP}.
The pin structure determines the topology of pinor states.
When the states $\ket{u_{m\bf k}}$ are orientable, transition functions can be confined to $\text{SO}(N_{\text{occ}})$ over all overlaps.
Then, we may define a spin structure rather than a pin structure.
However, real occupied states may not admit a pin or spin structure due to an obstruction coming from the global structure.
This obstruction for a pin$^+$ or spin structure (a pin$^-$ structure) is characterized by the topological invariant called the second Stiefel-Whitney class $w_2$ (the dual second Stiefel-Whitney class $\overline{w}_2$).
A pin$^+$ or spin structure (a pin$^-$ structure) exists if and only if $w_2=0$ ($\overline{w}_2=0$)~\cite{DeWitt-Morette_v2}.
Although there are two obstruction invariants $w_2$ and $\overline{w}_2$ in principle, when we consider 2D closed submanifolds of the 3D Brillouin zone, we have a unique obstruction invariant because the second Stiefel-Whitney class $w_2$ is identical to its dual $\overline{w}_2$ on those manifolds~
\footnote
{
The dual second Stiefel-Whitney class is defined by $\overline{w}_2=w_2+w_1\smile w_1$, where $\smile$ is the cup product~\cite{DeWitt-Morette_v2}.
We have $\overline{w}_2=w_2$ over orientable surfaces, because a closed orientable surface is a sphere, a torus with an arbitary genus, or a connected sum of them according to the classification theorem of surfaces~\cite{Munkres}, and the cup product $w_1\smile w_1=0$ is trivial on those surfaces~\cite{Hatcher_AT}.
Moreover, every closed surface in the 3D Brillouin zone is orientable, because non-orientable closed surfaces cannot be embedded in the three-torus $T^3$~\cite{embedding}.
It follows that $\overline{w}_2=w_2$ in the 3D Brillouin zone.
}.

\subsubsection{$Z_2$ monopole charge}

As the first Stiefel-Whitney class is associated with the one-dimensional topological invariant of a nodal line, the second Stiefel-Whitney class is also associated with the topological invariant of a nodal line.
It is the two-dimensional topological invariant called the $Z_2$ monopole charge.
We now show the $Z_2$ monopole charge of a nodal line corresponds to the second Stiefel-Whitney class, because the nontrivial $Z_2$ monopole charge forbids the existence of the spin structure on a sphere enclosing the nodal line.

The $Z_2$ monopole charge of a nodal line is defined over a sphere enclosing the nodal line.
It is defined by the winding number of the transition function at the overlapping region of two patches of the sphere~\cite{Z2line}.
Let $t^{AB}$ be the transition function between two patches $A$ and $B$ defined over the overlapping region: $\ket{u^B_n({\bf k})}=t^{AB}_{mn}({\bf k})\ket{u^A_m({\bf k})}$ for ${\bf k}\in A\cap B$.
We restrict the transition function to $\text{SO}(N_{\text{occ}})$, which is possible because every loop on a sphere is contractible to a point such that the first Stiefel-Whitney class is trivial.
Then we see that the winding number of $t^{AB}$ along a loop in $A\cap B$ gives a $Z_2$ number because $\pi_1(\text{SO}(N_{\text{occ}}))=Z_2$ for $N_{\text{occ}}>2$.
This $Z_2$ number is the $Z_2$ monopole charge.
When the number of occupied bands is two, the winding number is integer-valued because $\pi_1(\text{SO}(2))=Z$.
In this case, the $Z_2$ monopole charge is defined by the parity of the winding number.

We can show that this $Z_2$ invariant characterizes the obstruction to constructing a spin structure over the wrapping sphere.
For simplicity, we take a gauge where the transition function $t^{AB}({\bf k})$ is an identity at some ${\bf k}={\bf k}_0\in A\cap B$.
Then, it evolves from the identity to a $4\pi N$ rotation ($2\pi(2N+1)$ rotation) for an integer $N$ along a loop containing ${\bf k}_0$ in $A\cap B$ when the $Z_2$ monopole charge is trivial (nontrivial), because the generator of the homotopy group $\pi_1(\text{SO}(N_{\text{occ}}))$ is the path from the identity to a $2\pi$ rotation~\cite{pin_structure,Prasolov}.
While the $2\pi$ rotation and the identity are identical as $\text{SO}(N_{\text{occ}})$ elements, they are not identical as ${\rm }\text{Spin}(N_{\text{occ}})$ elements.
Therefore, the transition function is well-defined over the overlap $A\cap B$ only as a $\text{SO}(N_{\text{occ}})$ element when the $Z_2$ monopole charge is nontrivial.
On the other hand, no obstruction arises when the $Z_2$ monopole charge is trivial because a $4\pi$ rotation is identical to the identity element even as a $\text{Spin}(N_{\text{occ}})$ element.
Thus, the $Z_2$ monopole charge is identical to the second Stiefel-Whitney class over the enclosing sphere.

\subsection{\v{C}ech cohomology}

\begin{figure}[t!]
\includegraphics[width=8.5cm]{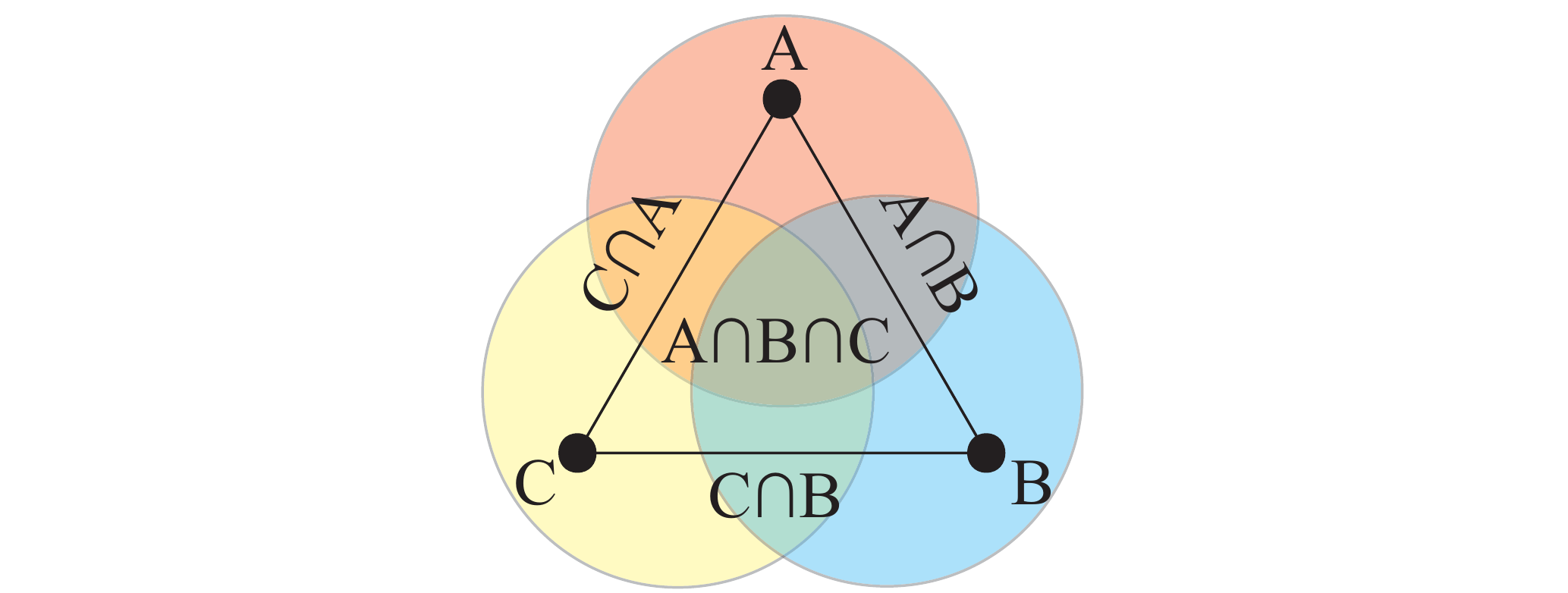}
\caption{Geometric structure of patches.
Patches, overlaps, and triple overlaps on a manifold can be interpreted as vertices, edges, and faces, respectively.
}
\label{Cech}
\end{figure}

Here we review how to formally define the second Stiefel-Whitney class as a \v{C}ech cohomology class~\cite{DeWitt-Morette_v2,Nakahara}.
In other words, we define it as a two-dimensional integral over the geometric structure (semi-simplicial complex) constructed from the patches and their overlaps on the original manifold [See Fig.~\ref{Cech}].
We consider a covering whose geometric structure is topologically equivalent to the original manifold. 
The value of the function to be integrated will be assigned according to whether the consistency conditions of transition functions are satisfied after they are lifted to the double covering group.
After we define the second Stiefel-Whitney class with the lifted transition functions, we will connect it to the original transition functions in two examples: on a sphere and on a torus.
This approach will be useful when we interpret the Wilson loop method in the next section.

In general, transition functions should satisfy the following consistency conditions.
\begin{align}
\label{consistency}
t^{AB}_{\bf k}t^{BA}_{\bf k}=1
\end{align}
for ${\bf k}\in A\cap B$ and
\begin{align}
\label{triple_consistency}
t^{AB}_{\bf k}t^{BC}_{\bf k}t^{CA}_{\bf k}=1
\end{align}
for ${\bf k}\in A\cap B\cap C$, where $A$, $B$, and $C$ are arbitrary patches.
The transition functions defined by $\ket{u^B_{n\bf k}}=t^{AB}_{mn}({\bf k})\ket{u^A_{m\bf k}}$ satisfy these consistency conditions automatically.

After well-defined transition functions are lifted at each overlap to the double covering group, the consistency conditions are not automatically satisfied in general.
Let us write $I$ and $-I$ to denote the $0$ and $2\pi$ rotation in the double covering group.
In general, after the lift $t^{AB}\rightarrow \tilde{t}^{AB}$, the lifted transition functions satisfy 
\begin{align}
\tilde{t}^{AB}_{\bf k}\tilde{t}^{BA}_{\bf k}=\pm I
\end{align}
for ${\bf k}\in A\cap B$ and
\begin{align}
f^{ABC}_{\bf k}
\equiv \tilde{t}^{AB}_{\bf k}\tilde{t}^{BC}_{\bf k}\tilde{t}^{CA}_{\bf k}
=\pm I,
\end{align}
for ${\bf k}\in A\cap B\cap C$.
The sign can be either $+$ or $-$ because both $I$ and $-I$ are projected to 1 by the two-to-one map from $\text{Spin}(N_{\text{occ}})$ to $\text{SO}(N_{\text{occ}})$ or from $\text{Pin}^+(N_{\text{occ}})$ to $\text{O}(N_{\text{occ}})$.
$f^{ABC}_{\bf k}$ is gauge-invariant as one can see from the transformation of the lifted transition functions 
$\tilde{t}^{AB}_{\bf k}\rightarrow (\tilde{g}^{A}_{\bf k})^{-1}\tilde{t}^{AB}_{\bf k}\tilde{g}^B_{\bf k}$ 
under $\ket{u^A_{n\bf k}}\rightarrow g^A_{mn \bf k}\ket{u^A_{m\bf k}}$, where $\tilde{g}$ is a lift of $g$.
Also, $f^{ABC}_{\bf k}$ has a unique value on each triple overlap, because it is fully symmetric with respect to the permutation of $A$, $B$, and $C$ and is independent on ${\bf k}$ within a triple overlap.
We will thus omit the subscript $\bf k$ and not care about the order of the superscripts $A$, $B$, and $C$ for $f^{ABC}_{\bf k}$ from now on.

Let us now see when we can find a lift satisfying the consistency conditions.
No obstruction arises for the first condition: we can always find a lift $\tilde{t}'^{AB}$ satisfying $\tilde{t}'^{AB}\tilde{t}'^{BA}=I$ from an arbitrary lift $\tilde{t}^{AB}$ at every overlap $A\cap B$.
We can see this as follows.
A lift $\tilde{t}'^{AB}$ is related to the lift $\tilde{t}^{AB}$ by
\begin{align}
\label{change_lift}
\tilde{t}'^{AB}_{\bf k}=(-I)^{c^{AB}}\tilde{t}^{AB}_{\bf k},
\end{align}
where $c^{AB}$ is an integer defined modulo two (the subscript ${\bf k}$ is omitted for $c$ because $c$ is uniform over each overlap.).
Let us take a lift $\tilde{t}'^{AB}$ such that $(-I)^{c^{AB}+c^{BA}}=\tilde{t}^{AB}_{\bf k}\tilde{t}^{BA}_{\bf k}$ at each overlap $A\cap B$.
Then, we have $\tilde{t}'^{AB}\tilde{t}'^{BA}=I$.
There is no obstruction here coming from topology because the constraint on $c^{AB}$ and $c^{BA}$ is local in the sense that the constraint is defined on a single overlap $A\cap B$.

If we require the first consistency condition, however, we may not be able to find a lift satifying the second consistency condition due to the global constraint.
$f^{ABC}$ transforms under the change of the lift by Eq.~(\ref{change_lift}) as
\begin{align}
f'^{ABC}=(-I)^{c^{AB}+c^{BC}+c^{CA}}f^{ABC}.
\end{align}
By this transformation, we can get $f'^{ABC}=I$ on a triple overlap $A\cap B\cap C$.
However, $f'^{ABC}=I$ is not a local condition because $c^{AB}$, $c^{BC}$, and $c^{CA}$ are defined on different overlaps, which implies that there can be a topological obstructure to taking $f'^{ABC}=I$ over every triple overlap.
In fact, the product 
\begin{align}
\label{Cech_SW2}
(-I)^{w_2}=\prod_{A\cap B\cap C}f^{ABC}.
\end{align}
over all triple overlaps in a closed manifold is invariant under the change of the lift.

We can observe this invariance from the geometric structure of patches shown in Fig.~\ref{Cech}, where patches, overlaps, and triple overlaps are considered as vertices, edges, and faces, respectively.
In this intepretation, if we let $f^{ABC}=(-I)^{g^{ABC}}$, $w_2$ is a two-dimensional integral defined by
\begin{align}
w_2=\sum_{A\cap B\cap C}g^{ABC}\equiv \oint_M g,
\end{align}
where $\int_{A\cap B\cap C}g\equiv g^{ABC}$, and $M$ is the surface formed by the union of all faces.
Because $g$ transforms under the change of the lift by Eq.~(\ref{change_lift}) as $g'=g+dc$, where $(dc)^{ABC}\equiv c^{AB}+c^{BC}+c^{CA}$, $w_2$ changes by $\oint_M dc=\oint_{\d M}c$, where $\d M$ is the boundary of $M$, and the line integral is defined by $\int_{A\rightarrow B} c=c^{AB}$ over each overlap $A\cap B$.
The Stokes' theorem is valied here because the line integral $\int c$ is well-defined modulo two: $\int_{A\rightarrow B}c=-\int_{B\rightarrow A}c$ modulo two, and this follows from the consistency condition $\tilde{t}^{AB}\tilde{t}^{BA}=1$.
Because we consider closed manifolds, i.e., $\d M=0$, we find $\oint_{\d M}c=0$.
Thus, $w_2$ is invariant under any change of the lift modulo two.

One can see that transition functions cannot be lifted to their double covering group when $w_2=1$ modulo two, because then there is at least one triple overlap $A\cap B\cap C$ where $f^{ABC}=-I$.
Therefore, the obstruction to the existence of a spin or pin structure is characterized by the $Z_2$ invariant $w_2$.
The obstruction invariant $w_2$ is the second Stiefel-Whitney class for spin and pin$^+$ structures, whereas it is the dual second Stiefel-Whitney class for pin$^-$ structures~\cite{DeWitt-Morette_v2}.
In our case, however, we can restrict our attension only to the second Stiefel-Whitney class, because $w_2=\overline{w}_2$ for 2D closed submanifolds in the 3D Brillouin zone as we mentioned in the previous subsection.
We will investigate more on Eq.~(\ref{Cech_SW2}) using a sphere and a torus as examples below.

\subsubsection{Sphere}

\begin{figure}[t!]
\includegraphics[width=8.5cm]{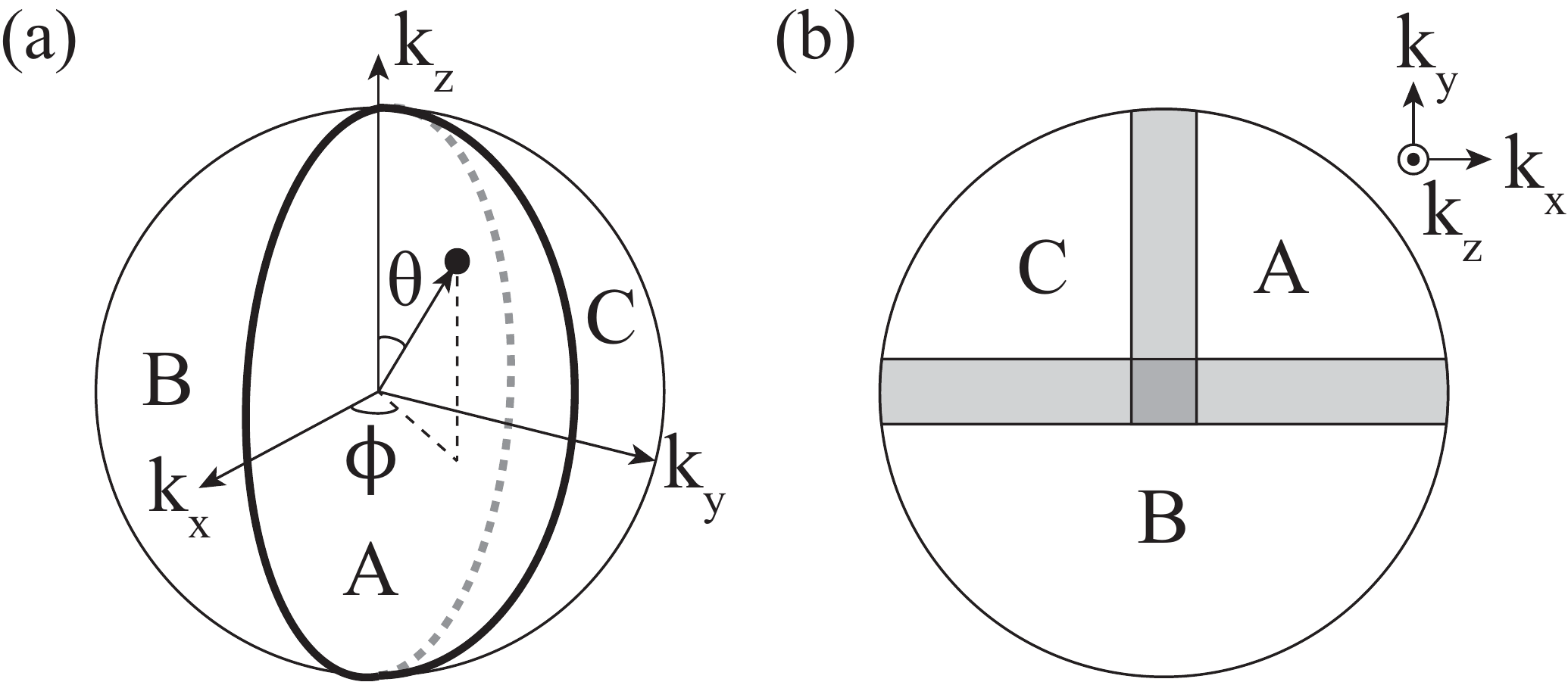}
\caption{Three patches covering a sphere.
(a) Orthographic view.
$\phi$ and $\theta$ are the azimuthal and polar angle.
$B$ and $A$ ($C$) overlap on $\phi=0$ ($\phi=\pi$), and $A$ and $C$ overlap on $\phi=\pi/2$.
The North pole $\theta=0$ and and the south pole $\theta=\pi$ are triple overlaps.
(b) Top view.
Overlapping regions are exaggerated to enhance visuality.
}
\label{sphere_patch}
\end{figure}

First we consider three patches $A$, $B$, and $C$ covering a sphere shown in Fig.~\ref{sphere_patch}.
In the spherical coordinates $(\phi,\theta)$, there are three overlaps $A\cap B$, $A\cap C$, and $B\cap C$ on $\phi=0$, $\phi=\pi/2$, and $\phi=\pi$, respectively.
We restrict all transition functions on the overlaps to $\text{SO}(N_{\text{occ}})$, which is possible because every loop on a sphere is contractible to a point such that the first Stiefel-Whitney class is trivial.
Then
\begin{align}
(-I)^{w_2}
&=f^{ABC}(0)f^{ABC}(\pi),
\end{align}
where $0$ and $\pi$ denotes the polar angle $\theta$.

The second Stiefel-Whitney class $w_2$ can be related to the winding number as follows.
Let us first define
\begin{align}
\tilde{W}(\theta)=\tilde{t}^{AB}(\theta)\tilde{t}^{BC}(\theta)\tilde{t}^{CA}(\theta),
\end{align}
where we omit $\phi$ in the argument of transition functions because they are uniquely specified by the overlapping region.
$\tilde{W}(\theta)$ is smooth for $0<\theta<\pi$ because $\tilde{t}$ is smooth within an overlap.
$\tilde{W}(0)=f^{ABC}(0)=\pm I$, and $\tilde{W}(\pi)=f^{ABC}(\pi)=\pm I$.
$w_2=1$ modulo two when the image of the map $\tilde{W}:[0,\pi]\rightarrow \text{Spin}(N_{\text{occ}})$ is an arc connecting $I$ and $-I$, while $w_2=0$ when the image is a closed loop containing $I$ or $-I$.
Next, we project $\tilde{W}$ by the two-to-one map $\text{Spin}(N_{\text{occ}})\rightarrow \text{SO}(N_{\text{occ}})$.
We have
\begin{align}
\label{winding_sphere}
W(\theta)=t^{AB}(\theta)t^{BC}(\theta)t^{CA}(\theta),
\end{align}
which is smooth for $0<\theta<\pi$, and $W(0)=W(\pi)=1$.
By this projection, an arc connecting $I$ and $-I$ projects to a loop winding the non-contractible cycle an odd number of times, whereas a closed loop projects to a contractible loop or a non-contractible loop winding the non-contractible cycles an even number of times~\cite{pin_structure}.
As a result, the second Stiefel-Whitney class is given by the winding number of $W({\theta})$ modulo two.
This definition of the second Stiefel-Whitney class corresponds to the $Z_2$ monopole charge enclosed by the sphere which is defined in Ref.~\onlinecite{Bzdusek} using the Wilson loop spectrum.
We will discuss more about it in the next section.

We can also connect the formalism given here to the definition of the $Z_2$ monopole charge in Ref.~\onlinecite{Z2line}, which was discussed in the previous subsection.
Let us choose a gauge where $t^{AC}=1$, and take a lift such that $\tilde{t}^{AC}=1$ and $f^{ABC}(\pi)=1$.
We have
\begin{align}
(-I)^{w_2}
&=f^{ABC}(0)\notag\\
&=\tilde{t}^{AB}(0)\tilde{t}^{BC}(0)\tilde{t}^{CA}(0)\notag\\
&=\tilde{t}^{AB}(0)(\tilde{t}^{CB}(0))^{-1}.
\end{align}
Let us define
\begin{align}
\label{Cech_sphere_second}
\tilde{W}(\varphi)
=
\begin{cases}
\tilde{t}^{AB}(\varphi) \quad&{\rm for}\; 0\le \varphi\le \pi, \\
\tilde{t}^{CB}(2\pi-\varphi) \quad&{\rm for}\; \pi\le \varphi\le 2\pi,
\end{cases}
\end{align}
where $0\le \varphi<2\pi$ parametrizes the overlap between the two hemispheres.
$\tilde{W}(\varphi)$ is smooth for $0<\varphi<2\pi$, and it satisfies the boundary condition $\tilde{W}(2\pi)=(-1)^{w_2}\tilde{W}(0)$.
The image of $\tilde{W}:[0,2\pi]\rightarrow \text{Spin}(N_{\text{occ}})$ is a closed loop when $w_2=0$ and is an arc connecting $\tilde{W}(0)$ and $-\tilde{W}(0)$ when $w_2=1$.
The former is project by $\text{Spin}(N_{\text{occ}})\rightarrow \text{SO}(N_{\text{occ}})$ to a contractible loop or a loop winding the non-contractible cycle an even number of times.
Let us see what is the projection of the latter, which is an arc connecting $\tilde{W}(0)$ and $-\tilde{W}(0)$.
The arc can be continuously deformed to an arc connecting $I$ and $-I$ by the homotopy $\tilde{W}_t(\varphi)= \tilde{W}(\varphi)\tilde{g}(t)$, where $0\le t\le 1$, $\tilde{g}(0)=1$ and $\tilde{g}(1)=\tilde{W}^{-1}(0)$.
Because $\tilde{W}_t$ is projected to a homotopy bewteen closed loops, the projection of an arc connecting $\tilde{W}(0)$ and $-\tilde{W}(0)$ is homotopically equivalent to the projection of an arc connecting $I$ and $-I$.
Accordingly, $w_2=1$ if and only if the projection of $\tilde{W}$, which is
\begin{align}
\label{winding_sphere2}
W(\varphi)
=
\begin{cases}
t^{AB}(\varphi) \quad&{\rm for}\; 0\le \varphi\le \pi, \\
t^{CB}(2\pi-\varphi) \quad&{\rm for}\; \pi\le \varphi\le 2\pi,
\end{cases}
\end{align}
winds the non-contractible cycle in $\text{SO}(N_{\text{occ}})$ an odd number of times.
The second Stiefel-Whitney class is thus given by the parity of the winding number of $W(\varphi)$.
This winding number corresponds to the definition of the $Z_2$ monopole charge in Ref.~\onlinecite{Z2line}, as one can see by noticing that $W(\varphi)$ is the transition function between two hemispheres defined by $0\le \phi\le \pi$ and $\pi \le \phi \le 2\pi$, respectively.

\subsubsection{Torus}

\label{subsubsec.torus}

\begin{figure}[t!]
\includegraphics[width=8.5cm]{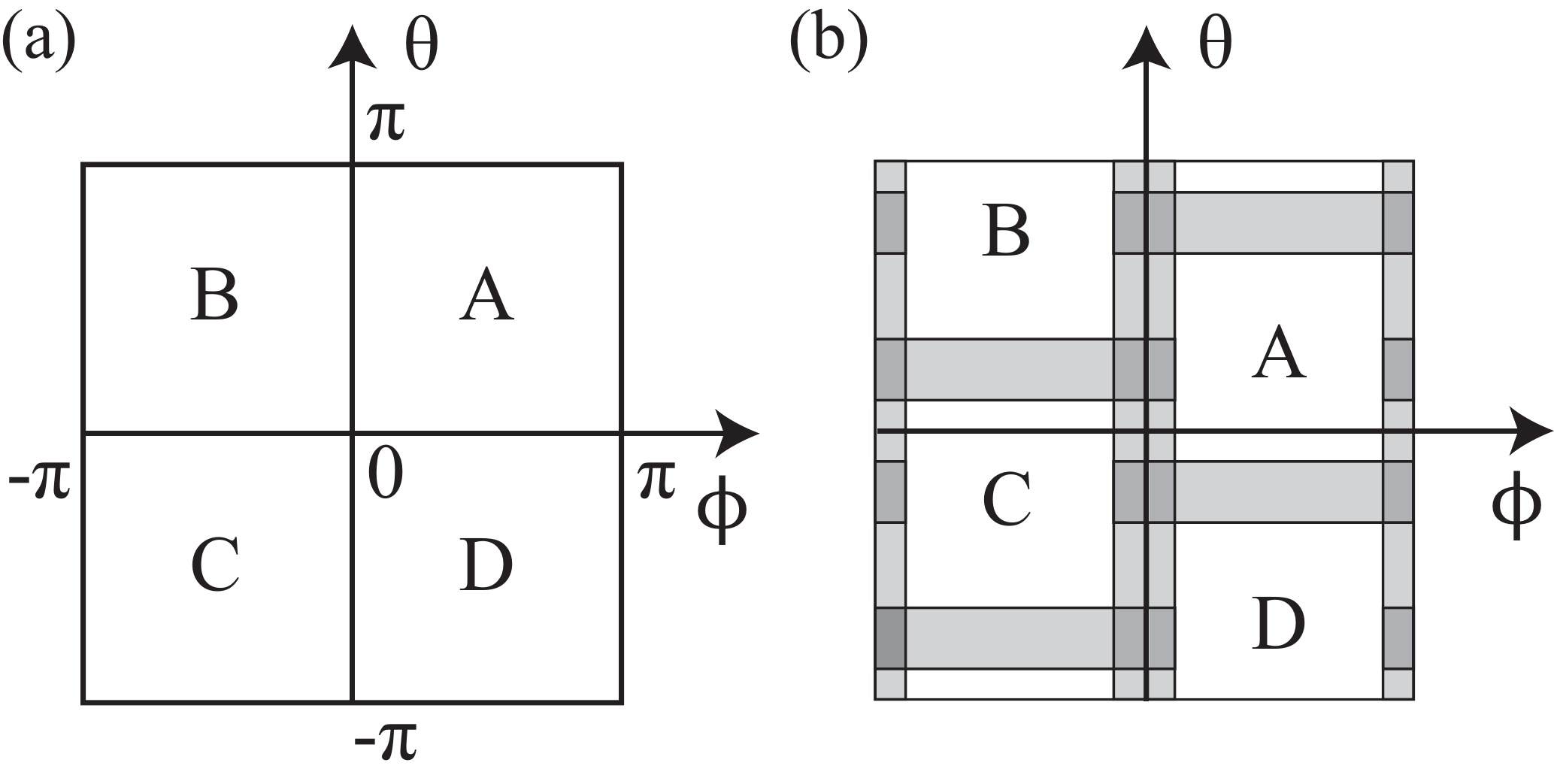}
\caption{Four patches covering a torus.
$\phi$ and $\theta$ are $2\pi$-periodic.
(a) is given by taking a vanishing overlap limit of (b).
}
\label{torus_patch}
\end{figure}

A torus has two cycles over which occupied states may be non-orientable.
If we consider the possibility of the non-orientability, we need at least four patches to cover a torus.
Here we cover a torus with four patches $A$, $B$, $C$, and $D$ as shown in Fig.~\ref{torus_patch}(a), which is a vanishing overlap limit of Fig.~\ref{torus_patch}(b).
For simplicity, we choose a gauge such that the $t^{BC}(\phi,\pi)=t^{AD}(\phi,\pi)=1$ and that $t^{BC}(\phi,0)$ and $t^{AD}(\phi,0)$ are the identity matrix (constant orientation-reversing matrices) when the occupied states are orientable (non-orientable) along $\theta$.
Moreover, we impose on the lift that the four triple overlaps near $\theta=\pi$ are trivial and that $\tilde{t}^{BC}(\phi,\pi)=\tilde{t}^{AD}(\phi,\pi)=1$.
We will consider $\tilde{M}^2=I$ for definiteness (i.e., we consider the pin$^+$ structure.).
The second Stiefel-Whitney class $w_2$ is then given by
\begin{align}
\label{Cech_2nd_SW_torus}
(-I)^{w_2}
&=\prod_{I\cap J\cap K}f^{IJK}\notag\\
&=f^{CDA}(0,0)f^{CAB}(0,0)f^{CBA}(\pi,0)f^{CAD}(\pi,0)\notag\\
&=(\tilde{t}^{CD}\tilde{t}^{DA}\tilde{t}^{AC})_{(0,0)}(\tilde{t}^{CA}\tilde{t}^{AB}\tilde{t}^{BC})_{(0,0)}\notag\\
&\quad \times(\tilde{t}^{CB}\tilde{t}^{BA}\tilde{t}^{AC})_{(\pi,0)}(\tilde{t}^{CA}\tilde{t}^{AD}\tilde{t}^{DC})_{(\pi,0)}\notag\\
&=(\tilde{t}^{CD}\tilde{t}^{DA}\tilde{t}^{AB}\tilde{t}^{BC})_{(0,0)}(\tilde{t}^{CB}\tilde{t}^{BA}\tilde{t}^{AD}\tilde{t}^{DC})_{(\pi,0)}\notag\\
&=(\tilde{t}^{CD}\tilde{t}^{DA}\tilde{t}^{AB})_{(0,0)}(\tilde{t}^{BA}\tilde{t}^{AD}\tilde{t}^{DC})_{(\pi,0)},
\end{align}
where we used in the last line that $\tilde{t}^{BC}(\phi,0)$ is independent on $\phi$, because $t^{BC}(\phi,0)$ is so.

Let us investigate how we can get $w_2$ using the original transition functions belonging to $\text{O}(N_{\text{occ}})$.
We define
\begin{align}
\tilde{W}(\theta)
=
\begin{cases}
\tilde{t}^{AB}(0,\theta)\tilde{t}^{BA}(\pi,\theta) \quad&{\rm for}\; 0\le \theta\le \pi, \\
\tilde{t}^{DC}(0,\theta)\tilde{t}^{CD}(\pi,\theta)  \quad&{\rm for}\; \pi\le \theta\le 2\pi.
\end{cases}
\end{align}
$\tilde{W}(\theta)$ is smooth for $0<\theta<2\pi$ due to the condition on the lift we have taken, and it satisfies the boundary condition
\begin{align}
\tilde{W}(2\pi)&=(-1)^{w_2}(\tilde{t}^{AD}(0,0))^{-1}\tilde{W}(0)\tilde{t}^{AD}(0,0),
\end{align}
where we used that $\tilde{t}^{AD}(\phi,0)$ is independent of $\phi$ such that $\tilde{t}^{AD}(0,0)=\tilde{t}^{AD}(\pi,0)$.
After we project $\tilde{W}$ by the two-to-one map $\text{Pin}^+(N_{\text{occ}})\rightarrow \text{O}(N_{\text{occ}})$, we have
\begin{align}
\label{winding_torus}
W(\theta)
=
\begin{cases}
t^{AB}(0,\theta)t^{BA}(\pi,\theta) \quad&{\rm for}\; 0\le \theta\le \pi, \\
t^{DC}(0,\theta)t^{CD}(\pi,\theta) \quad&{\rm for}\; \pi\le \theta\le 2\pi,
\end{cases}
\end{align}
and it follows that $\det W=1$ ($\det W=-1$) when the occupied states are orientable (non-orientable) along $\phi$.
$W$ satisfies the boundary condition
\begin{align}
\label{Wilson_torus_b.c.}
W(2\pi)=(t^{AD}(0,0))^{-1}W(0)t^{AD}(0,0).
\end{align}
By construction, $t^{AD}(0,0)$ is the identity (an orientation-reversing matrix) when the occupied bands are orientable (non-orientable) along $\theta$.
As the boundary condition depend on the orientability of the occupied states along $\theta$, we will consider two cases separately.

(1) When the occupied states are orientable along $\theta$ such that $t^{AD}=1$ and $\tilde{t}^{AD}=\pm I$, the boundary condition becomes the same as on a sphere [See Eq.~(\ref{Cech_sphere_second}) and explanations below it].
Then, the second Stiefel-Whitney classis given by the parity of the winding number of the projection of $\tilde{W}$, although $\tilde{W}$ belongs to the pin$^+$ group rather than the spin group in general.
$\text{Pin}^+(N_{\text{occ}})$ is the disjoint union of $\text{Spin}(N_{\text{occ}})$ and $\tilde{M}\text{Spin}(N_{\text{occ}})$ as a manifold, where $\tilde{M}$ is a lift of a mirror operation $M$ in $\text{O}(N_{\text{occ}})$.
Therefore, the image of $\tilde{W}$ is contained either in $\text{Spin}(N_{\text{occ}})$ or in $\tilde{M}\text{Spin}(N_{\text{occ}})$, which corresponds to the case where the occupied bands are orientable or non-orientable along $\phi$, respectively.
In the former, the second Steifel-Whitney class is given by the parity of the winding number of the projection of $\tilde{W}$ as we have already shown on a sphere.
Moreover, the same is true for the latter.
It is because $(-1)^{w_2}=[\tilde{W}(2\pi)]^{-1}\tilde{W}(0)=[\tilde{W}_o(2\pi)]^{-1}\tilde{W}_o(0)$ for $\tilde{W}=\tilde{M}\tilde{W}_o$, where $\tilde{W}_o\in \text{Spin}(N_{\text{odd}})$, such that $w_2$ is given by the parity of the winding number of $W_o$, which is the same as that of $W$.
We thus conclude that $w_2=0$ ($w_2=1$) when the image of $W$ winds the non-contractible loop an even (odd) number of times whether the occupied states are orientable along $\phi$ or not (i.e., whether $\det W=1$ or $\det W=-1$).

(2) When the occupied states are non-orientable along $\theta$ such that $\det t^{AD}=-1$, the boundary condition is totally different from that on a sphere.
The image of $\tilde{W}$ is an arc in $\text{Pin}^+(N_{\text{occ}})$ whose end point depends on $w_2$.
The arc for $w_2=1$ can be considered as a combination of an arc with $w_2=0$ and an arc connecting $(\tilde{t}^{AD})^{-1}\tilde{W}\tilde{t}^{AD}(0)$ to $-(\tilde{t}^{AD})^{-1}\tilde{W}\tilde{t}^{AD}(0)$.
Therefore, the images of $W$ for $w_2=0$ and $w_2=1$ are arcs which both satisfies the nontrivial boundary condition Eq.~(\ref{Wilson_torus_b.c.}) but differs by a loop winding the non-contractible cycle an odd number of times.
As we have found $w_2=0$ and $w_2=1$ are distinguished by the homotopy class of the image of $W$, let us now find how to specify the arc with $w_2=1$.

We first consider the case where the image of $\tilde{W}$ belongs to $\text{Spin}(N_{\text{occ}})$.
In this case, between the two homotopically inequivalent arcs, the one with $w_2=0$ is the arc which can be deformed to a point (i.e., the identity element) or to a loop which starts from the identity element and winds the non-contractible cycle an even number of times.
We can observe this by noticing that $\tilde{W}$ evolves from $I$ to $I$ or from $-I$ to $-I$ after such a deformation.
The other arc corresponds to $w_2=1$.
This arc is also contractible to a $t^{AD}$-invariant point, i.e., $W(0)=(t^{AD})^{-1}W(0)t^{AD}$, modulo an even number of winding the non-contractible cycle.
However, the invariant point here is not the identity element but a $\pi$ rotation whose rotation angle changes sign under the conjugation by $t^{AD}$.
We see that $w_2=1$ in this case because the $\pi$ and $-\pi$ rotation with a fixed plane of rotation are the different by $-I$ as elements of the spin group.

Similary, when the image of $\tilde{W}$ belongs to $\tilde{M}\text{Spin}(N_{\text{occ}})$, we find that $w_2=0$ corresponds to an arc which is contractible to $t^{AD}$ without breaking the boundary condition Eq.~(\ref{Wilson_torus_b.c.}) while $w_2=1$ corresponds to an arc which is contractible not to the identity but to the combination of $t^{AD}$ and a $\pi$ rotation modulo an even number of winding the non-contractible cycle.
In particular, when the number of occupied bands is odd, we can use $-1$ to substitute $t^{AD}$ in the above statement for contractibility because then $-1$ is an orientation-reversing element which commutes with $t^{AD}$.

Let us investigate a particular example of the case where the occupied bands are non-orientable along both $\phi$ and $\theta$.
It will demonstrate that the second Stiefel-Whitney class has an interesting property so-called the Whitney sum formula, which we will explain more in the next subsection.
Let us suppose we have diagonal transition functions
\begin{align}
t^{AB}(0,\theta)
&=t^{DC}(0,\theta)
=
\begin{pmatrix}
-1&0\\
0&1
\end{pmatrix},\notag\\
t^{AD}(\phi,0)
&=t^{BC}(\phi,0)
=
\begin{pmatrix}
1&0\\
0&-1
\end{pmatrix},
\end{align}
which are mirror operations in $\text{O}(2)$, and other transition functions are trivial.
Because the two occupied states are not mixed by transition functions in this gauge, topological invariants can be separately defined for each occupied band.
Both of the occupied bands have $w_2=0$, as it should for a single isolated band.
On the other hand, because $t^{AD}$ and $W(\theta)$ differs by a $\pi$ rotation in $\text{SO}(2)$, i.e.,
\begin{align}
(t^{AD}(0,0))^{-1}W(\theta)
=
\begin{pmatrix}
-1&0\\
0&-1
\end{pmatrix},
\end{align}
we should have $w_2=1$ according to the analysis in the previous subsection.
We can show this directly.
\begin{align}
(-I)^{w_2}
&=(\tilde{t}^{CD}\tilde{t}^{DA}\tilde{t}^{AB})_{(0,0)}(\tilde{t}^{BA}\tilde{t}^{AD}\tilde{t}^{DC})_{(\pi,0)}\notag\\
&=(\tilde{t}^{DA}\tilde{t}^{AB})_{(0,0)}(\tilde{t}^{BA}\tilde{t}^{AD}\tilde{t}^{DC})_{(\pi,0)}\tilde{t}^{CD}(0,0)\notag\\
&=(\tilde{t}^{DA})_{(0,0)}\tilde{W}(0)(\tilde{t}^{AD})_{(\pi,0)}\tilde{W}(2\pi)\notag\\
&=\left(\tilde{t}^{DA}(0,0)\tilde{W}(0)\right)^2\notag\\
&=-I.
\end{align}
where we used that $(\tilde{t}^{AD})^2=(\tilde{W})^2=I$ and that $\tilde{t}^{DA}\tilde{W}$ is a $\pi$ rotation.
We see that the second Stiefel-Whitney class is in contrast to all the other known 2D topological invariants, whose value for the total occupied bands is given by summing the invariant defined for each block.

\subsection{Whitney sum formula}

Whitney sum formula provides the rule for determining the total second Stiefel-Whitney class of blocks of bands from the Stiefel-Whitney classes of each block.
Let us suppose that the set of the occupied bands ${\cal B}$ is a direct sum of its $n$ subsets.
\begin{align}
{\cal B}=\bigoplus_{i}{\cal B}_i=B_1\oplus {\cal B}_2...\oplus {\cal B}_n.
\end{align}
The direct sum means that the transition function can be decomposed into blocks, i.e., we can find a basis where
\begin{align}
t^{AB}({\cal B})=
\begin{pmatrix}
t^{AB}({\cal B}_1)&0&\hdots&0\\
0&t^{AB}({\cal B}_2)&&0\\
\vdots&&\ddots&0\\
0&0&0&t^{AB}({\cal B}_n)
\end{pmatrix}
\end{align}
over all overlaps $A\cap B$.
For example, transition functions decomposes into blocks in the energy eigenstate basis when there are energy gaps between occupied bands, because energy eigenstates with different energies are not mixed by transition functions.
Then, the second Stiefel-Whitney class of the whole occupied states ${\cal B}$ is given by the Stiefel-Whitney classes of its subsets ${\cal B}_i$s according to the Whitney sum formula~\cite{Hatcher,DeWitt-Morette_v2}
\begin{align}
\label{Whitney_sum}
w_2\left(\oplus_{i}{\cal B}_i\right)=\sum_{i}w_2({\cal B}_i)+\sum_{i<j}w_1({\cal B}_i)\smile w_1({\cal B}_j),
\end{align}
where $\smile$ is the cup product
\footnote{
\label{terminology_SW}
Strictly speaking, we are abusing notations in two ways: 1. we have used the term Stiefel-Whitney classes to mean Stiefel-Whitney numbers. 2. we used the cup product notation for Stiefel-Whitney numbers, but the cup product is an operation for Stiefel-Whitney classes [See Ref.~\onlinecite{Hatcher,Hatcher_AT,DeWitt-Morette_v2}].
Let us clarify the terminology here to avoid a possible confusion for the ones who are familiar with algebraic topology.
In fact, $w_i$ is the $i$-th Stiefel-Whitney number which is given by the integration of the $i$-th Stiefel-Whitney class $\omega_i\in H^i({\cal M},Z_2)$.
For example, in \v{C}ech cohomology, $\omega_1$ and $\omega_2$ corresponds to $\det t^{AB}$ and $g^{ABC}$, respectively, and in de Rham cohomology, $\omega_1$ corresponds to the Berry connection computed in a smooth complex gauge.
We used the cup product notation for $w_1$s to mean the cup product of $\omega_1$s in the following sense:
$
w_1({\cal B}_i)\smile w_1({\cal B}_j)|_{T^2}
\equiv \oint_{T^2}\omega_1({\cal B}_i)\smile \omega_1({\cal B}_j).
$
}
, which can be understood as the exterior product as follows: on a 2D submanifold $\cal M$ in the Brillouin zone,
\begin{align}
w_1({\cal B}_i)\smile w_1({\cal B}_j)|_{\cal M}
&=\frac{1}{\pi^2}\oint_{\cal M} d{\bf S}\cdot {\bf A}_{i}\times {\bf A}_{j},
\end{align}
where ${\bf A}_i=\sum_{n\in {\cal B}_i}\braket{u_{n\bf k}|i\nabla_{\bf k}|u_{n\bf k}}$ is the Berry connection for the $i$-th block which is calculated in a smooth complex gauge
\footnote{
It is because the cup product in de Rham cohomology is provided by the exterior product of differential forms~\cite{Nakahara,DeWitt-Morette_v2}.
The de Rham cohomology class $[A_i]$ of the Berry connection 1-form $A_i$ is an element in the de Rham cohomology group with integer coefficients over $\cal M$, i.e., $[A_i]\in H^1({\cal M},Z)$.
After an equivalence relation is imposed between the Berry connections related by a gauge transformation, the new equivalence class $[[A_i]]$  becomes an element in $H^1({\cal M},Z_2)$.
This element is the first Stiefel-Whitney class $\omega_1({\cal B}_i)$ as we showed in Sec.~\ref{subsec.orientation}.
Then, we find $\omega_1({\cal B}_i)\smile \omega_1({\cal B}_j)=[[A_i]]\smile [[A_j]]=[[A_i\times A_j]]$ using $[A_i]\smile [A_j]=[A_i\times A_j]$ in de Rham cohomology.
Here, $[[A_i\times A_j]]$ means the de Rham cohomology class of $A_i\times A_j$ up to gauge transformations $A_i\rightarrow A_i+d\phi_i$.
}.
On a torus, this term is given by the Berry phase as
\begin{align}
\label{Kunneth}
\oint_{T^2} d{\bf S}\cdot {\bf A}_{i}\times {\bf A}_{j}
&=\Phi_{i,\phi}\Phi_{j,\theta}-\Phi_{i,\theta}\Phi_{j,\phi}
\end{align}
where $\Phi_{i,\psi}=\oint A_{i,\psi}d\psi$ is the Berry phase of the $i$-th block along $\psi=\theta$ or $\phi$.
It can be shown by noting that $\oint d{\bf S}\cdot {\bf A}_i\times {\bf A}_j$ is invariant under gauge transformations which does not mix blocks and taking a gauge where the Berry connection is constant over the torus.
Because the Berry phase $\Phi$ in a smooth complex gauge is the first Stiefel-Whitney class $w_1$ in a real gauge, Eq.~(\ref{Kunneth}) shows that the cup product of $w_1$ on torus is given by $w_1$ calculated along the cycles
\footnote{
This can be shown directly without using the relation between the Berry phase and the first Stiefel-Whitney class.
Let us first notice that the first Stiefel-Whitney class $\omega_1$ can be written as
$
\omega_1({\cal B}_i)=w_{1,\phi}({\cal B}_i)\phi+w_{1,\theta}({\cal B}_i)\theta
$
up to coboundary terms, where $\phi$ and $\theta$ are the generators of $H^1(T^2,Z_2)$.
Then, we have
$
w_1({\cal B}_i)\smile w_1({\cal B}_j)|_{T^2}
=\left(w_{1,\phi}({\cal B}_i)w_{1,\theta}({\cal B}_j)
-w_{1,\theta}({\cal B}_i)w_{1,\phi}({\cal B}_j)\right)\oint_{T^2}\phi\smile\theta
=w_{1,\phi}({\cal B}_i)w_{1,\theta}({\cal B}_j)
-w_{1,\theta}({\cal B}_i)w_{1,\phi}({\cal B}_j),
$
where we used $\phi\smile \phi=\theta\smile\theta=0$ on torus, and $\phi\smile \theta$ is the generator of $H^2(T^2,Z_2)$~\cite{Hatcher_AT}.
}
.
This kind of decomposition is in general known as the K$\ddot{\rm u}$nneth formula~\cite{Hatcher_AT} which applies when the base manifold is a direct product of two submanifolds (notice that $T^2=S^1\times S^1$).


\subsection{Linking number}
\label{sec.SW=Lk}

Now we consider the second Stiefel-Whitney class over a torus enclosing a nodal line.
Because the torus can be deformed to a sphere enclosing the line without closing the band gap, the second Stiefel-Whitney class over the torus is equal to that over the sphere which is the $Z_2$ monopole charge.
Here we show that the second Stiefel-Whitney class is identical to the linking number between the enclosed nodal line and the line of occupied band degeneracy.
It proves the equivalence of the linking number and the $Z_2$ monopole charge of a nodal line.

\begin{figure}[t!]
\includegraphics[width=5cm]{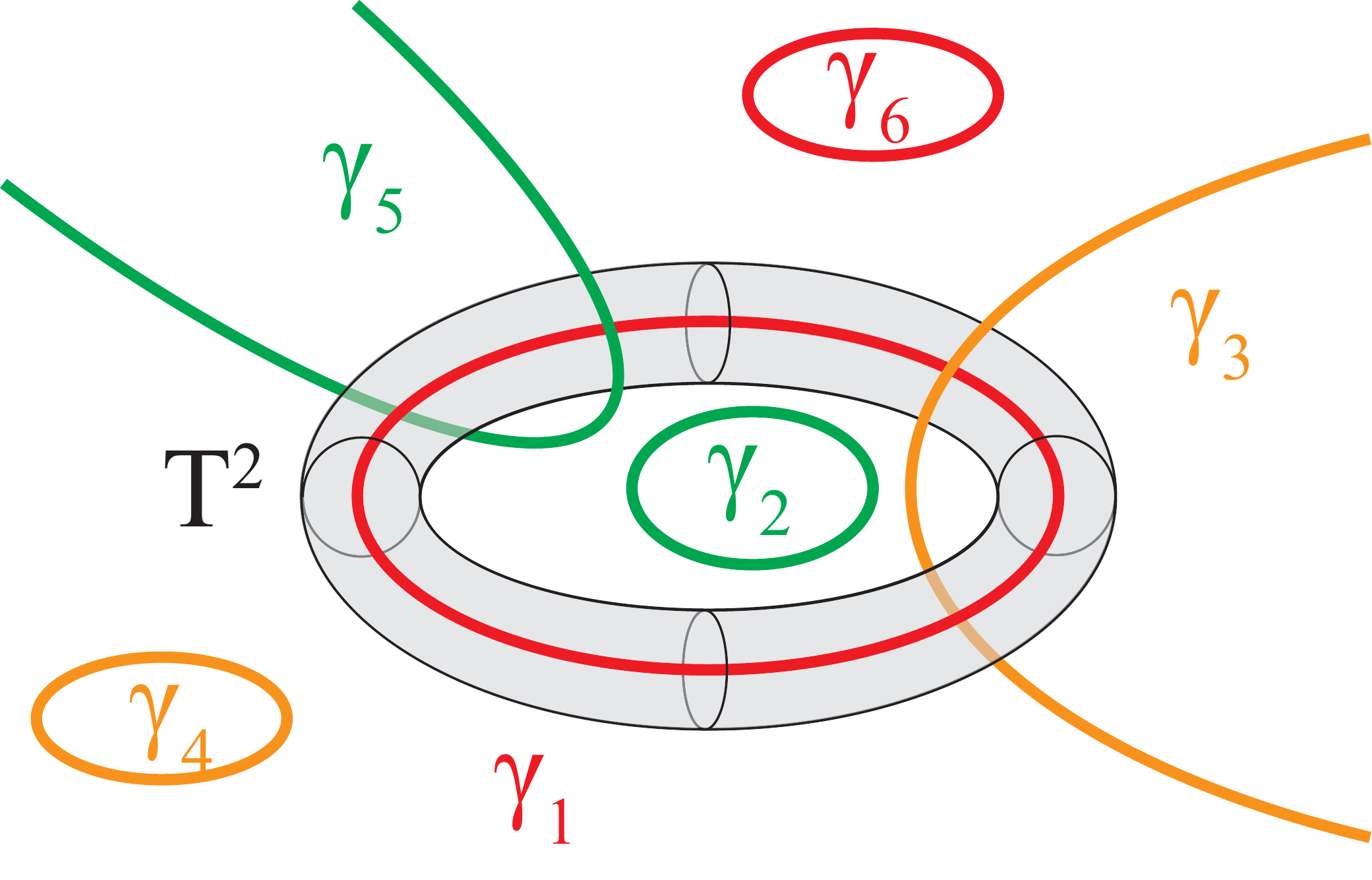}
\caption{Wrapping torus $T^2$ of nodal line $\gamma_1$ in the Brillouin zone.
$T^2$ is thin enough so that it does not intersect any other band degeneracies.
Red lines ($\gamma_1$ and $\gamma_6$) are lines of touching between the conduction and valence band.
Orange ($\gamma_3$ and $\gamma_4$) and green ($\gamma_2$ and $\gamma_5$) lines are lines of touching between the first and the second topmost occupied band and between the second and the third topmost occupied band, respectively.
Only the linking between $\gamma_1$ and $\gamma_3$ is protected as it generates the nontrivial second Stiefel-Whitney class on $T^2$, whereas the linking between $\gamma_1$ and $\gamma_5$ is not protected [See Eq.~(\ref{SW=linking})].
}
\label{wrapping_torus}
\end{figure}

Consider a very thin torus $T^2$ wrapping a nodal line $\gamma_1$.
As long as the line does not intersect any other nodal lines, we can take the torus thin enough so that all the occupied bands are non-degenerate everywhere on the torus as in Fig.~\ref{wrapping_torus}.
On this torus, we can use the Whitney sum formula to explicitly write down an analytic form of the second Stiefel-Whitney class as a two-dimensional integral
\begin{align}
\label{SW_torus}
w_{2}=&\sum_{n<m}\frac{1}{\pi^2}\oint_{T^2} d{\bf S}\cdot {\bf A}_{n}\times {\bf A}_{m},
\end{align}
where ${\bf A}_n=\braket{u_{n\bf k}|i\nabla_{\bf k}|u_{n\bf k}}$ is the Berry connection for the $n$th topmost occupied band, and $n$ and $m$  run over the occupied bands.
The Berry connection here is calculated in a complex smooth basis as in the discussion of the first Stiefel-Whitney class.
By noting that the quantization of the Berry phase, $\oint_C A_n=\pi$ or 0, resembles the Ampere's law in electromagnetics, we can solve the equation to get an analogy to the Bio-Savart law.

We start from the differential form of the quantization of the Berry phase.
It has the form of the differential form of the Ampere's law.
\begin{align}
\nabla_{\bf k}\times{\bf A}_n({\bf k})
=\sum_iI_n^i\oint_{\gamma_i} d{\bf k}_{i} \delta^{3}({\bf k}-{\bf k}_{i}),
\end{align}
where $\gamma_i$'s are lines of band touching points, and $I^i_n=\pi$ if the line $\gamma_i$ generates the Berry phase on $n$th band, and $I^i_n=0$ if the line does not.
One can check that $\oint_{\d \cal M} A_n=\sum_{\gamma_i\perp \cal M} I^i_n$ as required, where $\gamma_i\perp \cal M$ means $\gamma_i$ intersects the 2D manifold $\cal M$.
We will solve this equation for the Berry connection.
Applying curl to the equation and using an elementary vector identity $\nabla_{\bf k}\times(\nabla_{\bf k}\times{\bf A}_n)=\nabla_{\bf k}(\nabla_{\bf k}\cdot{\bf A}_n)-\nabla^2{\bf A}_n$, we have
$\nabla_{\bf k}(\nabla_{\bf k}\cdot{\bf A}_n)-\nabla^2{\bf A}_n=\nabla_{\bf k}\times{\bf F}_n,$
where ${\bf F}_n=\sum_iI_n^i\oint_{\gamma_i} d{\bf k}_{i} \delta^{3}({\bf k}-{\bf k}_{i})$.
Now we choose the Coulomb gauge
\begin{align}
\nabla_{\bf k}\cdot{\bf A}_n=0
\end{align}
to have a vector Poisson equation
\begin{align}
-\nabla_{\bf k}^2{\bf A}_n=\nabla_{\bf k}\times{\bf F}_n.
\end{align}
We solve this equation in the extended $k$-space (i.e., $-\infty<k_x,k_y,k_z<\infty$) using a Green's function $G({\bf k},{\bf k}')$ by
\begin{align}
{\bf A}_n({\bf k})
&=\int d^3k' G({\bf k,k'})\nabla_{\bf k'}\times {\bf F}_n({\bf k'})\notag\\
&=-\int d^3k' \nabla_{\bf k'}G({\bf k,k'}) \times {\bf F}_n({\bf k'}),
\end{align}
where we integrated by parts.
A surface term does not arise because it vanishes as $|{\bf k}'|\rightarrow \infty$.
Using the Green's function in free space
$G({\bf k,k'})=1/4\pi|{\bf k}-{\bf k'}|,$
the Berry connection is given by
\begin{align}
{\bf A}_n({\bf k})
&=\frac{1}{4\pi}\sum_{i}\oint_{\gamma_i}\frac{I^i_nd{\bf k}_i\times ({\bf k}-{\bf k}_i)}{|{\bf k}-{\bf k}_i|^3}.
\end{align}

Using this solution for the Berry connection, we have
\begin{align}
\label{linking_derivation}
&\frac{1}{\pi^2}\oint_{T^2} d{\bf S}\cdot {\bf A}_{n}\times {\bf A}_{m}\notag\\
&=\frac{1}{\pi^2}\oint_{S^1\times D^2} d{\bf k} \nabla_{\bf k}\cdot {\bf A}_{n}\times {\bf A}_{m}\notag\\
&=\frac{1}{\pi^2}\oint_{S^1\times D^2} d{\bf k} \left(\nabla_{\bf k}\times {\bf A}_{n}\right)\cdot {\bf A}_{m}-\left(\nabla_{\bf k}\times {\bf A}_{m}\right)\cdot {\bf A}_{n}\notag\\
&=\frac{1}{\pi^2}\oint_{S^1\times D^2} d{\bf k} \sum_{i} I_n^i\oint_{\gamma_i} d{\bf k}_i \delta^{3}({\bf k}-{\bf k}_{i})\cdot {\bf A}_{m}-(n\leftrightarrow m)\notag\\
&=\sum_{\gamma_i\in S^1\times D^2} \frac{I_n^i}{\pi^2}\oint_{\gamma_i} d{\bf k}_{i}\cdot {\bf A}_m-(n\leftrightarrow m)\notag\\
&=\sum_{\gamma_i\in S^1\times D^2}\sum_{\gamma_j}\bigg[\frac{I_{[n}^iI_{m]}^j}{\pi^2}\frac{1}{4\pi}\oint_{\gamma_i}\oint_{\gamma_j} \frac{d{\bf k}_{i}\cdot d{\bf k}_j\times ({\bf k}_i-{\bf k}_j)}{|{\bf k}_i-{\bf k}_j|^3}\bigg]\notag\\
&=\sum_{\gamma_i\in S^1\times D^2}\sum_{\gamma_j}\left[\frac{I_{n}^iI_{m}^j-I_{m}^iI_{n}^j}{\pi^2}{\rm Lk}(\gamma_i,\gamma_j)\right]\notag\\
&=\sum_{\gamma_j}\left(\delta_{1n}\frac{I_m^j}{\pi}-\delta_{1m}\frac{I_n^j}{\pi}\right){\rm Lk}(\gamma_1,\gamma_j),
\end{align}
where $T^2$ is a torus wrapping lines, $S^1\times D^2$ is the solid torus whose boundary is $T^2$, $I_{[n}^iI_{m]}^j=I_{n}^iI_{m}^j-I_{m}^iI_{n}^j$, and
\begin{align}
{\rm Lk}(\gamma_i,\gamma_j)
=\frac{1}{4\pi}\oint_{\gamma_i}\oint_{\gamma_j}\frac{d{\bf k}_i\times d{\bf k}_j\cdot ({\bf k}_i-{\bf k}_j)}{|{\bf k}_i-{\bf k}_j|^3}
\end{align}
is the Gauss' linking integral of the closed lines $\gamma_i$ and $\gamma_j$.
In the last line we used that $\gamma_1$ is the only line inside the torus and that $\gamma_1$ produces $\pi$ Berry phase only for the valence band.

Eq.~(\ref{SW_torus}) then becomes
\begin{align}
\label{SW=linking}
w_2(T^2)
&=\sum_{n<m}\sum_{\gamma_j}\left(\delta_{1n}\frac{I_m^j}{\pi}-\delta_{1m}\frac{I_n^j}{\pi}\right){\rm Lk}(\gamma_1,\gamma_j)\notag\\
&=\sum_{\gamma_j}\sum_{m>1}\frac{I_m^j}{\pi}{\rm Lk}(\gamma_1,\gamma_j)\notag\\
&=\sum_{\tilde{\gamma}_j}{\rm Lk}(\gamma_1,\tilde{\gamma}_j),
\end{align}
where $\tilde{\gamma}_j$'s are line of band touching between the first and second topmost occupied bands.
In the last line, the contribution coming from the band touching below the the second topmost occupied band was canceled because a line of band touching generates $\pi$ Berry phase to both bands touching at the line. 
If $\gamma_j$ is formed by the crossing of $a$-th and $(a+1)$-th occupied bands, $I^j_a=I^j_{a+1}=\pi$ and $I^j_{m\ne a,a+1}=0$.

Let us finish this subsection with two remarks.
First, the expression Eq.~(\ref{SW_torus}) still gives the linking number in smooth complex gauges which are not the Coulomb gauge.
This is because the expression is invariant under diagonal gauge transformations ${\bf A}_n\rightarrow {\bf A}'_n={\bf A}_n+\nabla_{\bf k}\phi_n$.
Second, we could also derive the linking number starting from the expression $\Phi_{n,\phi}\Phi_{m,\theta}-\Phi_{m,\phi}\Phi_{n,\theta}$ rather than from $\oint_{T^2} {\bf A}_n\times {\bf A}_m$ because they are identical.
In fact, the fifth line in Eq.~(\ref{linking_derivation}) corresponds to the Kunneth formula Eq.~(\ref{Kunneth}) after some additional algebra as follows.
\begin{align}
&\frac{1}{\pi^2}\oint_{T^2} d{\bf S}\cdot {\bf A}_{n}\times {\bf A}_{m}\notag\\
&=\sum_{\gamma_i\in S^1\times D^2} \frac{I_n^i}{\pi^2}\oint_{\gamma_i} d{\bf k}_{i}\cdot {\bf A}_m-(n\leftrightarrow m)\notag\\
&=\sum_{\gamma_i\in S^1\times D^2} \frac{I_n^i}{\pi^2}\oint_{S^1_{\phi}} d{\bf k}\cdot {\bf A}_m-(n\leftrightarrow m)\notag\\
&=\frac{1}{\pi^2}\oint_{S^1_{\theta}} d{\bf k}\cdot {\bf A}_n\oint_{S^1_{\phi}} d{\bf k}\cdot {\bf A}_m-(n\leftrightarrow m)\notag\\
&=\frac{1}{\pi^2}\left(\Phi_{n,\theta}\Phi_{m,\phi}-\Phi_{m,\theta}\Phi_{n,\phi}\right),
\end{align}
where $S^1_{\phi/\theta}$ in the fourth line is a poloidal/toroidal cycle on the torus, and $\Phi_{n,\phi/\theta}$ is the Berry phase of $n$th occupied band along the toroidal/poloidal cycle.

\subsection{Flux integral}
\label{sec.Euler}

The universal integral form of the second Stiefel-Whitney class has not been known yet~\cite{Gomi}.
Integral forms are given only in some special cases, where Eq.~(\ref{SW_torus}) is an example.
There is another important integral form of the second Stiefel-Whitney class, which was used by Zhao and Lu in Ref.~\onlinecite{realDirac} (they called it real Chern number).
This integral form gives an intuitive understanding of the $Z_2$ monopole charge as a ``monopole charge'' in terms of the Gauss' law.
It is also useful in the analytic calculation of the invariant.
We use the integral form in section~\ref{sec.inversion} when we derive the relation between the inversion eigenvalues and the Stiefel-Whitney class.
Here, we review the flux integral form for two occupied bands and discuss its extension to the case with more than three occupied bands.
Finally, we show by using the flux integral form that the $Z_2$ monopole charge of a Dirac point is given by the skyrmion number of its Hamiltonian.

\subsubsection{Euler class for two occupied bands}

For two orientable occupied bands, the second Stiefel-Whitney class is identical to the Euler class $e_2$ modulo two~\cite{realDirac,Hatcher,Nakahara}.
The Euler class is defined by
\begin{align}
\label{Euler}
e_{2}=\frac{1}{2\pi}\oint_{{\cal M}}d{\bf S}\cdot {\bf F}_{12}^R({\bf k}),
\end{align}
where $\cal M$ is a two-dimensional manifold,
$
F_{mn}^R({\bf k})
=\nabla_{\bf k}\times A_{mn}^R({\bf k})
$
and
$
A^R_{mn}({\bf k})
=\braket{u_m({\bf k})|\nabla_{\bf k}|u_n({\bf k})}
$
are $2\times 2$ antisymmetric real Berry curvature and connection defined by real occupied states $\ket{u_n({\bf k})}$.
This integral form is valid in any real smooth gauge where transition functions are restricted to $\text{SO}(2)$.
It is invariant under any $\text{SO}(2)$ gauge transformation.
One can check that this definition coincides with the one above as the obstruction to the spin structure.
As an example, on a sphere covered by two patches $N$ and $S$ which respectively cover the northern and southern hemisphere,
\begin{align}
e_2
&=\frac{1}{2\pi}\oint_{S^2} d{\bf S}\cdot {\bf F}_{12}^R\notag\\
&=\frac{1}{2\pi}\int_{N} d{\bf S}\cdot {\bf F}_{12}^R+\frac{1}{2\pi}\int_{S} d{\bf S}\cdot {\bf F}_{12}^R\notag\\
&=\frac{1}{2\pi}\oint_{S^1}d{\bf k}\cdot \left({\bf A}_{N,12}^R- {\bf A}_{S,12}^R\right)\notag\\
&=\frac{1}{2\pi}\oint_{S^1}d{\bf k}\cdot \nabla_{\bf k}\phi_{NS},
\end{align}
where $S^1$ is the equator, and $\exp(-i\sigma_y\phi_{NS})$ is the transition function between $N$ and $S$.
The parity of the winding number of the transition function gives the obstruction to the spin structure as we showed above.

\subsubsection{Flux integral for more occupied bands}

For more than three occupied bands, however, there is no flux integral form which is invariant under all orientation-preserving gauge transformations.
Although Zhao and Lu suggested a flux integral in Ref.~\onlinecite{realDirac} for arbitrary number of occupied bands, the expression is not gauge invariant, and it is not clear in which gauge the expression is valid.
We now show precisely in which gauge we have an flux integral form for $N_{\text{occ}}>2$ orientable states.

As we know the flux integral form of two orientable occupied bands, we have a flux integral form when the occupied states decompose into $2\times 2$ blocks, according to the Whitney sum formula.
Suppose that we can find a basis where the transition function is block-diagonal and each $2\times 2$ block ${\cal B}_i$ is orientable ($w_1({\cal B}_i)=0$) over a 2D manifold $\cal M$.
Then, the Whitney sum formula reduces to $w_2(\oplus_{i=1}{\cal B}_i)=\sum_{i}w_2({\cal B}_i)$ in that basis, and thus we have
\begin{align}
w_{2}
=\sum_i\frac{1}{2\pi}\oint_{{\cal M}} d{\bf S}\cdot \nabla_{\bf k}\times{\bf A}_{2i-1,2i}^R \quad ({\rm mod} \;2),
\end{align}
where ${\cal M}$ is a 2D manifold where occupied states are orientable, and $i$ runs over the $2\times 2$ blocks,
If the number of occupied bands is odd, we have an additional $1\times 1$ block, but it does not contribute to the second Stiefel-Whitney class.

We can indeed find a gauge where the transition function is diagonalized into orientable $2\times 2$ blocks.
Let $A$ and $B$ be two patches covering a sphere.
By an orthogonal gauge transformation $\ket{u^{A/B}_n({\bf k})}=O^{A/B}_{mn}({\bf k})\ket{u^{A/B}_m({\bf k})}$, the transition function transforms as
\begin{align}
t^{AB}({\bf k})\rightarrow \left(O^{A}({\bf k})\right)^{-1}t^{AB}({\bf k})O^B({\bf k}).
\end{align}
At each ${\bf k}\in A\cap B$, the transition function $t^{AB}$ can be diagonalized by this gauge transformation as we can see by taking $O^A=O^B$ and noting that a special orthogonal matrix$\in \text{SO}(N_{\text{occ}})$ can be diagonalized into the canonical form, i.e., into blocks of $\text{SO}(2)$ and $\text{SO}(1)$ matrices~\cite{Gelfand}.
Moreover, we can block-diagonalize $t^{AB}$ not only locally but also globally using a well-defined gauge transformation, such that the occupied states remain smooth within $A$ and $B$ after the transformation.
It is because $t^{AB}$ can be block-diagonalized using a homotopically trivial matrix.
Let us suppose that $t^{AB}({\bf k})$ is block-diagonalized by a homotopically nontrivial matrix $S({\bf k})$. 
Then, we can find a homotopically trivial matrix by multiplying a topologically nontrivial block-diagonal matrix $R({\bf k})$.
The resulting matrix $O({\bf k})=S({\bf k})R({\bf k})$ still diagonalizes the transition function.
We can thus take $O^A=O^B=O$ to diagonalize $t^{AB}$.

In conclusion, we can write down a flux integral for the second Stiefel-Whitney class in a gauge where the transition functions are diagonalized into orientable $2\times 2$ blocks (and one $1\times 1$ block if the number of occupied bands is odd).

\subsubsection{Example: Dirac point}
\label{skyrmion}

Let us now explicitely show that the $Z_2$ monopole charge is given by the skyirmion number for the Dirac-type Hamiltonian, which was conjectured in Ref.~\onlinecite{Fang_inversion}.
First, consider a Dirac Hamiltonian
\begin{align}
H
&=|{\bf k}|(\sin\theta\cos\phi\Gamma_1+\sin\theta\sin\phi\Gamma_2+\cos\phi\Gamma_3)
\end{align}
where
$\Gamma_1=\sigma_x$,
$\Gamma_2=\sigma_ys_y$,
$\Gamma_3=\sigma_z$.
Its eigenstates are
\begin{align}
u_1
&=
\left(
0
,\sin\frac{\theta}{2}
,-\cos\frac{\theta}{2}\sin\phi
,-\cos\frac{\theta}{2}\cos\phi
\right)^T,
\notag\\
u_2
&=
\left(
\sin\frac{\theta}{2}
,0
,-\cos\frac{\theta}{2}\cos\phi
,\cos\frac{\theta}{2}\sin\phi
\right)^T,
\notag\\
u_3
&=
\left(
0
,\cos\frac{\theta}{2}
,\sin\frac{\theta}{2}\sin\phi
,\sin\frac{\theta}{2}\cos\phi
\right)^T,
\notag\\
u_4
&=
\left(
\cos\frac{\theta}{2}
,0
,\sin\frac{\theta}{2}\cos\phi
,-\sin\frac{\theta}{2}\sin\phi
\right)^T,
\end{align}
whose eigenvalues are $E_{1,2}=-|{\bf k}|$ and $E_{3,4}=|{\bf k}|$.
Then, the Euler class $e_2$ is given by
\begin{align}
e_2
&=\frac{1}{2\pi}\oint_{S^2}d{\bf S}\cdot F^R_{12}({\bf k})\notag\\
&=\frac{1}{2\pi}\oint_{S^2}d{\bf S} \cdot \nabla_{\bf k} \times \braket{u_1({\bf k})|\nabla_{\bf k}|u_2({\bf k})}\notag\\
&=\frac{1}{2\pi}\oint_{S^2}d{\bf S}\cdot \sum_{n=3,4}\frac{\braket{u_1|\nabla_{\bf k} H|u_n}\times\braket{u_n|\nabla_{\bf k} H|u_2}}{(E_n-E_1)(E_n-E_2)}\notag\\
&=\frac{1}{2\pi}\oint_{S^2}d{\bf S}\cdot \frac{2\hat{\bf k}}{4|{\bf k}|^2}\notag\\
&=\frac{1}{4\pi}\oint_{S^2}d\Omega.
\end{align}

In general, for
\begin{align}
H
=f_1({\bf k})\Gamma_1+f_2({\bf k})\Gamma_2+f_3({\bf k})\Gamma_3,
\end{align}
we have
\begin{align}
e_2
&=\frac{1}{2\pi}\oint_{S^2}dS \hat{\bf k}\cdot \nabla_{\bf k} f_i\times \nabla_{\bf k} f_j\notag\\
&\quad \times\sum_{n=3,4}\frac{\braket{u_1|\nabla_{f_i} H|u_n}\braket{u_n|\nabla_{f_j} H|u_2}}{(E_n-E_1)(E_n-E_2)}\notag\\
&=\frac{1}{4\pi}\oint_{S^2}dS \hat{\bf k}\cdot \nabla_{\bf k} f_i\times \nabla_{\bf k} f_j \epsilon^{ijk} \frac{f_k}{2|{\bf f}|^3}\notag\\
&=\frac{1}{8\pi}\oint_{S^2}d{\bf S}\cdot \epsilon^{ijk}\hat{f}_i\cdot \nabla_{\bf k}\hat{f}_j\times \nabla_{\bf k}\hat{f}_k.
\end{align}
The parity of this invariant is the second Stiefel-Whitney class, which is the $Z_2$ monopole charge of the Dirac point.

\section{Wilson Loop Method}
\label{sec.Wilson}

Wilson loop methods have been introduced in Ref.~\onlinecite{Z2line,realDirac,Bzdusek} to calculate the $Z_2$ monopole charge of Dirac points or nodal lines in the Brillouin zone.
In particular, Bzdu\v{s}ek and Sigrist developed a gauge-invariant method in Ref.~\onlinecite{Bzdusek}.
However, they could not explicitly connect the Wilson loop method to the bulk topological invariant, which is the second Stiefel-Whitney class.
Here, we explain in detail how to get the second Stiefel-Whitney class using the Wilson loop method both on a sphere and on a torus.
This method can be used to gauge-invariantly calculate the $Z_2$ monopole charge or the $Z_2$ invariant of two-dimensional insulators.

\subsection{Wilson loop and transition functions}
The Wilson loop operator is defined by~\cite{Wilson_loop,inversion_Wilson_loop,group_cohomology}
\begin{align}
\label{Wilson_discrete}
W_{(\phi_0+2\pi,\theta)\leftarrow (\phi_0,\theta)}
=\lim_{N\rightarrow \infty}F_{N-1}F_{N-2}...F_{1}F_{0},
\end{align}
where $(\phi,\theta)$ parametrizes a sphere or a torus in the Brillouin zone or the (2D) Brillouin zone itself, and $F_j$ is the overlap matrix at $\phi_j=\phi_0+2\pi j/N$ whose matrix elements are given by $[F_{j}]_{mn}=\braket{u_{m\phi_{j+1}}|u_{n\phi_j}}$, and $\phi_{N+1}=\phi_0$.
We will show that the homotopy class of the Wilson loop operator gives the second Stiefel-Whitney class in a special gauge, the so-called the parallel-transport gauge~\cite{Vanderbilt}.

\begin{figure}[t!]
\includegraphics[width=8.5cm]{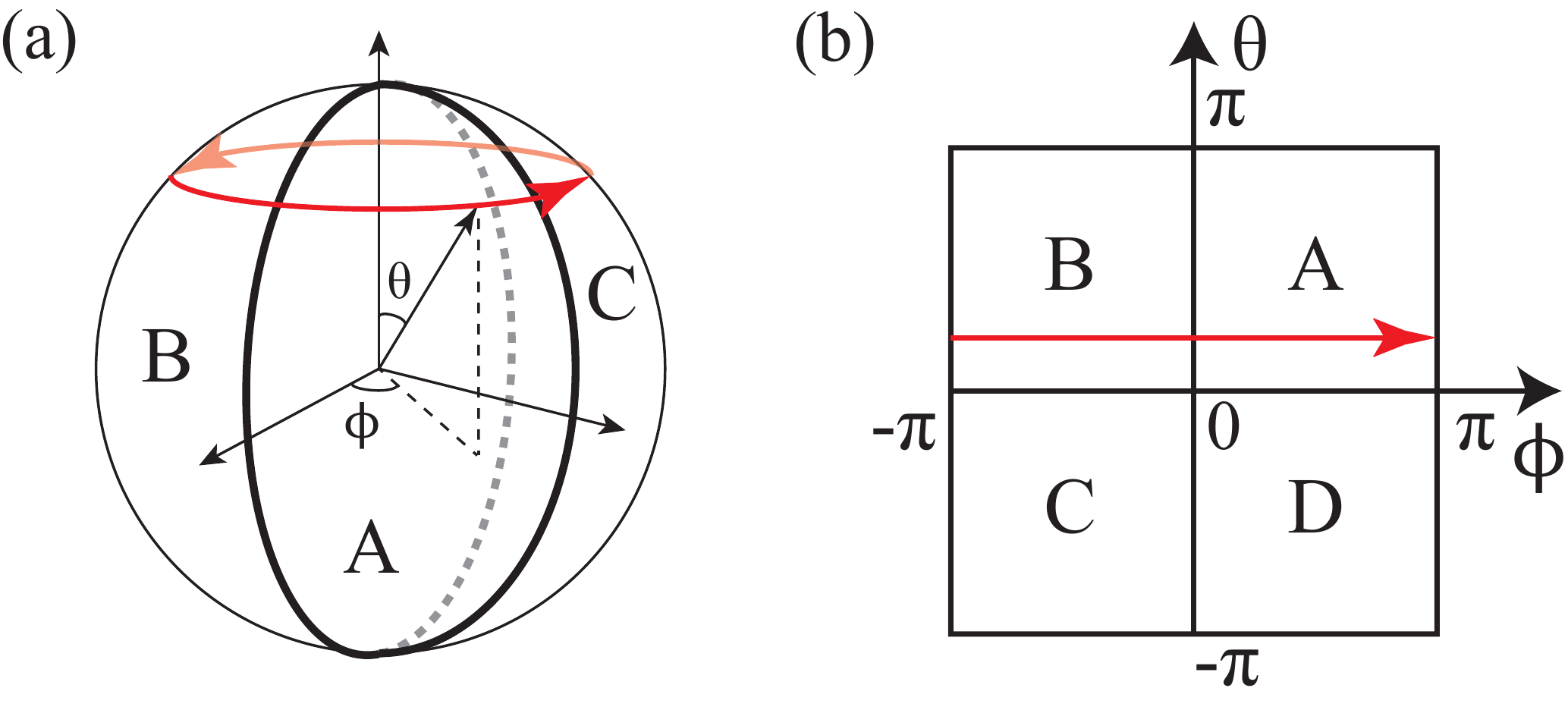}
\caption{Patches covering a sphere and a torus.
Wilson loop operators are calculated along the red arrows.
(a) Sphere covered with two patches which overlap on $\phi=0$ and $\phi=\pi$.
(b) Torus covered with four patches.
$\phi$ and $\theta$ are $2\pi$-periodic.
Transition functions are taken to be nontrivial only over $\theta=0$ and $\phi=0$ lines.
}
\label{Wilson_patch}
\end{figure}

First, we consider a sphere in the Brillouin zone, which is covered by three patches $A$, $B$, and $C$ whose overlap $A\cap C$, $C\cap B$, and $B\cap A$ is $\phi=\pi/2$, $\phi=\pi$, and $\phi=2\pi$ for all $0\le \theta\le \pi$ [See Fig.~\ref{Wilson_patch}(a)].
Here $\phi$ and $\theta$ are the azimuthal and the polar angle of the sphere.
Real occupied states $\ket{u_{n\bf k}}$ are smooth within each patch.
The Wilson loop operator $W_{0}(\theta)\equiv W_{(2\pi,\theta)\leftarrow (0,\theta)}$ is then
\begin{align}
W_{0}(\theta)
=
&\braket{u^A(0,\theta)|u^B(2\pi,\theta)}W_{(2\pi,\theta)\leftarrow (\pi,\theta)}\notag\\
&\braket{u^B(\pi,\theta)|u^C(\pi,\theta)}W_{(\pi,\theta)\leftarrow (\pi/2,\theta)}\notag\\
&\braket{u^C(\pi/2,\theta)|u^A(\pi/2,\theta)}W_{(\pi/2,\theta)\leftarrow (0,\theta)}\notag\\
=
&t^{AB}(\theta){\cal P}e^{-i\int^{2\pi}_{\pi} d\phi'  A^B_{\phi}(\theta,\phi')}\notag\\
&t^{BC}(\theta){\cal P}e^{-i\int^{\pi}_{\pi/2} d\phi'  A^C_{\phi}(\theta,\phi')}\notag\\
&t^{CA}(\theta){\cal P}e^{-i\int^{\pi/2}_{0} d\phi'  A^A_{\phi}(\theta,\phi')},
\end{align}
where we used that $W_{(\theta,\phi_2)\leftarrow (\theta,\phi_1)}={\cal P}e^{-i\int^{\phi_2}_{\phi_1} d\phi'  A_{\phi}(\theta,\phi')}$ when the states $\ket{u_{n(\phi,\theta)}}$ are smooth for $\phi_1<\phi<\phi_2$, and $A_{nm,\phi}=\braket{u_{m(\theta,\phi)}|i\d_{\phi}|u_{n(\theta,\phi)}}$ is the $\phi$ component of the Berry connection.
If we take the parallel-transport gauge which is defined by
\begin{align}
\label{parallel-transport_gauge_sphere}
\ket{u^A_{p,n{(\phi,\theta)}}}
&=\left[{\cal P}e^{-i\int^{\phi}_{0} d\phi'  A^A_{\phi}(\theta,\phi')}\right]_{mn}\ket{u^A_{m(\phi,\theta)}},\notag\\
\ket{u^B_{p,n{(\phi,\theta)}}}
&=\left[{\cal P}e^{-i\int^{\phi}_{\pi} d\phi'  A^B_{\phi}(\theta,\phi')}\right]_{mn}\ket{u^B_{m(\phi,\theta)}},\notag\\
\ket{u^C_{p,n{(\phi,\theta)}}}
&=\left[{\cal P}e^{-i\int^{\phi}_{\pi/2} d\phi'  A^C_{\phi}(\theta,\phi')}\right]_{mn}\ket{u^C_{m(\phi,\theta)}},
\end{align}
the Wilson loop operator is then
\begin{align}
W_{0}(\theta)
=W_{p,0}(\theta)
=t^{AB}_p(\theta)t^{BC}_p(\theta)t^{CA}_p(\theta),
\end{align}
where $W_p$ and $t_p$ are the Wilson loop operator and the transition function in the parallel-transport gauge.
Notice that it takes the form of Eq.~(\ref{winding_sphere}). 
Therefore, the winding number of $W_{0}(\theta)$ gives the second Stiefel-Whitney class.

Similarly, the Wilson loop operator on a torus can be related to the transition function.
Let us consider a torus covered with the patches shown in Fig.~\ref{Wilson_patch}(b).
We take a gauge where transition functions are trivial over $\theta=\pi$ line.
Also, we impose on $\theta=0$ that $t^{AD}$ and $t^{BC}$ are the identity (constant orientation-reversing matrices) when the occupied states are orientable (non-orientable) along $\theta$.
Then, we move to the parallel-transport gauge which is defined by
\begin{align}
\label{parallel-transport_gauge_torus}
\ket{u^{A/D}_{p,n{(\phi,\theta)}}}
&=\left[{\cal P}e^{-i\int^{\phi}_{0} d\phi'  A^{A/D}_{\phi}(\theta,\phi')}\right]_{mn}\ket{u^{A/D}_{m(\phi,\theta)}},\notag\\
\ket{u^{B/C}_{p,n{(\phi,\theta)}}}
&=\left[{\cal P}e^{-i\int^{\phi}_{\pi} d\phi'  A^{B/C}_{\phi}(\theta,\phi')}\right]_{mn}\ket{u^{B/C}_{m(\phi,\theta)}},
\end{align}
where $0\le \phi \le \pi$ in the first line, $\pi\le \phi \le 2\pi$ in the second line.
In this gauge, the Wilson loop operator is related to transition functions as
\begin{align}
W_{0}(\theta)
=
\begin{cases}
t_p^{AB}(0,\theta)t_p^{BA}(\pi,\theta) \quad&{\rm for}\; 0\le \theta\le \pi, \\
t_p^{DC}(0,\theta)t_p^{CD}(\pi,\theta) \quad&{\rm for}\; \pi\le \theta\le 2\pi.
\end{cases}
\end{align}
$W_0(2\pi)$ is smoothly defined in the range $0\le \theta<2\pi$, but its periodic condition is nontrivial:
\begin{align}
\label{Wilson_p.c.}
W_{0}(2\pi)
&=t_p^{DC}(0,0)t_p^{CD}(\pi,0)\notag\\
&=t_p^{DA}(0,0)t_p^{AB}(0,0)t_p^{BC}(0,0)\notag\\
&\quad \times t_p^{CB}(\pi,0)t_p^{BA}(\pi,0)t_p^{AD}(\pi,0)\notag\\
&=t_p^{DA}(0,0)W_{0}(0)t_p^{BC}(0,0)\notag\\
&=(t_p^{AD}(0,0))^{-1}W_{0}(0)t_p^{AD}(0,0),
\end{align}
where we used that $t_p^{AD}(\phi,0)=t^{AD}(0,0)$ and $t_p^{BC}(\phi,0)=t^{BC}(\pi,0)$ are independent of $\phi$, which can be shown by
\begin{align}
t^{AD}_p(\phi,0)
&=(W^{A}_{(\phi,0)\leftarrow (0,0)})^{\dagger}t^{AD}(\phi,0)W^D_{(\phi,0)\leftarrow (0,0)}\notag\\
&=(W^{A}_{(\phi,0)\leftarrow (0,0)})^{\dagger}t^{AD}(\phi,0)\notag\\
&\times (t^{AD}(\phi,0))^{\dagger}W^A_{(\phi,0)\leftarrow (0,0)}t^{AD}(0,0)\notag\\
&=t^{AD}(0,0)
\end{align}
and similary for $t^{BC}$.
The homotopy class of the Wilson loop operator $W_0(\theta)$ determines the second Stiefel-Whitney class according to the discussion on Eq.~(\ref{winding_torus}) in the previous section.
We can read off the homotopy class of the Wilson loop operator gauge-invariantly from its spectrum as we will show below.

\subsection{$Z$ and $Z_2$ topology of the Wilson loop}
We have used a particular gauge to relate the Wilson loop operator to the second Stiefel-Whitney class.
Nevertheless, the relation is still meaningful in other gauges because the spectrum of the Wilson loop operator is gauge-invariant~\cite{Wilson_loop}.
One can immediately observe from Eq.~(\ref{Wilson_discrete}) that the Wilson loop operator transforms as
\begin{align}
W_{\phi_0}(\theta)
\rightarrow U^{-1}(\phi_0,\theta)W_{\phi_0}(\theta)U(\phi_0,\theta)
\end{align} 
under any transformation $\ket{u_{n(\phi,\theta)}}\rightarrow U_{mn}(\theta,\phi))\ket{u_{n(\phi,\theta)}}$ which may be discontinuous or complex-valued, where we used a short-hand notation $W_{\phi_0}(\theta)$ for $W_{(\phi_0+2\pi,\theta)\leftarrow (\phi_0,\theta)}$.
Moreover, the change of the initial azimuthal angle is also a similarity transformation because
\begin{align}
W_{\tilde{\phi}_0}(\theta)
&=W_{(\tilde{\phi}_0+2\pi,\theta)\leftarrow (\phi_0+2\pi,\theta)}W_{\phi_0}(\theta)W_{(\phi_0,\theta)\leftarrow (\tilde{\phi}_0,\theta)}\notag\\
&=W_{(\tilde{\phi}_0,\theta)\leftarrow (\phi_0,\theta)}W_{\phi_0}(\theta)W_{(\phi_0,\theta)\leftarrow (\tilde{\phi}_0,\theta)}\notag\\
&=W_{(\phi_0,\theta)\leftarrow (\theta,\tilde{\phi}_0)}^{-1}W_{\phi_0}(\theta)W_{(\phi_0,\theta)\leftarrow (\tilde{\phi}_0,\theta)}
\end{align}
in a smooth complex gauge where the transition function is trivial everywhere on the sphere (we can take this gauge because the Chern number is zero due to the $PT$ symmetry).
Thus, we can work in any gauge with any $\phi_0$ as far as spectrum is concerned because a similarity transformation does not change the spectrum.
We will show below how we can extract the winding number of the Wilson loop operator from its eigenvalues.
In other words, we show how to get the second Stiefel-Whitney class from the eigenvalues of the Wilson loop operator.

In our analysis, it will be convenient to use the exponentiated form of the Wilson loop.
In the real parallel-transport gauges we constructed above, the Wilson loop belongs to the orthogonal group.
We can further restrict it to the special orthogonal group if we calculate the Wilson loop along the orientable cycle.
Then
\begin{align}
W(\theta)=e^{iV({\theta})},
\end{align}	
where we omitted $\phi_0=0$ in the notation, and $V$ is an anti-symmetric imaginary matrix so-called the $PT$-symmetric Wilson Hamiltonian~\cite{multipole}.
We will consider the base manifold first as a sphere and then as a torus.
In both cases, the second Stiefel-Whitney class is given by the parity of the number of the linear crossing of the eigenvalues $\Theta$ of $V$ which occurs on $\Theta=\pi$.

\subsubsection{Sphere}

(1) {\it Integer classification for two occupied bands.|}
When the number of occupied bands is two, the Wilson loop operator is parametrized by only one parameter (rotation angle of the $\text{SO}(2)$ element), i.e.,
\begin{align}
V=
\begin{pmatrix}
0&-ia\\
ia&0
\end{pmatrix},
\end{align}
such that the Wilson loop operator one-to-one corresponds to $a$ modulo $2\pi$.
Accordingly, the winding number of the Wilson loop operator can be read off from the winding of its eigenvalues
\begin{align}
\exp(i\Theta_{\pm})=\exp(\pm ia).
\end{align}
By taking one of the phase eigenvalues (either with $+a$ or $-a$) and counting its winding number by smoothly following it as $\theta$ is varied, we have the winding number of the Wilson loop operator.

\begin{figure}[t!]
\includegraphics[width=8.5cm]{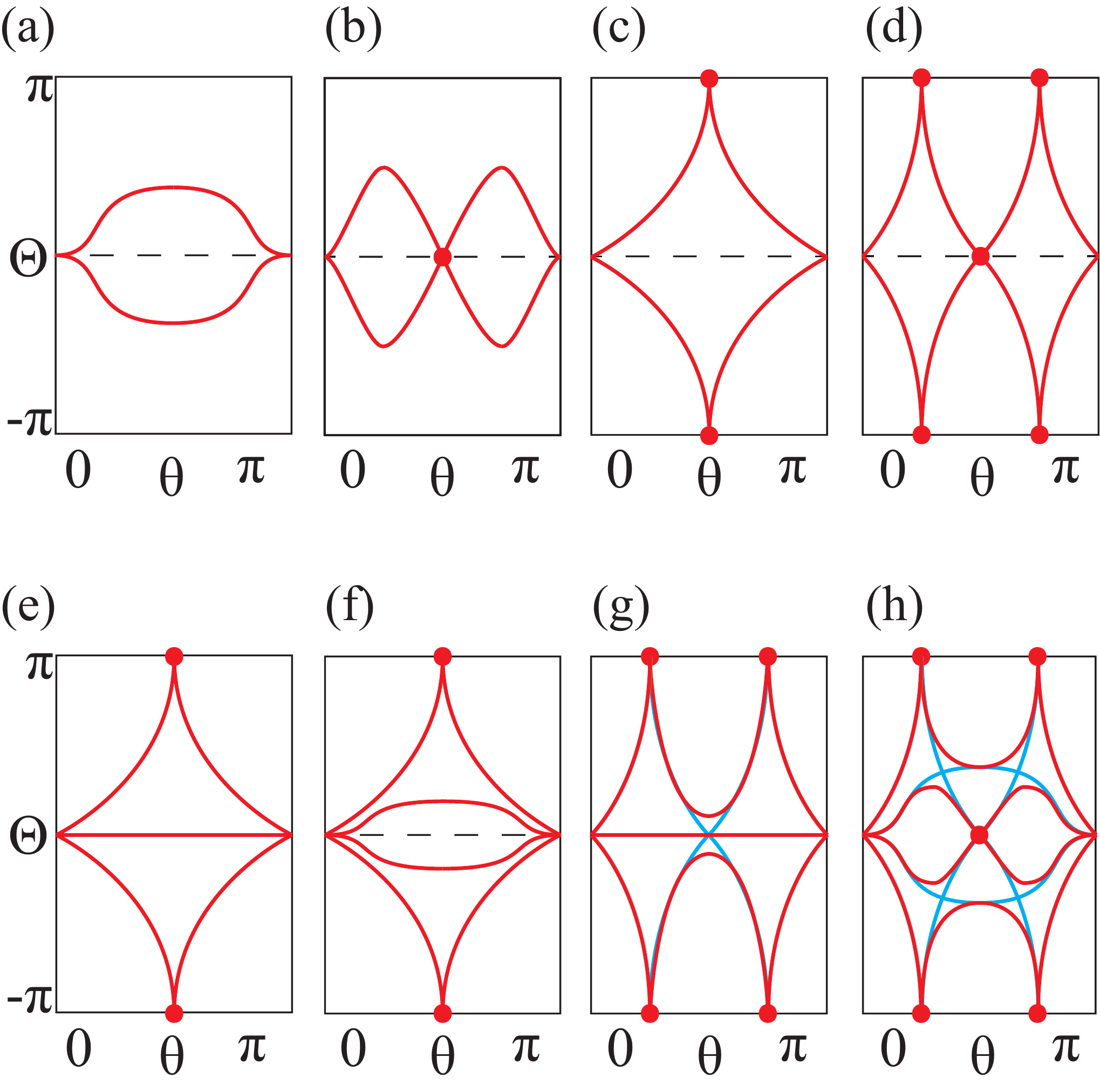}
\caption{Wilson loop spectrum on a sphere.
The Wilson loop operator is calculated along the azimuthal direction with a fixed polar angle $\theta$.
(a-d) two occupied bands.
The winding number is (a,b) zero, (c) one, and (d) two.
(b) can be adiabatically deformed to (a) as we push the crossing point on $\Theta=0$ out of the boundary at $\theta=0$ or $\theta=\pi$.
(e-h) Three and four occupied bands.
(e) shows the spectrum of the generator of the non-trivial homotopy in Eq.~(\ref{WL_generator}).
It is given by adding flat spectra on $\Theta=0$ in addition to (c).
Any nontrivial Wilson loop spectrum is given by smoothly deforming (e).
Adding a small perturbation to (e) does not deform the spectrum for three occupied bands, while it deforms the spectrum to (f) for four occupied bands.
(g) and (h) shows the $Z_2$ nature of the Wilson loop spectrum for three and four occupied bands, respectively.
Blue lines in (g) and (h) are given by adding one and two flat bands to (d), respectively, whereas red lines are the spectrum after adding a $PT$-preserving deformation which eliminates non-protected crossing points.
The crossing points on $\Theta=\pi$ can always be pair-annihilated after this elimination.
}
\label{Wilson_sphere}
\end{figure}

Figure~\ref{Wilson_sphere}(a-d) shows four Wilson loop spectra of two occupied bands.
As (a) and (b) both have zero winding number, (a) can be smoothly deformed to (b), and vice versa.
It is because the crossing point on $\Theta=0$ can be annihilated at the boundary $\theta=0$ or $\theta=\pi$.
One can suppose that the crossing point moves out of the boundary. 
However, (b) cannot be adiabatically deformed to (c) and (d) because they have different winding numbers.

Let us notice that the parity of the winding number, which is the second Stiefel-Whitney class, is given by the parity of the number of crossing points on $\Theta=\pi$.
Thus, we can get the second Stiefel-Whitney class by counting the crossing points on the line.
It is also true when the number of occupied bands is larger than two because the crossing points are stable against adding trivial bands as we show below.

(2) {\it $Z_2$ classification for more than three occupied bands.|}
When the number of occupied bands exceed two, the generator of the nontrivial homotopy $\pi_1(\text{SO}(N_{\text{occ}}))=Z_2$ is given by the generator of $\pi_1(\text{SO}(2))=Z$ defined in a two-band subspace~\cite{Prasolov}.
In other words, a Wilson loop operator is homotopically nontrivial when it can be continuously deformed to
\begin{align}
\label{WL_generator}
W[\theta]
=
\begin{pmatrix}
0&e^{-2i\theta}&\dots&0&0\\
e^{2i\theta}&0&&0&0\\
\vdots&&\ddots&&\\
0&0&&1&0\\
0&0&&0&1
\end{pmatrix},
\end{align}
where $0\le \theta<\pi$.
One can interpret this as follows: by adding topologically trivial occupied bands to an insulator with two occupied bands, we get an insulator with more occupied bands characterized by the same Stiefel-Whitney classes.
Based on this homotopy equivalence, we see that the homotopy class of the Wilson loop operator is given by the winding number of its eigenvalues as in the case of two occupied bands.
We explain below why the winding number is meaningful only modulo two for more than three occupied bands.
We will study the stability of the crossing points because the nontrivial connectivity of the Wilson loop eigenvalues arises from the crossing between the eigenvalues.

The spectrum of the Wilson loop of the type Eq.~(\ref{WL_generator}) has a linear crossing on the line $\Theta=\pi$ [See Fig.~\ref{Wilson_sphere}(e)].
This crossing point is locally stable against continuous deformations as we show now.

Let us consider the most general form of the $2\times 2$ effective Wilson Hamiltonian near a crossing point at $(\theta_0,\Theta_0)$
\begin{align}
V_{\Theta_0}
=v_1\delta \theta\sigma_x+v_2\delta \theta\sigma_y+v_3\delta \theta\sigma_z,
\end{align}
where $\delta \theta=\theta-\theta_0$.
The effective Wilson Hamiltonian should satisfy
\begin{align}
\label{PT_effective}
V^*_{\Theta_0}=-V_{-\Theta_0}
\end{align}
if we take real basis states, because of the $PT$ symmetry constraint $V^*=-V$.
Let us explain more about this.
To avoid the complication coming from choosing a basis, we introduce the basis-independent notation using brakets
$
\hat{V}
=\sum_{\alpha,\beta}\ket{w_{\alpha}}V_{\alpha\beta}\bra{w_{\beta}}$
and
$
\hat{P}_{\Theta_0}
=\sum_{\Theta_{\alpha}=\Theta_0}\ket{w_{\alpha}(\theta_0)}\bra{w_{\alpha}(\theta_0)},
$
where $\hat{V}\ket{w_{\alpha}}=\Theta_{\alpha}\ket{w_{\alpha}}$, and then 
$
\hat{V}_{\Theta_0}
=\hat{P}_{\Theta_0}\hat{V}\hat{P}_{\Theta_0}.
$
In this notation, the basis independent form of $V^*=-V$ is
$
\hat{V}^*=-\hat{V}.
$
Because of this,
$
\hat{V}\ket{w^*_{\alpha}}
=-\Theta_{\alpha}\ket{w^*_{\alpha}}.
$
Then, it follows that
$
\hat{V}^*_{\Theta_0}
=-\hat{V}_{-\Theta_0}.
$
Accordingly, we have the matrix equation Eq.~(\ref{PT_effective}) in the real basis.
At a generic level $\Theta_0\ne n\pi$, where $n$ is any integer, $V_{\Theta_0}$ is not constrained by Eq.~(\ref{PT_effective}).
The constraint Eq.~(\ref{PT_effective}) just tells us that, when there is a crossing point at $\Theta_0$, there is also a crossing point at $-\Theta_0$ .
Then $V_{\Theta_0}$ has two eigenvalues $\delta\Theta=\pm\sqrt{\sum_{i}(v_i\delta\theta)^2}$.
A small perturbation such as $m\sigma_y$ opens a gap.
On the other hand, when $\Theta_0=n\pi$ for an integer $n$, such that $\Theta_0=-\Theta_0$ mod $2\pi$,
\begin{align}
V_{n\pi}=v\delta \theta\sigma_y.
\end{align}
$V_{n\pi}$ has two eigenvalues $\delta\Theta=\pm v\delta\theta$.
The crossing points are stable against adding a perturbation because a small perturbation $m\sigma_y$ just moves the point along the line $\Theta=n\pi$.

While a single linear crossing point is locally stable on the line $\Theta=n\pi$ for any integer $n$, two linear crossing points may be gapped by a pair-annihilation when they are on the same level.
The difference between two occupied bands and more occupied bands comes from whether the pair-annihilation is always possible.
In the case of two occupied bands, two linear crossing points on the line $\Theta=\pi$ at $\theta=\theta_1$ and $\theta=\theta_2$ cannot be pair-annihilated if there is a linear crossing point on the line $\Theta=0$ at $\theta=\theta_0$ such that $\theta_1<\theta_0<\theta_2$ [See Fig.~\ref{Wilson_sphere}(d)].
Because both of the phase eigenvalues are on $\Theta=0$ at $\theta=\theta_0$, no eigenvalues exist on $\Theta=\pi$ at $\theta_0$.
Accordingly, the crossing points at $\theta_1$ and $\theta_2$ on $\Theta=\pi$ cannot be pair-annihilated because they can never reach a polar angle $\theta_0$ which is between $\theta_1$ and $\theta_2$.
On the other hand, a pair-annihilation is always possible on $\Theta=\pi$ for more than three occupied bands.
When the number of the occupied bands is odd, the crossing on $\Theta=0$ involves at least three bands because there is a flat spectrum at $\Theta=0$.
The $3\times 3$ effective $V$ on the $\Theta=0$ line has the form
\begin{align}
V_{0}=\delta\theta
\begin{pmatrix}
0&-iv_1&-iv_2&\\
iv_1&0&-iv_3\\
iv_2&iv_3&0
\end{pmatrix},
\end{align}
whose three eigenvalues are
\begin{align}
\Theta=\pm\sqrt{(v_1\delta\theta)^2+(v_2\delta\theta)^2+(v_3\delta\theta)^2},\; 0.
\end{align}
A gap $m$ opens at the crossing point by a small perturbation $v_1\delta\theta\rightarrow v_1\delta\theta+m$.
No crossing points are locally protected on $\Theta=0$ [See Fig.~\ref{Wilson_sphere}(g)].
Then there is no obstruction to a pair-annihilation on $\Theta=\pi$.
When the number of the occupied bands is even, a linear crossing point can exist on $\Theta=0$ at $\theta_0$ between $\theta_1$ and $\theta_2$.
Nevertheless, if more than four occupied bands exist, the crossing points at $\theta_1$ can always be directly connected to those at $\theta_2$.
Figure~\ref{Wilson_sphere}(h) shows that the crossing points on $\Theta=0$ can be isolated from the crossing points on $\Theta=\pi$ by a continuous deformation, because the crossing points away from $\Theta=n\pi$ are not protected against small perturbations.
A pair-annihilation on $\Theta=\pi$ is not obstructed by the crossing point on $\Theta=0$.

As a consequence, only the parity of the winding number is topologically meaningful when the number of occupied bands exceeds two.
This parity is the second Stiefel-Whitney class and it is given by counting the crossing points on $\Theta=\pi$ modulo two.


\subsubsection{Torus}
\label{subsubsec.Wilson_torus}

\begin{figure}[t!]
\includegraphics[width=8.5cm]{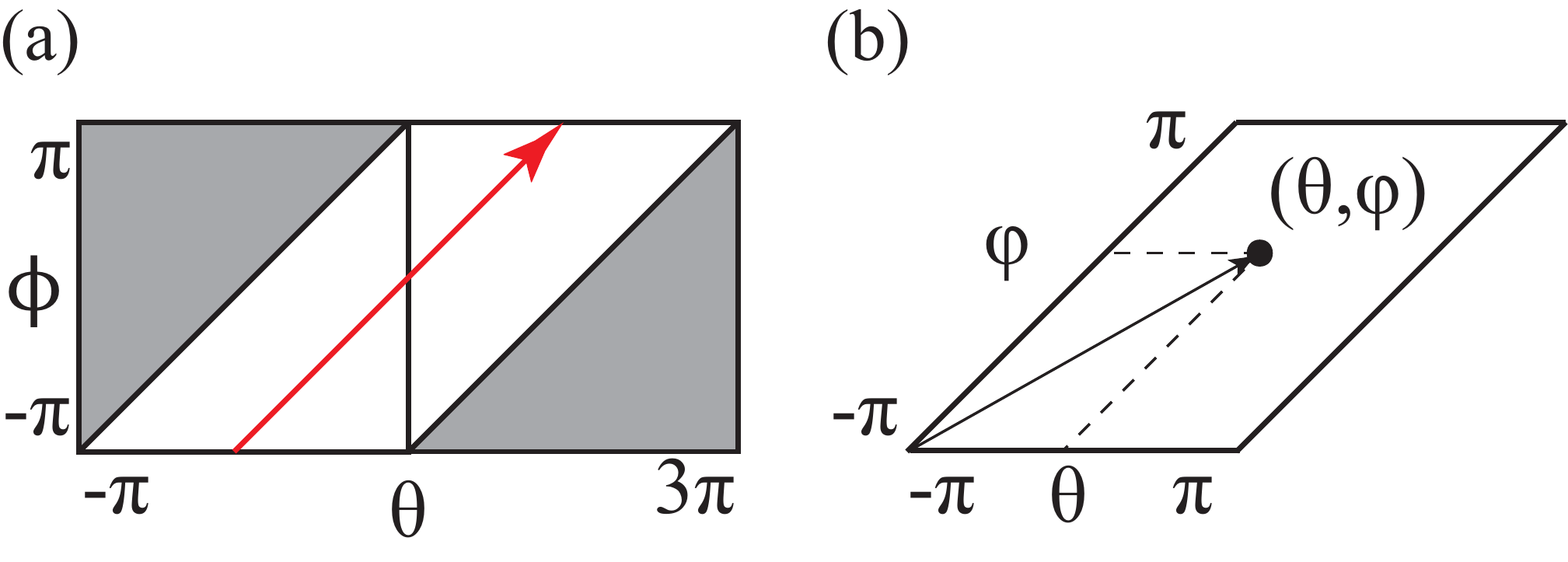}
\caption{Diagonal cycle of a torus.
When the occupied states are non-orientable along both $\phi$ and $\theta$, the diagonal cycle shown as a red arrow in (a) is orientable.
We calculate the Wilson loop operator along the arrow.
It corresponds to choosing new coordinates $(\varphi,\theta)$ in (b) and calculating the Wilson loop along $\varphi$.
}
\label{tilt_torus}
\end{figure}

A torus is described by two periodic cycles $(\phi,\theta)$ along which occupied states may be non-orientable, whereas occupied states are always orientable over a sphere.
We should be cautious due to the potential non-orientability.
If the Wilson loop operator is calculated along a non-orientable cycle, its spectrum cannot show the topological feature of the operator itself when $N_{\text{occ}}$ is even.
It is because the spectrum has two flat spectrum on both $\Theta=0$ and $\Theta=\pi$ as one can see as follows. 
Because $PT$ symmetry requires for a Wilson loop operator $W$ that $W^*=G^{\dagger}WG$ where $G$ is the sewing matrix,
the set $\{\Theta_j\}$ of eigenvalues is equal to the set $\{-\Theta_j\}$ of negative eigenvalues~\cite{group_cohomology}.
Eigenvalues form a pair ($\Theta_j,-\Theta_j$) or take invariant values $\Theta_j=0$ and $\pi$ modulo $2\pi$.
Also, $\exp(i\sum_j\Theta_j)=\det W=e^{i\Phi}=e^{iw_1}=-1$, where  $\Phi$ and $w_1$ is the Berry phase in a complex smooth gauge and the first Stiefel-Whitney class in a real gauge along the non-trivial cycle.
It shows that there is one (mod 2) invariant eigenvalues $\Theta_j=\pi$.
As $N_{\text{occ}}$ is even, there should be one (mod 2) more invariant eigenvalues, which is $\Theta_j=0$.
Thus, we have flat spectrum on both $\Theta=0$ and $\Theta=\pi$ if we calculate the Wilson loop operator along a non-orientable cycle.
Any degeneracy on $\Theta=0$ and $\pi$ then can be lifted after hybridized with the flat spectrum.

For this reason, we will consider Wilson loop operators calculated along the orientable cycle.
It is possible even if both of the cycles $\phi$ and $\theta$ are non-orientable, because then we can calculate the Wilson loop operator along the orientable diagonal direction as shown in Fig.~\ref{tilt_torus}(a).
In other words, we calculate the Wilson loop along the orientable cycle of a different but equivalent torus parametrized by $(\varphi,\theta)$ as shown in Fig.~\ref{tilt_torus}(b), where $(\varphi,\theta)$ in new coordinates correspond to $(\varphi/2,\theta+\varphi/2)$ in the original coordinates.

\begin{figure}[t!]
\includegraphics[width=8.5cm]{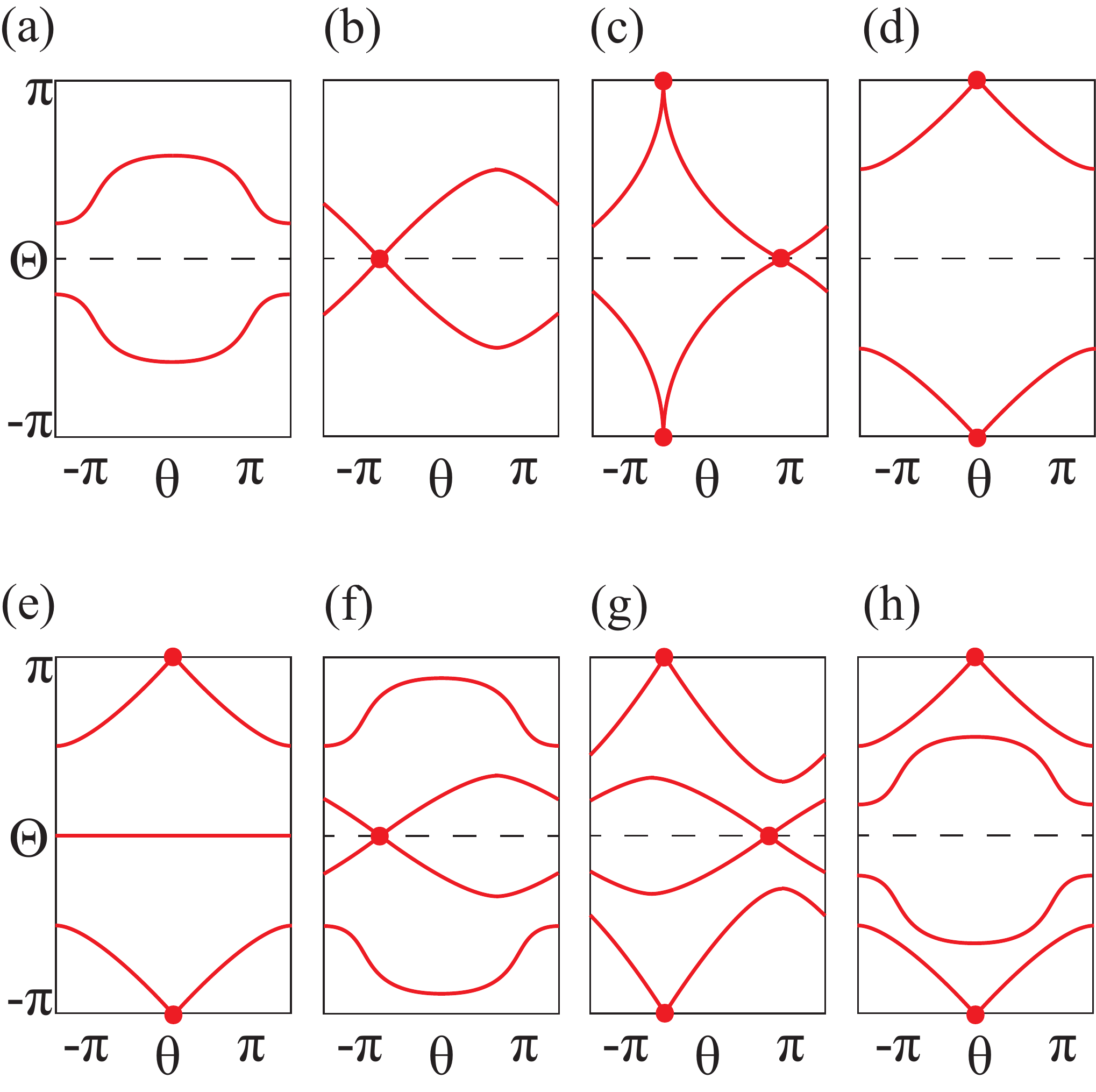}
\caption{Wilson loop spectrum on a torus.
The Wilson loop operator is calculated along the orientable cycle at a fixed $\theta$.
(a-d) Spectrum for two occupied bands.
$(w_{1,\theta},w_2)$=
(a) $(0,0)$,
(b) $(1,0)$,
(c) $(0,1)$,
(d) $(1,1)$.
(e-h) Nontrivial spectrum for three and four occupied bands.
(e) Three occupied bands with $w_2=1$. 
When the number of occupied bands is odd, we can only get $w_2$ if we calculate the Wilson loop operator along a orientable cycle.
(f-h) Four occupied bands.
$(w_{1,\theta},w_2)$=
(f) $(1,0)$,
(g) $(0,1)$,
(h) $(1,1)$.
Here, $w_{1,\theta}$ is the first Stiefel-Whitney class along $\theta$.
}
\label{Wilson_torus}
\end{figure}

Now we analyze the topology of the Wilson loop spectrum on a torus.
As we discussed in Sec.~\ref{subsubsec.torus}, the second Stiefel-Whitney class on a torus is determined according to whether the Wilson loop operator can be continuously deformed into the identity operator (modulo winding the non-contractible cycle even number of times).
Accordingly, the parity of the number of crossing points on $\Theta=\pi$ gives the second Stiefel-Whitney class as it does on a sphere.
For example, $w_2=0$ in Fig.~\ref{Wilson_torus}(a,b,f), and $w_2=1$ in Fig.~\ref{Wilson_torus}(c,d,e,g,h).

What makes the spectrum on a torus distinguished from that on a sphere is the boundary condition of the Wilson loop operator: $W=1$ at $\theta=0$ and $\pi$ on a sphere, and the periodic boundary condition Eq.~(\ref{Wilson_p.c.}) on a torus.
Because the boundary condition on a torus does not require that all eigenvalues are degenerate at a point on $\Theta=0$, an odd number of the crossing points on $\Theta=\pi$ does not necessarily mean that the eigenvalues winds as shown in Fig.~\ref{Wilson_torus}(d,e,g,h).
Moreover, when $N_{\text{occ}}$ is even, crossing points are protected on $\Theta=0$ not only locally but also globally.
As crossing points are protected on both $\Theta=0$ and $\pi$, there are three nontrivial topological phases which is characterized by an odd number of crossing points on $\Theta=0$ only, on $\Theta=\pi$ only, and on both.
This does not apply for an odd $N_{\text{occ}}$ because the crossing points on $\Theta=0$ can be lifted due to the flat spectrum.

Let us investigate on the four topological phases for an even $N_{\text{occ}}$.
We first consider two occupied bands.
Because a crossing point on $\Theta=0$ is topologically stable, the spectrum in Fig.~\ref{Wilson_torus}(a) and (b) [(c) and (d)] is distinct although $w_2=0$ [$w_2=1$] for both.
It is the first Stiefel-Whitney class along $\theta$ ($w_{1,\theta}=1$) which distinguish the two spectrum.
Recall that the periodic condition for the Wilson loop operator is nontrivial when $w_{1,\theta}=1$, and it is given by
\begin{align}
\label{Wilson_winding_torus}
W(2\pi)
&=M^{-1}W(0)M,
\end{align}
where $\det M=-1$, in the parallel-transport gauge defined above.
For two occupied bands, it becomes
\begin{align}
\exp(i\Theta(2\pi)\sigma_y)
&=M^{-1}\exp(i\Theta(0)\sigma_y)M\notag\\
&=\exp(-i\Theta(0)\sigma_y),
\end{align}
which shows that eigenvalues are interchanged as $\theta$ goes from $0$ to $2\pi$ such that an odd number of crossing points occur.
As one can adiabatically diagonalize a Wilson loop operator into $2\times 2$ blocks, this applies to any even $N_{\text{occ}}$.
Three nontrivial topological phases of four occupied bands are shown in Fig.~\ref{Wilson_torus}(f,g,h) which corresponds to $(w_{1,\theta},w_2)=(1,0), (0,1)$, and $(1,1)$, respectively.

\begin{figure}[t!]
\includegraphics[width=8.5cm]{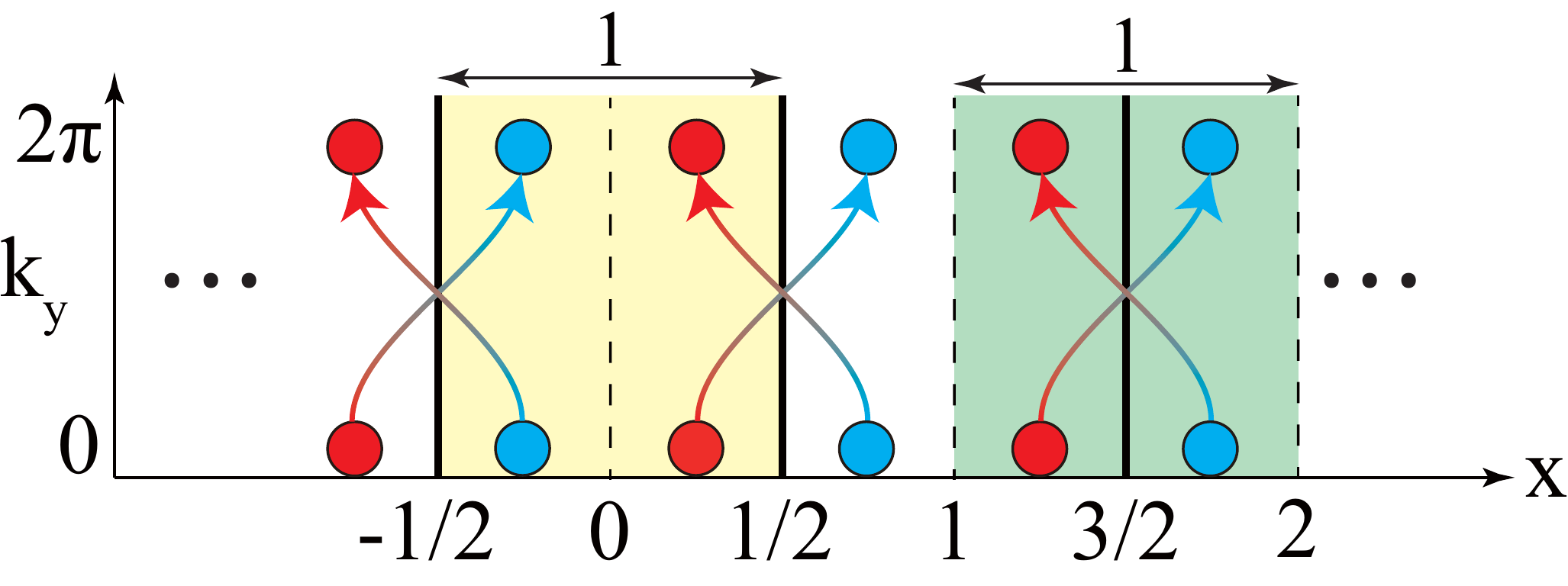}
\caption{Wannier center flows in a $PT$-symmetric system.
Wannier centers relative to the unit cell position are determined, modulo lattice translations, by Wilson loop eigenvalues.
The second Stiefel-Whitney class is therefore nontrivial (trivial) when the left (right) unit cell is chosen.
}
\label{Wannier_flow}
\end{figure}

The spectrum in Fig.~\ref{Wilson_torus} shows that the second Stiefel-Whitney class of a 2D Brillouin zone depends on the choice of the unit cell in real space.
Notice that the spectrum in Fig.~\ref{Wilson_torus}(b) and (d) differ by a constant shift  by $\pi$ while they have different second Stiefel-Whitney class.
The same is true for Fig.~\ref{Wilson_torus}(f) and (h).
Let us use $(k_x,k_y)$ to parametrize the Brillouin zone.
Because the eigenstates of the Wilson loop operator calculated along $k_x$ are Wannier states localized in the $x$-direction~\cite{inversion_Wilson_loop}, such that the eigenvalues are Wannier centers, Fig.~\ref{Wilson_torus}(b) and (d) (also (f) and (h)) shows that a uniform shift of the Wannier centers changes the second Stiefel-Whitney class.
In other words, the second Stiefel-Whitney class can change if we shift the unit cell a half lattice constant.
Figure~\ref{Wannier_flow} shows two choices of unit cells which give different second Stiefel-Whitney class.

On the other hand, the second Stiefel-Whitney class is independent on the choice of the unit cell when $N_{\text{occ}}$ is odd, although the translation of the unit cell can remove the crossing of Wannier centers across its boundary.
According to Sec.~\ref{subsubsec.torus}, if the Wilson loop operator is calculated along the non-orientable cycle, we have $w_2=0$ ($w_2=1$) when the Wilson loop operator can (cannot) be continuously deformed to $-1$ for an odd $N_{\text{occ}}$.
Thus, $w_2$ is determined by the parity of the number of crossing points on $\Theta=0$ if the occupied bands are non-orientable along $k_x$ for an odd $N_{\text{occ}}$.
Consequently, $w_2=0$ and $w_2=1$ transforms to $w_2=0$ and $w_2=1$ by the change of the unit cell, respectively.

This peculiar property of the second Stiefel-Whitney class shows its quadrupole-moment-like character in the following sense.
Consider the electron charge $Q$, dipole moment $P_x$, and quadrupole moment $Q_{xy}$ of a unit cell $v$ in the classical limit.
\begin{align}
Q
&=\int_v d^2r \rho({\bf r}),\notag\\
P_{x}
&=\int_v d^2r \rho({\bf r})x,\notag\\
Q_{xy}
&=\int_v d^2r \rho({\bf r})xy.
\end{align}
When we translate electrons by $(D_x,D_y)$, the quadrupole moment transforms as
\begin{align}
\label{ambiguity_quadrupole}
Q_{xy}'
&=\int_v d^2r \rho({\bf r})(x+D_x)(y+D_y)\notag\\
&=Q_{xy}+D_xP_y+P_xD_y+QD_xD_y.
\end{align}
Quantum mechanically, the charge and dipole moment are defined by
\begin{align}
Q
&=\sum_{n\in {\rm occ}}1,\notag\\
P_{x}
&=\sum_{n\in {\rm occ}}p_{x,n},
\end{align}
where
$
p_{x,n}
=\oint_{BZ} \frac{d^2k}{(2\pi)^2}{\rm Tr}A
=\frac{1}{2}w_{1,k_x}({\cal B}_n)
$
is the dipole moment for the $n$th band.
If we define a {\it pseudo quadrupole moment} by
\begin{align}
Q_{xy}
&\equiv \sum_{n\in {\rm occ}}p_{x,n}p_{y,n},
\end{align}
it conforms to the transformation rule in Eq.~(\ref{ambiguity_quadrupole}).
Then $P_xP_y-Q_{xy}$ transforms as
\begin{align}
P'_xP'_y-Q'_{xy}
=P_xP_y-Q_{xy}+\Delta,
\end{align}
where
\begin{align}
\label{Delta}
\Delta=(Q-1)(D_xP_y+P_xD_y+QD_xD_y).
\end{align}
Consider the case where $D_x$ and $D_y$ are half-integers.
We immediately see that $\Delta=0$ when $N_{\text{occ}}=Q$ is odd.
Moreover, $\Delta=1/4$ modulo $1/2$ when $N_{\text{occ}}$ is even and $D_xP_y=1/4$ for $P_x=0$.
This is exactly the way the second Stiefel-Whitney class transforms under half-lattice translations.
Indeed, we can show
\begin{align}
w_2
&=4\left(P_xP_y-Q_{xy}\right)
\end{align}
for non-degenerate bands.
It follows from the Whitney sum formula and the K$\ddot{\rm u}$nneth formula:
\begin{align}
w_2({\cal B})
&=\sum_{i<j}w_1({\cal B}_i)\smile w_1({\cal B}_j)\notag\\
&=\sum_{i<j}w_{1,k_x}({\cal B}_i)w_{1,k_y}({\cal B}_j)-w_{1,k_x}({\cal B}_j)w_{1,k_y}({\cal B}_i)\notag\\
&=\sum_{i<j}w_{1,k_x}({\cal B}_i)w_{1,k_y}({\cal B}_j)+w_{1,k_x}({\cal B}_j)w_{1,k_y}({\cal B}_i)\notag\\
&=\sum_{i\ne j}w_{1,k_x}({\cal B}_i)w_{1,k_y}({\cal B}_j)\notag\\
&=\sum_{i,j}w_{1,k_x}({\cal B}_i)w_{1,k_y}({\cal B}_j)-\sum_{i}w_{1,k_x}({\cal B}_i)w_{1,k_y}({\cal B}_i)\notag\\
&=4P_xP_y-4Q_{xy}.
\end{align}

Let us notice that the pseudo quadrupole moment is different from the physical quadrupole moment in general.
The gauge-invariant form of $Q_{xy}$ is given by $P_xP_y-w_2/4$.
Because $4P_xP_y$ and $w_2$ are $Z_2$ topological invariants defined modulo two, $4P_xP_y-w_2$ is also a $Z_2$ topological invariant~\footnote{This topological invariant is a real gauge analog of the second Chern character ${\rm Ch}_2$, which is related to the first Chern class $c_1$ and second Chern class $c_2$ by ${\rm Ch}_2=\frac{1}{2}c_1\smile c_1-c_2$~\cite{Nakahara}.}, which reduces to $4Q_{xy}$ for non-degenerate occupied bands.
Nevertheless, it is not a physical quantized quadrupole moment when there is no addition symmetry, because $PT$ symmetry alone does not quantize the quadrupole moment~\cite{multipole}.
The physical quantized quadrupole moment arises when the Wilson loop spectrum is gapped~\cite{multipole} while in our case the spectrum has gap-closing points when $P_x=P_y=0$ and $4Q_{xy}=1$ (i.e., $w_2=1$ mod 2).

\subsection{Numerical calculations}

Using simple models, we demonstrate the nontrivial patterns of the Wilson loop spectrum which we have studied above.
We will consider two $4\times 4$ model Hamiltonians and their extension to $6\times 6$ models.
They are all real because we take $PT=K$.

One is a model of a nodal line with an integer monopole charge $e_2=n$ (See Eq.~(\ref{Euler}) for the definition of this integer monopole charge.).
\begin{align}
\label{monopole_model}
H({\bf k})
&={\rm Re}(k_+^n)\Gamma_1+{\rm Im}(k_+^n)\Gamma_2+k_z\Gamma_3+m\Gamma_{15},
\end{align}
where $k_+=k_x+ik_y$, $\Gamma_{(1,2,3)}=(\sigma_x,\tau_y\sigma_y,\sigma_z)$, $\Gamma_{(4,5)}=(\tau_x\sigma_y, \tau_z\sigma_y)$, and $\Gamma_{ij}=[\Gamma_i,\Gamma_j]/2i$.
Its energy spectrum is given by $E=\pm\sqrt{f_1^2+\left(f_\rho\pm |m|\right)^2}$, where $f_{\rho}=\sqrt{f_2^2+f_3^2}$, where $(f_{1},f_2,f_3)=({\rm Re}(k_+^n),{\rm Im}(k_+^n),k_z)$.
Af half-filling ($E_F=0$), this model describes a nodal line with monopole charge $n$, as we can see by taking the zero-size limit $m\rightarrow 0$ to have a nodal point with the same monopole charge.
In this zero-size limit, the monopole charge is given by $e_2=(1/8\pi)\oint_{S^2}d{\bf k}\cdot\sum_{i,j,k}\epsilon^{ijk} \hat{f}_i\nabla_{\bf k}\hat{f}_j\times \nabla_{\bf k}\hat{f}_k$ for $H=\sum_{i=1,2,3}f_i\Gamma_i$ as shown in Sec.~\ref{skyrmion} [See Ref.~\onlinecite{Fang_inversion} also].
Therefore, $e_2=n$, and $w_2=n$ modulo two in this model.

The other is a lattice model which can describe pair-creations of $Z_2$ monopoles.
\begin{align}
\label{lattice_model}
H({\bf k})
&=\sum_{i=1}^3 f_i({\bf k})\Gamma_i+m\Gamma_{15},
\end{align}
where $f_1=2\sin k_x$, $f_2=2\sin k_y$, $f_3=M+2(\cos k_x-1)+2(\cos k_y-1)+\cos k_z-1$.
This system is an insulator at $M<-m$, and a single trivial nodal line is created from ${\bf k}={\bf 0}$ when $M$ exceeds the critical value $M=-m$.
As we increase $M$ further, the system evolves in two ways according to the value of $m$.
(1) If $|m|<1$, the single trivial line is separated into a pair of  $Z_2$ monopole lines at $M=m$.
The plane $k_z=0$ then has a unit $Z_2$ flux (i.e., $w_2=1$) because it is between the two $Z_2$ monopoles.
(2) If $|m|>1$, the single trivial line is not separated but increases its size, and it crosses the Brillouin zone boundary $(k_x,k_y,k_z)=(0,0,\pi)$ at $M=2-m$.
The evolution from $M<-m$ to $M>2-m$ induces a Berry phase transition on the plane $k_y=0$ such that the plane has a nontrivial Berry phase along $k_x$ for $2-m<M<m$ (gap closes again at $M=m$).

We can extend these two models into $6\times 6$ models by adding two trivial bands and hybridizing them with other bands as follows
\begin{align}
\label{6by6}
H_{6\times 6}=
\begin{pmatrix}
\multicolumn{4}{c}{\multirow{4}{*}{$H_{4\times 4}$}}&p_{1}&p_{2}\\
\multicolumn{4}{c}{}&p_{3}&p_{4}\\
\multicolumn{4}{c}{}&p_{5}&p_{6}\\
\multicolumn{4}{c}{}&p_{7}&p_{8}\\
p_{1}&p_{3}&p_{5}&p_{7}&\mu_5&p_{9}\\
p_{2}&p_{4}&p_{6}&p_{8}&p_{9}&\mu_6
\end{pmatrix},
\end{align}
where $p_{i=1,...,9}$ and $\mu_{i=5,6}$ are real.
For $E_F=0$, we have three (four) occupied bands by considering $\mu_{5}=-\mu_{6}=0.5$ ($\mu_{5}=\mu_{6}=-0.5$).

\begin{figure}[t!]
\includegraphics[width=8.5cm]{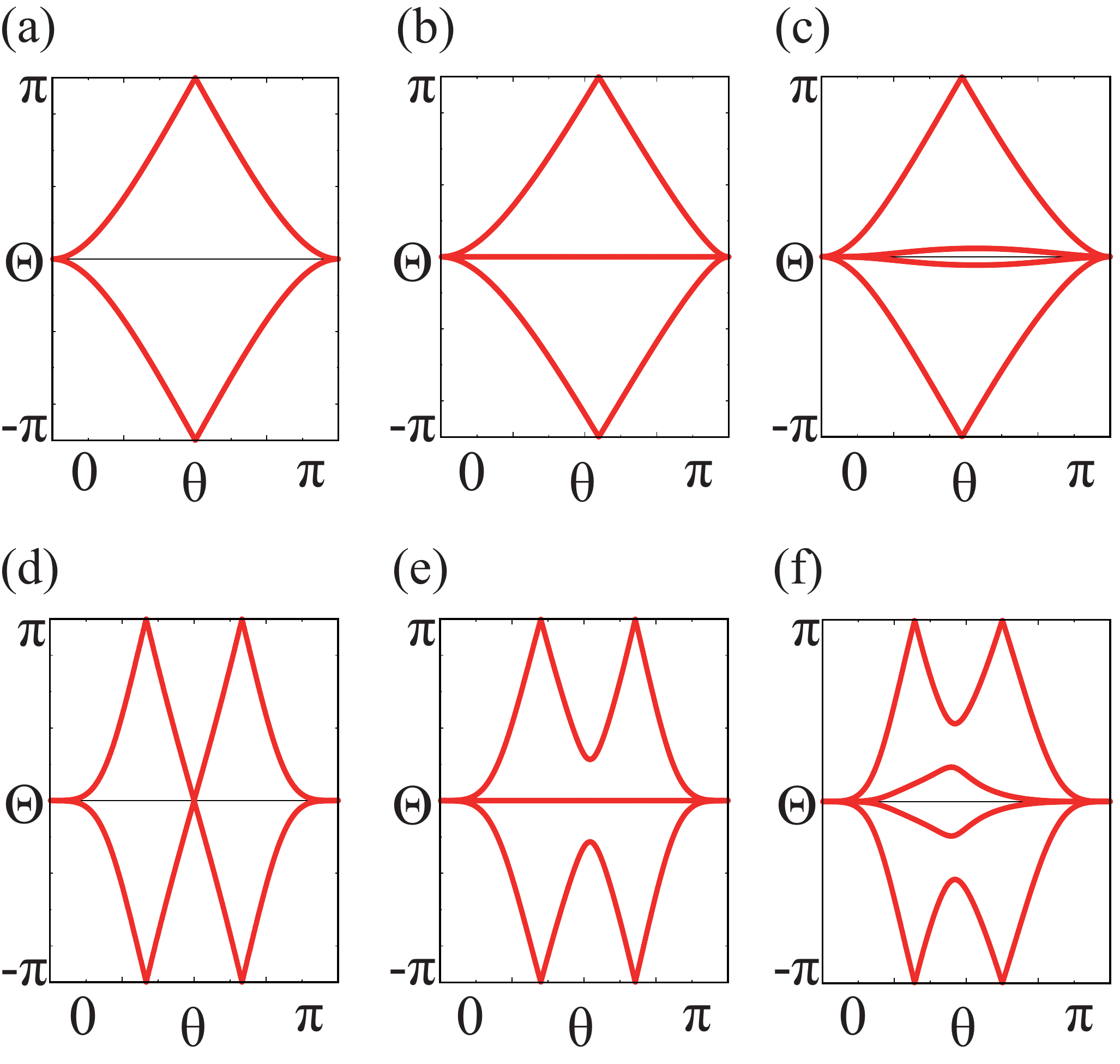}
\caption{Numerical calculations of the Wilson loop spectrum on spheres.
The Wilson loop operator is calculated along $\phi$ at each $\theta$ over the unit sphere $|{\bf k}|=1$, where $(k_x,k_y,k_z)=|{\bf k}|(\sin\theta\cos\phi,\sin\theta\sin\phi,\cos\theta)$, using the model in Eq.~(\ref{monopole_model}) with $m=0.5$ for (a,d) and its $6\times 6$ extension through Eq.~(\ref{6by6}) for (b,c,e,f).
The Fermi level is given by $E_F=0$, which is half-filling for the four-band model.
One and two more bands are occupied in the six-band model by choosing (b,e) $\mu_{5}=-\mu_{6}=0.5$ and (c,f) $\mu_{5}=\mu_{6}=-0.5$, respectively.
$p_6=0.5\sin k_x$, $p_7=\sin k_x$, and $p_{i\ne 6,7}=0$ for (b,e,c,f).
(a,b,c) $n=1$.
The winding number of the Wilson loop eigenvalues is identical to $n$ independent of the number of occupied bands.
(d,e,f) $n=2$.
The double winding number in (d) deforms to the zero winding number in (e) and (f) after one and two more occupied bands are added.
The winding number of the Wilson loop operator is stable only modulo two for more than three occupied bands.
}
\label{Wilson_numerics_sphere}
\end{figure}

Figure~\ref{Wilson_numerics_sphere} shows the Wilson loop spectrum on the sphere which encloses the gap-closing object described by Eq.~(\ref{monopole_model}) and its extension through Eq.~(\ref{6by6}).
For two occupied bands, the (integer) monopole charge is identical to the winding number of the Wilson loop eigenvalues as shown in Fig.~\ref{Wilson_numerics_sphere}(a) and (d).
On the other hand, for three or four occupied bands, the monopole charge is identical to the winding number only modulo two.
After we add occupied bands, the double winding in Fig.~\ref{Wilson_numerics_sphere}(d) can deform to the trivial winding in Fig.~\ref{Wilson_numerics_sphere}(e,f), whereas the single winding in Fig.~\ref{Wilson_numerics_sphere}(a) remains nontrivial as in Fig.~\ref{Wilson_numerics_sphere}(b,c).

\begin{figure}[t!]
\includegraphics[width=8.5cm]{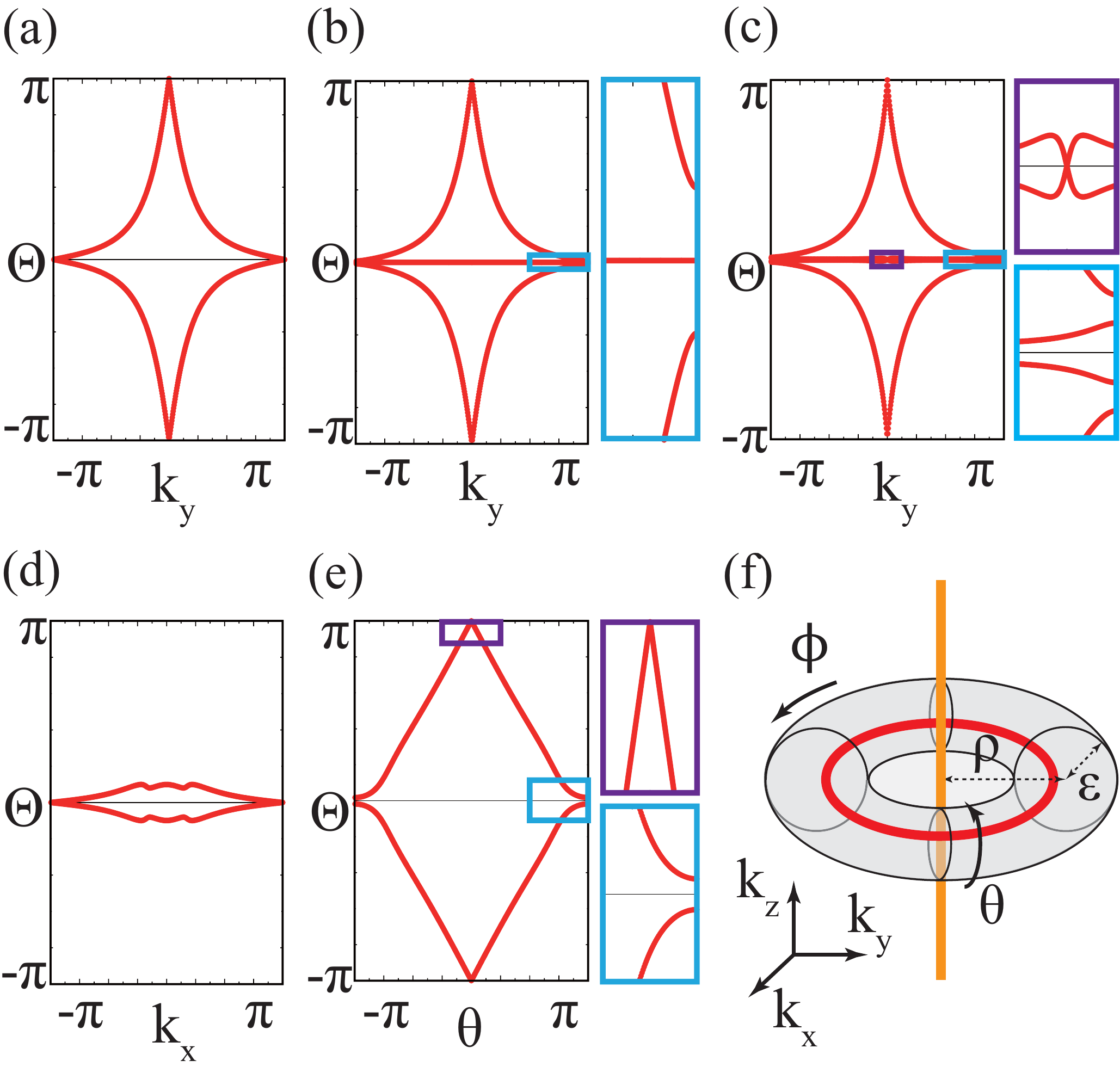}
\caption{Numerical calculations of the Wilson loop spectrum on tori.
(a,b,c) Spectrum of the orientable occupied bands.
The Wilson loop operator is calculated along $k_x$ for each $k_y$ on the plane $k_z=0$ using the model Eq.~(\ref{lattice_model}) with $m=0.5$ and $M=1$ in (a) and its $6\times 6$ extension by Eq.~(\ref{6by6}) in (b) and (c).
As we take $E_F=0$, three and four bands are occupied for (b) $\mu_{5}=-\mu_{6}=0.5$ and (c) $\mu_{5}=\mu_{6}=-0.5$, respectively.
$p_6=0.5$ and $p_7=1$ for both (b) and (c).
(d) $w_2=0$ and $w_{1,k_x}=1$.
The Wilson loop operator is calculated along $k_z$-direction at each $k_x$ on the plane $k_y=0$ using Eq.~(\ref{lattice_model}) with $m=1.2$ and $M=1$.
(e) $w_2=1$ and $w_{1,\theta}=1$.
The Wilson loop operator is calculated along the toroidal cycle $\phi$ at each polar angle $\theta$ on a torus enclosing a $Z_2$ monopole line.
We used Eq.~(\ref{monopole_model}) with $n=1$ and $m=0.1$ plus an additional term $\frac{1}{2}\Gamma_{35}$, and the enclosing torus is parametrized by $\rho=0.6$ and $\epsilon=0.2$, where $(k_x,k_y,k_z)=((\rho +\epsilon\cos\theta)\cos\phi, (\rho+\epsilon\cos\theta) \sin\phi, \epsilon\sin\theta)$.
Here $w_{2,i}$ is the first Stiefel-Whitney class calculated along the $i$-cycle.
}
\label{Wilson_numerics_torus}
\end{figure}

Now we move to the spectrum on torus shown in Fig.~\ref{Wilson_numerics_torus}.
Figure~\ref{Wilson_numerics_torus}(a,b,c) shows the Wilson loop spectrum for orientable occupied bands on torus.
For two orientable occupied bands, the spectrum on torus has the same form as the spectrum on sphere.
Figure~\ref{Wilson_numerics_torus}(a) shows the Wilson loop spectrum on the two-dimensional Brillouin zone $k_z=0$ described by Eq.~(\ref{lattice_model}) with $m=0.5$ and $M=1$.
As the first Stiefel-Whitney class is trivial over the Brillouin zone, $w_2=1$ is characterized by the single winding of the Wilson loop eigenvalues.
However, if we add one and two occupied bands, the spectrum for $w_2=1$ is not characterized by the winding of the eigenvalues as shown in Fig.~\ref{Wilson_numerics_torus}(b) and (c).
What characterizes $w_2=1$ phase on torus is the odd number of crossing on $\Theta=\pi$.
In Fig.~\ref{Wilson_numerics_torus}(c), the eigenvalues cross on $\Theta=0$ also because the sum of the number of crossing points on $\Theta=0$ and $\Theta=1$ is even when the first Stiefel-Whitney class is trivial along $k_y$.

When the first Stiefel-Whitney class is nontrivial along the cycle along which Wilson loop operator evolves, the total number of the crossing points is odd as shown in Fig.~\ref{Wilson_numerics_torus}(d) and (e).
The spectrum in Fig.~\ref{Wilson_numerics_torus}(d) is calculated with Eq.~(\ref{lattice_model}) with $m=1.2$ and $M=1$, which described an insulator with $w_1=1$ only along $k_x$, and $w_2=0$.
There is one crossing point on $\Theta=0$ at $k_x=\pi$.
Figure~\ref{Wilson_numerics_torus}(e) shows the spectrum over a torus enclosing the nodal line described by Eq.~(\ref{monopole_model}) with $m=0.1$ and an additional term $\frac{1}{2}\Gamma_{35}$.
The torus is defined by $\rho=0.6$ and $\epsilon=0.2$, where $(k_x,k_y,k_z)=((\rho +\epsilon\cos\theta)\cos\phi, (\rho+\epsilon\cos\theta) \sin\phi, \epsilon\sin\theta)$ [See Fig.~\ref{Wilson_numerics_torus}(f)].
While the spectrum has a crossing point on $\Theta=0$, it is gapped on $\Theta=0$.
It is because not only $w_2=1$ but also $w_1=1$ along $\theta$ on the torus.

\section{Inversion Symmetry}
\label{sec.inversion}

We have developed our theory by requiring only the combination of inversion and time reversal symmetries.
Many $PT$-symmetric systems, however, have both inversion and time reversal symmetries.
Inversion symmetry helps us to identify the topological phase of matter because the phase is partially determined by the inversion eigenvalues at the inversion-invariant momenta (which is commonly called time-reversal-invariant momenta or TRIM in short)~\cite{Fu-Kane_inversion,inversion,inversion_response,Kim,indicator,Fang_inversion}.
Here we derive the formula for calculating the second Stiefel-Whitney class using inversion eigenvalues, which is
\begin{align}
\label{second_inversion}
(-1)^{w_2}
&=\prod_{i=1}^4(-1)^{[N^-_{\text{occ}}(\Gamma_i)/2]},
\end{align}
where $\Gamma_{i=1,2,3,4}$ are four TRIM on the inversion-invariant plane where $w_2$ is evaluated, $N_{\text{occ}}^-(\Gamma_i)$ is the number of occupied bands with negative inversion eigenvalues at $\Gamma_i$, and the bracket means the greatest integer function, i.e., $[n+x]=n$ for $n\in {\bb Z}$ and $0\le x<1$.
After we derive Eq.~(\ref{second_inversion}), we discuss about using the formula to count the $Z_2$ monopole charges in the Brillouin zone.
Finally, we show that the quantized magnetoelectric polarization can be induced by applying a magnetic field on a nodal line semimetal with odd pairs of $Z_2$ monopoles.


\subsection{The first Stiefel-Whitney class from parity}
Before we move on, let us briefly review the relation between the Berry phase and inversion eigenvalues following Ref.~\onlinecite{Kim,inversion}, because it is needed in the derivation of our formula.
Here the Berry phase is calculated in a smooth complex gauge.
As shown in Sec.~\ref{subsec.orientation}, this Berry phase corresponds to the first Stiefel-Whitney class $w_1$ of real gauges.

Inversion symmetry imposes a constraint on the Berry connection by
\begin{align}
{\rm Tr}A({\bf k})+{\rm Tr}A(-{\bf k})=-i\nabla_{\bf k}\log \det B({\bf k}),
\end{align}
where $B_{mn}({\bf k})=\bra{u_{m\bf -k}}P\ket{u_{n\bf k}}$ is the sewing matrix for inversion symmetry.
Accordingly, the Berry phase along an inversion-invariant line is given by~\cite{inversion,Kim}
\begin{align}
\int^{\pi}_{-\pi} dk{\rm Tr}A(k)
&=\int^{\pi}_{0}dk \left({\rm Tr}A(k)+{\rm Tr}A(-k)\right)\notag\\
&=\int^{\pi}_{0}dk -i\nabla_k\log \det B(k)\notag\\
&=-i\log \frac{\det B(\pi)}{\det B(0)}.
\end{align}
That is,
\begin{align}
\label{Berry_inversion}
\exp\left[i\int^{\pi}_{-\pi} dk{\rm Tr}A(k)\right]
&=\frac{\det B(\pi)}{\det B(0)}\notag\\
&=\det B(\pi)\det B(0)\notag\\
&=\prod_{i=1}^2\prod_{n=1}^{N_{\text{occ}}}\xi_n(\Gamma_i)\notag\\
&=\prod_{i=1}^2\xi(\Gamma_i),
\end{align}
where we used $\det B(0)=\pm 1$, $\Gamma_1=0$ and $\Gamma_2=\pi$, $\xi_n=\pm 1$ is the inversion eigenvalue of the $n$th occupied band, and $\xi(\Gamma_i)$ is the product of all inversion eigenvalues of occupied states over the TRIM $\Gamma_i$ for simplicity of notation.
The product of $\xi$ at two TRIM gives the Berry phase along the inversion-invariant line passing through the two TRIM.

By applying the above relation, we can investigate the inversion-required band degeneracies~\cite{Kim}.
Let us recall that, in $PT$-symmetric systems, the nontrivial Berry phase along a contractible loop indicates that band degeneracies are enclosed by the loop.
The band degeneracies appear as nodal lines in the 3D Brillouin zone.
If we consider the Berry phase $\Phi$ around the half of an inversion-invariant plane, which we parametrize by $0\le k_x\le \pi$ for simplicity, it is given by
\begin{align}
\label{count_line}
(-1)^{\Phi/\pi}
&=\exp\left[i\oint_{\d (\rm hIP)} d{\bf k}\cdot {\rm Tr}{\bf A}({\bf k})\right]\notag\\
&=\exp\left[i\oint dk_y {\rm Tr}A_y(\pi,k_y)-i\oint dk_y {\rm Tr}A_y(0,k_y)\right]\notag\\
&=\prod_{i=1}^2\xi(\Gamma_i)\left[\prod_{i=3}^4\xi(\Gamma_i)\right]^{-1}\notag\\
&=\prod_{i=1}^4\xi(\Gamma_i),
\end{align}
where $\d (\rm hIP)$ is the boundary of the half invariant plane, and $\Gamma_i$'s are contained in the invariant plane.
Therefore, when the product of all inversion eigenvalues over four TRIM is $-1$ ($+1$), an odd (even) number of nodal lines penatrating through the half of the inversion-invariant plane containing the four momenta.
Here only the parity of the number of nodal lines can be counted because an even number of nodal lines can be removed from an invariant plane without changing the inversion eigenvalues.

\subsection{The second Stiefel-Whitney class from parity}

We are now ready to derive an analogous formula for the second Stiefel-Whitney class.
While we have taken a complex smooth gauge to associate the first Stiefel-Whitney class with inversion eigenvalues, we now take a real gauge to associate the second Stiefel-Whitney class charge with inversion eigenvalues.

\subsubsection{Two occupied bands}

First we consider two orientable occupied bands over an invariant plane.
Because the first Stiefel-Whitney class is trivial for orientable occupied bands, there are  two cases according to Eq.~(\ref{Berry_inversion}): the product of two inversion eigenvalues at each TRIM is all negative or all positive.

In the former, the second Stiefel-Whitney class is trivial because the occupied bands can be deformed to topologically trivial bands without closing the band gap between the conduction and valence band as follows.
We invert the occupied bands such that the topmost occupied band has the positive inversion eigenvalue and the other has the negative eigenvalue at each TRIM.
Then, by applying Eq.~(\ref{count_line}) to the lowest occupied band rather than the whole occupied bands, we find that inversion eigenvalues do not require a degeneracy between the occupied bands because the product of inversion eigenvalues is positive for the lowest occupied bands.
In other words, all degeneracies between the occupied bands can be removed without changing the inversion eigenvalues.
After removing all the accidental degeneracies, we have two non-degenerate orientable occupied bands.
The Whitney sum formula shows that the second Stiefel-Whitney class is trivial for the resulting bands.
As the band gap is not closed during the deformation we have described, the second Stiefel-Whitney class of the original phase is also trivial.

In the latter case where the inversion eigenvalues are the same at each TRIM, it may be impossible to isolate the two occupied bands from each other without closing the band gap between the conduction and valence band.
For example, when the eigenvalues are $-1$ at one or three TRIM and $+1$ at the other TRIM, the degeneracy of the occupied bands is required by Eq.~(\ref{count_line}) applied to the lowest occupied band.
The second Stiefel-Whitney class is nontrivial in this case.
We will show this by associating the flux integral form of the second Stiefel-Whitney class with inversion eigenvalues.
As we did when reviewing the result for the Berry phase, we begin by defining an 1D integral with inversion eigenvalues and then extend the result to 2D integrals.

Inversion symmetry gives the following constraints on the real Berry connection and curvature by
\begin{align}
\label{symmetry-constraint}
A^R({\bf k})
&=-B^{T}({\bf k})A^R({\bf -k})B({\bf k}) -B^{T}({\bf k})\nabla_{\bf k}B({\bf k}),\notag\\
F^R({\bf k})
&=B^{T}({\bf k})F^R({\bf -k})B({\bf k}),
\end{align}
where $A^R_{mn}=\braket{u_{m\bf k}|\nabla_{\bf k}|u_{n\bf k}}$, and $\ket{u_{n\bf k}}$ is real.

We consider an inversion-invariant line and take a smooth real gauge over the line where the sewing matrix is also smooth over the line.
It is always possible because we consider orientable occupied bands.
Then, because the sewing matrix is smooth and $\det B=1$ at TRIM over the line, the sewing matrix belongs to $\text{SO}(2)$, i.e.,
\begin{align}
B({\bf k})
&=
\exp
\begin{pmatrix}
0&\phi({\bf k})\\
-\phi({\bf k})&0
\end{pmatrix}
\end{align}
It follows that the constraint equation for the real Berry connection becomes
\begin{align}
\label{inversion-connection}
A^R({\bf k})+ A^R(-{\bf k})
&=-(B^{T}({\bf k})\nabla_{\bf k}B({\bf k}))\notag\\
&=
\begin{pmatrix}
0&-\nabla_{\bf k}\phi({\bf k})\\
\nabla_{\bf k}\phi({\bf k})&0
\end{pmatrix},
\end{align}
from which we find
\begin{align}
\int^{\pi}_{-\pi}dk A^R_{12}(k)
&=\int^{\pi}_{0}dk \left(A^R_{12}(k)+A^R_{12}(-k)\right)\notag\\
&=-\int^{\pi}_{0}dk \nabla_k\phi(k),
\end{align}
and so
\begin{align}
\exp\left[i\int^{\pi}_{-\pi}dk A^R_{12}(k)\right]
&=\exp\left[-i\int^{\pi}_{0}dk \nabla_k\phi(k)\right]\notag\\
&=\exp\left[-i\left(\phi(\pi)-\phi(0)\right)\right]\notag\\
&=\prod_{i=1}^2\xi_1(\Gamma_i),
\end{align}
where $\xi_1({\Gamma_i})$ is the eigenvalue of the sewing matrix, i.e., $B({\Gamma_i})=\xi_1({\Gamma_i})I_{2\times 2}$.

Next, we consider an inversion-invariant plane.
As the occupied states may not be smooth over the whole plane, the sewing matrix also may not be smooth.
The sewing matrix defined on $C$ and $D$ patches are related to the one defined on $A$ and $B$ patches as
\begin{align}
B^{CD}({\bf k})=(t^{AC}({\bf -k}))^{-1}B^{AB}({\bf k})t^{BD}({\bf k}),
\end{align}
where $A$ and $C$ covers $-\bf k$, and $B$ and $D$ covers $\bf k$, and $t^{AB}$ and $t^{CD}$ are the transition functions defined by
$\ket{u^C_{n\bf -k}}= t^{AC}_{mn}({\bf  -k})\ket{u^A_{n\bf - k}}$ and $\ket{u^D_{n\bf k}}= t^{BD}_{mn}({\bf k})\ket{u^B_{n\bf k}}$.
We require all the transition functions be orientation-preserving.
Then, the above relation shows that the sewing matrix belongs to $\text{SO}(2)$ everywhere on the plane because it belongs to $\text{SO}(2)$ within the patches covering a TRIM.
The symmetry constraint on $F^R$ becomes
\begin{align}
F^R({\bf k})
&=F^R({\bf -k}).
\end{align}
The second Stiefel-Whitney class over the invariant plane is then given by
\begin{align}
\exp\left[i\pi w_2\right]
&=\exp\left[\frac{i}{2}\int^{\pi}_{-\pi} dk_x\int^{\pi}_{-\pi}dk_y F^R_{12}(k_x,k_y)\right]\notag\\
&=\exp\left[i\int^{\pi}_{0} dk_x\int^{\pi}_{-\pi}dk_y F^R_{12}(k_x,k_y)\right]\notag\\
&=\exp\left[i\oint dk_y A^R_{12}(\pi,k_y)-i\oint dk_y A^R_{12}(0,k_y)\right]\notag\\
&=\prod_{i=1}^2\xi_1(\Gamma_i)\left[\prod_{i=3}^4\xi_1(\Gamma_i)\right]^{-1}\notag\\
&=\prod_{i=1}^4\xi_1(\Gamma_i),
\end{align}
where we applied the Stokes' theorem on a patch fully covering the half Brillouin zone to get the second line.

Next, we consider two non-orientable occupied bands.
Because the invariant plane should be gapped in order that the second Stiefel-Whitney class is defined, $\xi(\Gamma_i)$ can be negative at an even number of TRIM.
When $\xi$ is negative at none or all the four TRIM, the occupied bands are orientable, which we have discussed.
The remaining is the case where $\xi(\Gamma_i)$ is negative at two TRIM, which we take as $\Gamma_1$ and $\Gamma_2$.
There are three configuration of inversion eigenvalues up to the permutation $\Gamma_1\leftrightarrow \Gamma_2$ and $\Gamma_3\leftrightarrow \Gamma_4$ and up to a band inversion between the occupied bands.
They are
\begin{align}
&\Gamma_1\Gamma_2\Gamma_3\Gamma_4
\qquad\Gamma_1\Gamma_2\Gamma_3\Gamma_4
\qquad \Gamma_1\Gamma_2\Gamma_3\Gamma_4 \notag\\
&--++ 
\;\qquad++--  
\;\qquad+-+-    \notag\\
&++++ 
\;\qquad----
\;\qquad-++- 
\end{align}
After removing all the accidental degeneracies, the lowest occupied band can be isolated by a band gap from the other band in all three configurations.
In the first and second case, the second Stiefel-Whitney class for the occupied bands is trivial according to the Whitney sum formula.
Here, the Whitney sum formula is $w_2({\cal B}_1\oplus {\cal B}_2)=w_{1\phi}({\cal B}_1)w_{1\theta}({\cal B}_2)-w_{1\theta}({\cal B}_1)w_{1\phi}({\cal B}_2)$ modulo two, where ${\cal B}_1$ and ${\cal B}_2$ is the bottom and top occupied bands.
Because $w_1({\cal B}_1)=0$ for trivial ${\cal B}_1$, we find $w_2=0$.
On the other hand, the second Stiefel-Whitney class is nontrivial in the last case according to the Whitney sum formula, because $(w_{1\phi}({\cal B}_1),w_{1\theta}({\cal B}_1),w_{1\phi}({\cal B}_2),w_{1\theta}({\cal B}_2))=(\pi,\pi,0,\pi)$ or $(0,\pi,\pi,0)$ or their permutation by $\phi\leftrightarrow \theta$.

Let us notice that the second Stiefel-Whitney class is nontrivial only when there is an odd number of TRIM at which inversion eigenvalues for the two occupied bands are both $-1$.
Thus, we can summarize the result for two occupied bands as
\begin{align}
(-1)^{w_2}
&=\prod_{i=1}^4(-1)^{[N^-_{\text{occ}}(\Gamma_i)/2]},
\end{align}
where the bracket is the greatest integer function.

\subsubsection{General occupied bands}

We now extend the above derivation to the case with $N_{\text{occ}}>2$.
We do this by decomposing occupied states into two-level blocks and applying the Whitney sum formula.

Let us consider $2N+1$ or $2N+2$ occupied bands with a non-negative integer $N$ on an inversion-invariant plane.
At each TRIM, we decompose the occupied bands into $N$ pairs which have inversion eigenvalues $--$ or $++$ and the remaining band or bands through the band inversion only between the occupied bands.
Also, we re-order the energy level of the occupied bands such that the remaining block is at the highest level.
We have, for example, the following pattern.
We have $N$ blocks with degenerate inversion eigenvalues e.g.,
\begin{align}
\label{parity-order}
&\Gamma_1\Gamma_2\Gamma_3\Gamma_4 \notag\\
2N+0:&-+-+ \notag\\
2N-1:&-+-+ \notag\\
&\qquad\vdots\notag\\
2:&++-- \notag\\
1:&++-- 
\end{align}
for the $1,2,...,2N$th lowest energy level, and we have a remaining block at the highest energy level, which consists of one and two bands for $N_{\text{occ}}=2N+1$ and $2N+2$, respectively, e.g.,
\begin{align}
\label{parity-order}
&\Gamma_1\Gamma_2\Gamma_3\Gamma_4
\qquad\qquad\quad\; \Gamma_1\Gamma_2\Gamma_3\Gamma_4 \notag\\
2N+1:&-++-
\qquad 2N+2: -++- \notag\\
&
\qquad\qquad\qquad 2N+1:+-+- 
\end{align}
where $\Gamma_{1},...,\Gamma_4$ are TRIM on the inversion-invariant plane.
After this decomposition, the product of all inversion eigenvalues is positive within each of $N+1$ blocks.
This is obvious for the lowest $N$ blocks, and for the $N+1$ block it follows from that the band gap between the conduction and valence band is open.
Hence, each block can be isolated from the other bands by removing all the accidental degeneracies without changing the inversion eigenvalues.
After lifting all the accidental degeneracies, the second Stiefel-Whitney class of the whole occupied bands is given by summing up the second Stiefel-Whitney class of $N+1$ blocks according to the Whitney sum formula [See Eq.~(\ref{Whitney_sum})].
From the relation between the inversion eigenvalues and the second Stiefel-Whitney class for each block, we find that
\begin{align}
(-1)^{w_2}
&=\prod_{i=1}^4(-1)^{[N^-_{\text{occ}}(\Gamma_i)/2]}.
\end{align}


\subsubsection{$Z_2$ monopole charge}

Suppose that the band gap is open on two invariant planes which do not intersect each other~\footnote{This condition is not satisfied when $\prod_{i=1}^8\xi(\Gamma_i)=-1$ because the band gap is closed on at least one of the two plane according to Ref.~\onlinecite{Kim} [See Eq.~(\ref{count_line})].}.
In such a case, we can detect the $Z_2$ monopole charges in the Brillouin zone using the relation between the inversion eigenvalues and the second Stiefel-Whitney class.
It is because the difference of the second Stiefel-Whitney classes measure the $Z_2$ monopole charges in the half-Brillouin zone.
Let two planes $k_z=0$ and $k_z=\pi$ be gapped for convenience.
We have
\begin{align}
\label{monopole_inversion}
\exp\left[i\pi N_{mp}\right]
&=\exp\left[i\pi \left(w_2(\pi)-w_2(0)\right)\right]\notag\\
&=\prod_{i=1}^4(-1)^{[N^-_{\text{occ}}(\Gamma_i)/2]}\left(\prod_{i=5}^8(-1)^{[n^-_{\text{occ}}(\Gamma_i)/2]}\right)^{-1}\notag\\
&=\prod_{i=1}^8(-1)^{[N^-_{\text{occ}}(\Gamma_i)/2]},
\end{align}
where $N_{mp}$ is the number of $Z_2$ monopoles in the half Brillouin zone.
Eq.~(\ref{monopole_inversion}) can be intuitively understood as counting the number of double band inversions modulo two.
Suppose we start with a trivial insulator whose inversion eigenvalues are all positive.
In order to have $[N^-_{\text{occ}}(\Gamma_i)/2]$ pairs of negative inversion eigenvalues at $\Gamma_i$, we need $[N^-_{\text{occ}}(\Gamma_i)/2]$ times of double band inversions at $\Gamma_i$ modulo two.
As a double band inversion creates a pair of $Z_2$ monopoles in the Brillouin zone, it creates one $Z_2$ monopole in the half Brillouin zone in the presence of inversion symmetry.
The parity of the number of $Z_2$ monopoles in the half Brillouin zone is then given by summing $[N^-_{\text{occ}}(\Gamma_i)/2]$ over all TRIM.
This is exactly what Eq.~(\ref{monopole_inversion}) shows.

\subsection{Magnetoelectric polarization induced by a magnetic field}

Interestingly, our formula for the number of $Z_2$ monopoles is identical to the formula for the magnetoelectric polarizability of inversion symmetric systems~\cite{inversion,inversion_response}.
Let us shortly review the parity formula for the magnetoelectric polarizability following Ref.~\onlinecite{inversion}.
The magnetoelectric polarizability $P_3$ is defined by the integration of the Chern-Simons 3-form~\cite{QHZ}.
\begin{align}
\label{magnetoelectric_polarizability}
P_3
&=\frac{1}{8\pi^2}\int_{\rm BZ}d^3k\epsilon^{ijk}{\rm Tr}\left[A_i\d_jA_k-\frac{2i}{3}A_iA_jA_k\right],
\end{align}
where $A_{i,nm}({\bf k})=\braket{u_{n\bf k}|i\d_{k_i}|u_{m{\bf k}}}$.
In the presence of inversion symmetry, it has the following form~\cite{inversion}
\begin{align}
P_3
&=\frac{1}{48\pi^2}\int_{BZ}d^3k\epsilon^{ijk}{\rm Tr}\left[B^{\dagger}\d_iBB^{\dagger}\d_jBB^{\dagger}\d_kB\right]\notag\\
&=\frac{1}{2}\deg(B),
\end{align}
where $\deg(B)$ is the ``winding number'' of the map $B({\bf k}):T^3\rightarrow \text{U}(N_{\text{occ}})$.
It is a $Z_2$ topological invariant because it is gauge-invariant only modulo one.
After a decomposition into $1\times 1$ and $2\times 2$ blocks as in Eq.~(\ref{parity-order}), $\deg(B)$ is given by the sum of the winding number over all blocks.
Because $T^3\rightarrow \text{U}(1)\approx S^1$ is homotopically trivial, only $2\times 2$ blocks contribute to $\det(B)$.
Within each $2\times 2$ block, the parity of the winding number is given by the parity of the number of points satisfying $B({\bf k})=-1$~\cite{3D_TI_CS}.
Moreover, $B({\bf k})=-1$ occurs even times for ${\bf k}\ne$TRIM because $B({\bf k})=-B^T(-{\bf k})$.
Therefore, the number of TRIM points $\Gamma_i$ satisfying $B({\Gamma_i})=-1$ determines $\deg(B)$ on an isolated $2\times 2$ block with degenerate inversion eigenvalues as in Eq.~(\ref{parity-order}).
As this counting corresponds to calculating $[N^-_{\text{occ}}(\Gamma_i)/2]$, we see that $P_3$ is given by the formula
\begin{align}
\label{P3_inversion}
\exp\left[i 2\pi  P_3\right]
&=\prod_{i=1}^8(-1)^{[N^-_{\text{occ}}(\Gamma_i)/2]},
\end{align}
which is identical to the formula for counting $Z_2$ monopoles in the Brillouin zone.

There are two implications of this identification.
First, $P$- and $T$-symmetric spinless systems do not have an axion insulator phase.
While $P_3$ is well-defined only when the system is gapped, the system becomes gapless when $P_3=1$ according to Eq.~(\ref{P3_inversion}) and Eq.~(\ref{monopole_inversion}).
Therefore, no insulating phase with $P_3=1$ exists in the presence of both inversion and time reversal symmetries.
Instead, there are odd pairs of $Z_2$ monopoles in the Brillouin zone.
Second, we can get an axion insulator by breaking time reversal symmetry of the nodal line semimetal.
If we apply a magnetic field on the $P$- and $T$-symmetric nodal line semimetal hosting odd pairs of $Z_2$ monopoles, time reversal symmetry is broken while inversion symmetry is kept.
Then, after $PT$ symmetry is broken, the nodal lines are fully gapped in general~\cite{Okugawa}.
The resulting insulating phase has a nontrivial magnetoelectric polarizability according to the parity formula.

\begin{figure}[t!]
\includegraphics[width=8.5cm]{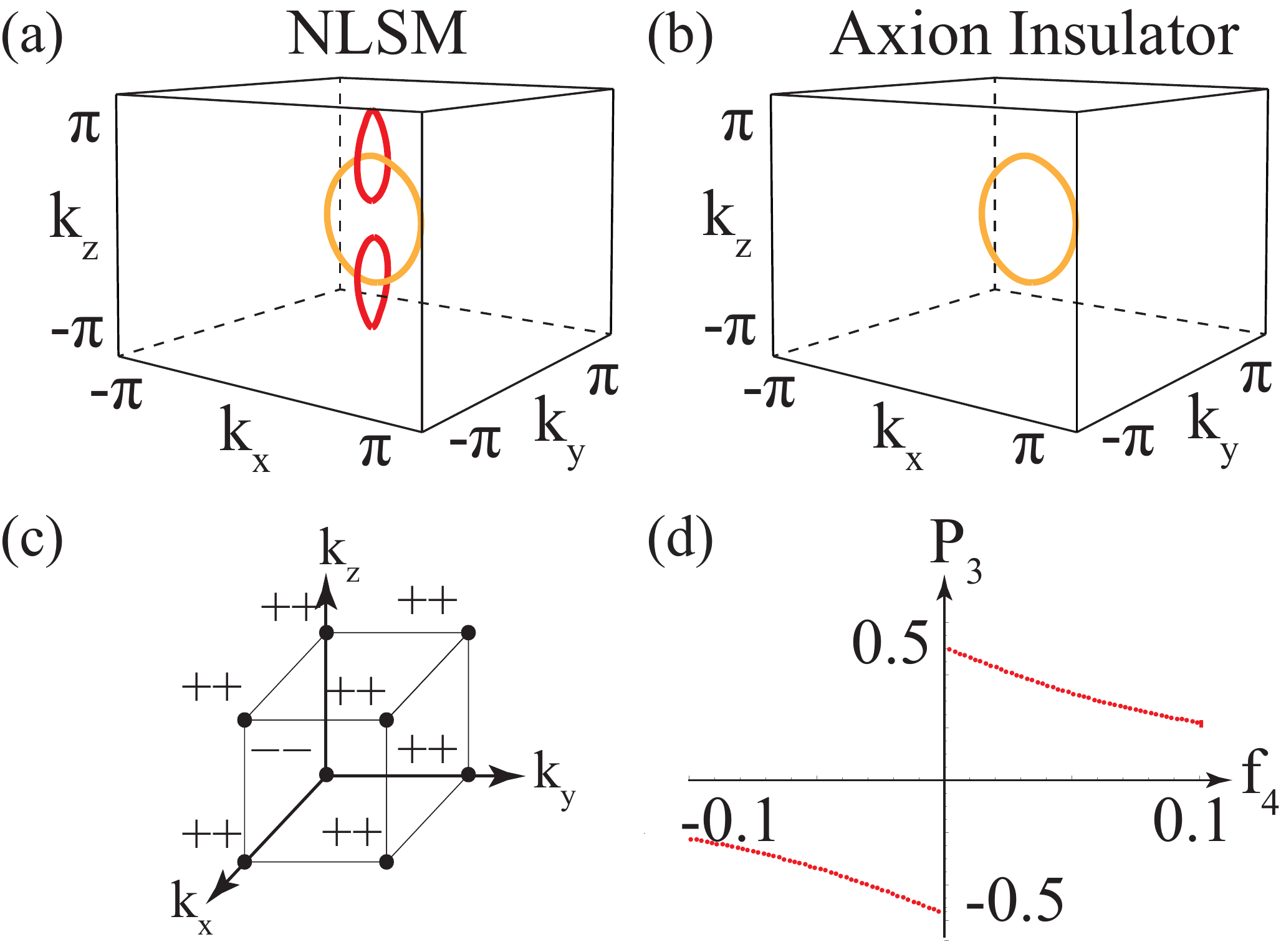}
\caption{Topological phase transition from a nodal line semimetal (NLSM) to an axion insulator in an inversion symmetric system.
(a,b) Nodal structures calculated using Eq.~(\ref{TMEI_transition}) with $m=0.9$, and
$b=0$ in (a) and $b=0.1$ in (b).
Two nodal loops (red) in (a) are fully gapped by breaking time reversal symmetry in (b), while the degeneracy of the occupied bands (orange) remain.
(c) Inversion eigenvalues of occupied bands at eight TRIM for both (a) and (b).
(d) Numerical calculation of magnetoelectric polarizability.
$P_3$ is calcaulted using Eq.~(\ref{P3_numerical_formula}) for the model Eq.~(\ref{TMEI_transition}) with  $b=0.1$, $m=0$.
$P_3$ approaches to $\pm 1/2$ as the constant inversion breaking term $f_4$ vanishes.
}
\label{NLSMtoTMEI}
\end{figure}

For definiteness, let us demonstrate the transition from a nodal line semimetal to an axion insualtor.
We consider the model in the main text with $M=-1$, $r=1/2$, and an additional time reversal breaking term $f_5\Gamma_5$.
\begin{align}
\label{TMEI_transition}
H({\bf k})
&=\sum_{i=1}^3 f_i({\bf k})\Gamma_i+m\Gamma_{15}+ f_5({\bf k})\Gamma_5,
\end{align}
where $\Gamma_{(1,2,3)}=(\sigma_x,\tau_y\sigma_y,\sigma_z)$, $\Gamma_{(4,5)}=(\tau_x\sigma_y, \tau_z\sigma_y)$,
$\Gamma_{ij}=[\Gamma_i,\Gamma_j]/2i$, $f_1=2\sin k_x$, $f_2=2\sin k_y$, $f_3=-4+2\cos k_x+2\cos k_y+\cos k_z$, and $f_5=b\sin k_z$.
In this model, band gap can close only when $b=0$ because the energy eigenvalues are given by
\begin{align}
E=\pm\sqrt{f_1^2+\left((\sqrt{f_2^2+f_3^2}(\pm) m\right)^2+f_5^2}.
\end{align}
When $b=0$, the Hamiltonian is invariant under inversion $P=\Gamma_3$, and time reversal $T=\Gamma_{3}K$.
It describes a nodal line semimetal with two nodal lines with $Z_2$ monopole charges as shown in Fig.~\ref{NLSMtoTMEI}(a).
The presence of $Z_2$ monopoles can be captured from inversion eigenvalues shown in  Fig.~\ref{NLSMtoTMEI}(c) according to our formula Eq.~(\ref{monopole_inversion}).
When $b\ne 0$, time reversal symmetry is broken while inversion symmetry is preserved.
Figure~\ref{NLSMtoTMEI}(b) shows that the nodal lines are fully gapped by this term.
Because inversion eigenvalues are not changed by the time reversal breaking, the resulting insulator has a nontrivial magnetoelectric polarizability according to Eq.~(\ref{P3_inversion}).

We can explicitly calculate the magnetoelectric polarizability gauge-invariantly after we adiabatically deform $m$ to zero.
As the band gap is open in this process as long as $b\ne 0$, the magnetoelectric polarizability is not changed.
Then we include a small constant inversion-breaking perturbation $f_4\Gamma_4$ to have a definite sign of $P_3$.
With this construction, the magnetoelectric polarizability can be calculated using the formula in Ref.~\onlinecite{Zhang_axion}:
\begin{align}
\label{P3_numerical_formula}
P_3
&=\frac{1}{8\pi^2}\int_{BZ} d^3k \frac{2|f|+f_3}{(|f|+f_3)^2|f|^{3/2}}\epsilon^{ijkl}f_i\d_{k_x}f_j \d_{k_y}f_k\d_{k_z}f_l,
\end{align}
where $|f|^2=f_1^2+f_2^2+f_3^2+f_4^2+f_5^2$.
The magnetoelectric polarizability converges to $\pm 1/2$ as $f_4$ approaches to zero as shown in Fig.~\ref{NLSMtoTMEI}(d).

Let us shortly comment on the experimental observation of magnetoelectric effect.
Based on our analysis above, one may expect a bulk polarization induced the magnetoelectric effect when magnetic field is applied.
Nonetheless, charges cannot be polarized as long as the inversion symmetry is preserved by the termination.
Instead, it has been recently found that this system belongs to a class of second-order topological insulators, which has chiral hinge states living on an inversion-invariant 1D subsystem of the inversion-symmetric surface~\cite{RotationAnomaly,Khalaf,Khalaf-Po-Vishwanath-Watanabe}~\footnote{We thank Haruki Watanabe for pointing out this issue.}.
The chiral hinge states can be observed by angle-resolved photoemission spectroscopy to verify the second-order topological nature of this system.
If we add some degrees of freedom having nontrivial Chern numbers on the surface, such that inversion symmetry is broken, a half-integer magnetoelectric polarization can be observed when magnetic field is applied.

\section{Pair Creation of Z2 Monopoles}
\label{sec.pair_creation}

\begin{figure}[t!]
\includegraphics[width=8.5cm]{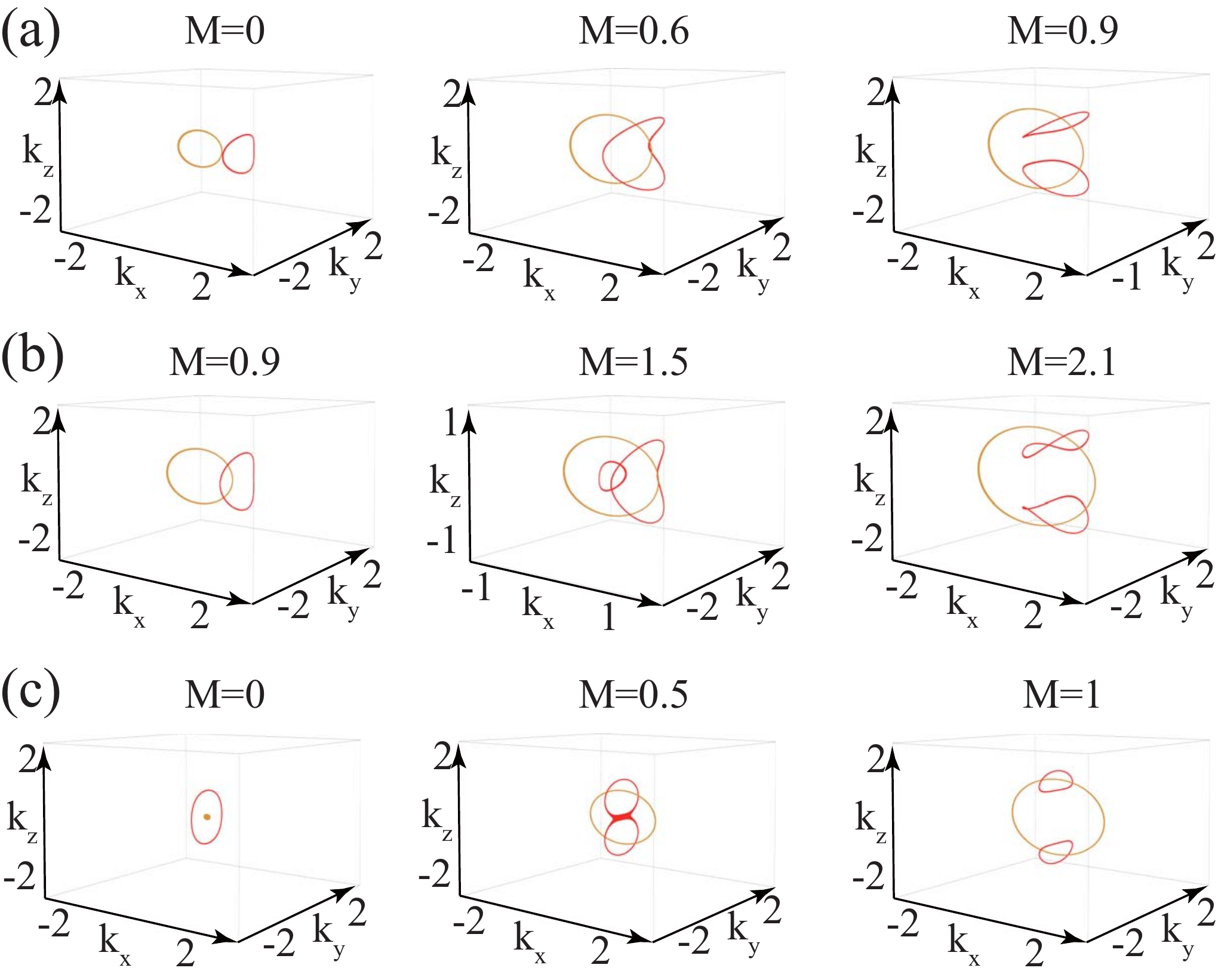}
\caption{Pair creation of nodal lines carrying nontrivial $Z_2$ monopole charges.
Red (Orange) points and lines indicate the crossing between the conduction and valence bands (two occupied bands).
The nodal structures are numerically calculated using Eq.~(\ref{pair_creation_Hamiltonian}).
(a,b) Inversion-asymmetric pair creation.
$(A,m_{15},m_{35})=(1,0.5,0.5)$, $(2,0.5,0.5)$.
(c) inversion symmetric version of (a).
$(A,m_{15},m_{35})=(1,0.5,0)$.
In all cases, $Z_2$ monopoles are pair-created as $M$ is increased.
}
\label{pair_creation-numerics}
\end{figure}

Here we show some inversion-asymmetric pair creations of nodal lines carrying nontrivial $Z_2$ monopole charges, which is generic in systems with only $PT$ symmetry.
We will also illustrate how linking structures are created by a DBI in more detail than we did in the main text.

Let us use the following effective Hamiltonian to demonstrate typical pair creations of $Z_2$ monopoles. 
\begin{align}
\label{pair_creation_Hamiltonian}
H({\bf k})
&=k_x\Gamma_1+k_y\Gamma_2+(M-k_x^2-Ak_y^2-k_z^2)\Gamma_3\notag\\
&+m_{15}\Gamma_{15}+m_{35}\Gamma_{35},
\end{align}
where $\Gamma_{(1,2,3)}=(\sigma_x,\tau_y\sigma_y,\sigma_z)$, $\Gamma_{(4,5)}=(\tau_x\sigma_y, \tau_z\sigma_y)$,
$\Gamma_{ij}=[\Gamma_i,\Gamma_j]/2i$, and $\tau_{x,y,z}$ and $\sigma_{x,y,z}$ are Pauli matrices.

Figure~\ref{pair_creation-numerics} shows typical pair creations of $Z_2$ monopoles Here the transition is driven by varying $M$.
$(A,m_{15},m_{35})=(1,0.5,0.5)$, $(2,0.5,0.5)$ ,$(1,0.5,0)$ in Fig.~\ref{pair_creation-numerics}(a), Fig.~\ref{pair_creation-numerics}(b), Fig.~\ref{pair_creation-numerics}(c), respectively.
Nodal lines with $Z_2$ monopole charges are created in pair when a trivial nodal line [See Fig.~\ref{pair_creation-numerics}(a)] or multiple trivial lines [See Fig.~\ref{pair_creation-numerics}(b)] surround the line of occupied band degeneracy such that a linking structure is created.
The process shown in Fig.~\ref{pair_creation-numerics}(c) corresponds to an inversion-symmetric version of that shown in Fig.~\ref{pair_creation-numerics}(a).
In case (c), one may find that the effective Hamiltonian is symmetric under inversion $P=\sigma_z$ which acts as $\bf k\rightarrow -k$.

\begin{figure}[t!]
\includegraphics[width=8.5cm]{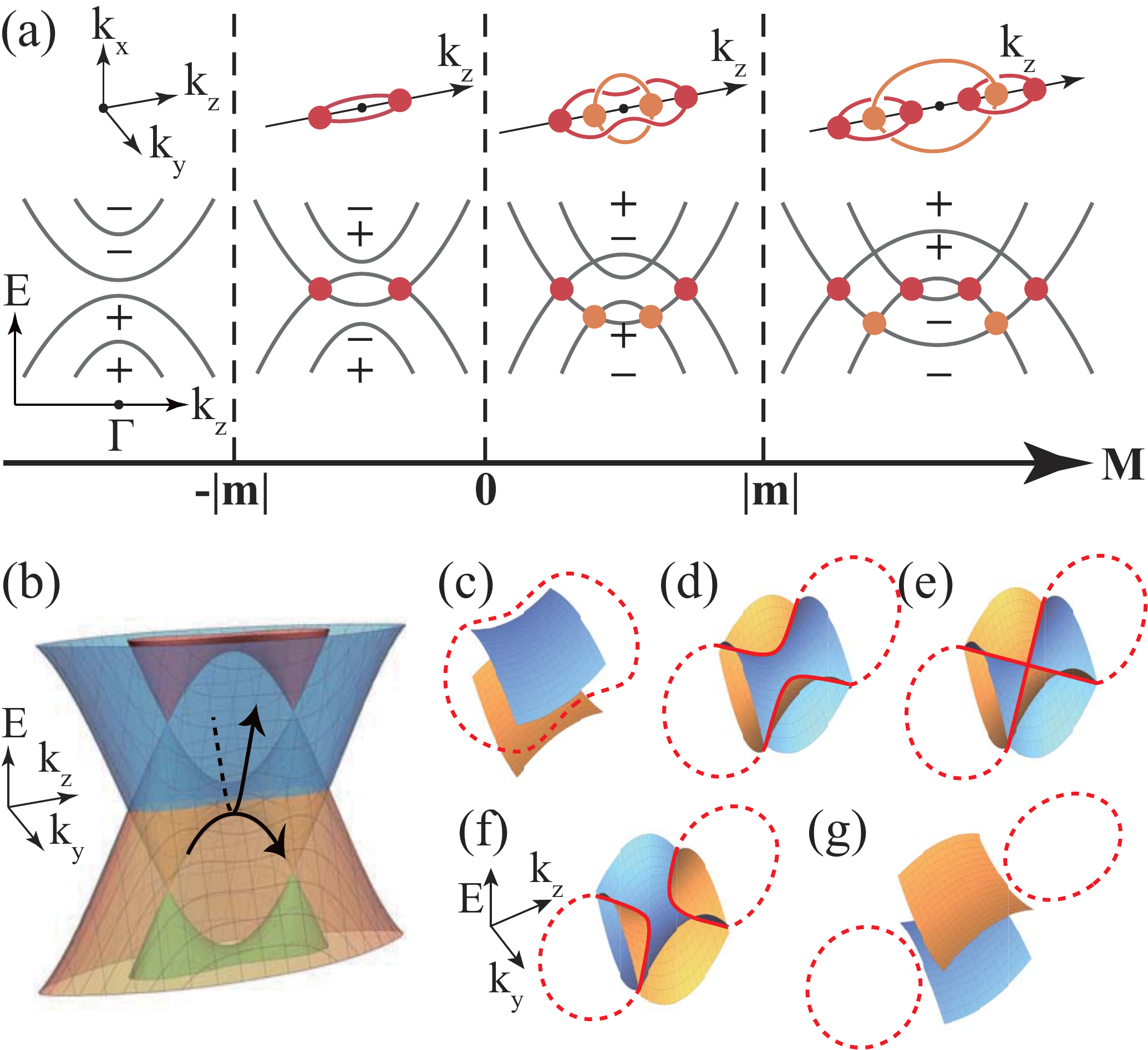}
\caption{Pair creation of $Z_{2}$NLs via a double band inversion.
$M$ and $m$ ($\equiv m_{15}$) are parameters defined in Eq.~(\ref{pair_creation_Hamiltonian}). 
$A=1$ and $m_{35}=0$.
(a) Evolution of band structure during double band inversion.
Red (Orange) points and lines indicate the crossing between the conduction and valence bands (two occupied bands).
(b) Saddle-shaped band structure when $0<M<|m|$.
(c-g) Change in nodal line structure when two saddle-shaped bands cross.
}
\label{saddle_inversion}
\end{figure}

The process described in Fig.~\ref{pair_creation-numerics}(c) is a part of the process we described in the main text, which we called double band inversion [See Fig.~\ref{saddle_inversion}(a)].
Eq.~(\ref{monopole_inversion}) implies that a change of two inversion eigenvalues at a TRIM, i.e., a double band inversion, generates a pair of $Z_2$ monopoles in the Brillouin zone.
Though, it may be confusing why double band inversion necessarily creates a linking structure.
Specifically, the second band gap closing at TRIM splits a trivial loop into two rather than creating a new trivial nodal loop, as one can naively expect.

The splitting of a trivial nodal line can be understood by noting that the second band gap closing (i.e., band gap closure at $M=|m|$ in Fig.~\ref{saddle_inversion}(a)) occurs between two saddle surfaces rather than two convex surfaces.
Let us clarify how a linking structure is created in double band inversion by describing the band structure carefully. 
For concreteness, we use the Hamiltonian in Eq.~(\ref{pair_creation_Hamiltonian}) with $A=1$, $m_{15}=m$, and $m_{35}=0$, which is the one that we used to describe the double band inversion in main text.
The evolution of the band structure during the double band inversion is illustrated in Fig.~\ref{saddle_inversion} (a) as a function of the parameter $M$.
As we increase $M$ from $M<-|m|$, the first band inversion occurs at $M=-|m|$ between the top valence and bottom conduction bands, creating a trivial nodal line.
Then, the inversion at $M=0$ between two occupied (unoccupied) bands generates another nodal line below (above) $E_F$.
After this band inversion, the band structure near ${\bf k}=0$ develops saddle-shape around $E_F$ [Fig.~\ref{saddle_inversion}(b)]. 
The last band inversion at $M=|m|$ between the two saddle-shaped bands induces a Lifshitz transition [Fig.~\ref{saddle_inversion}(c-g)], during which the trivial nodal line splits into two nodal lines carrying nontrivial $Z_2$ monopole charges, which are linked by another nodal line existing between the occupied bands.

\section{Material Realization}
\label{sec.material}


\begin{figure}[!ht]
\centering
\includegraphics[width=.5\textwidth]{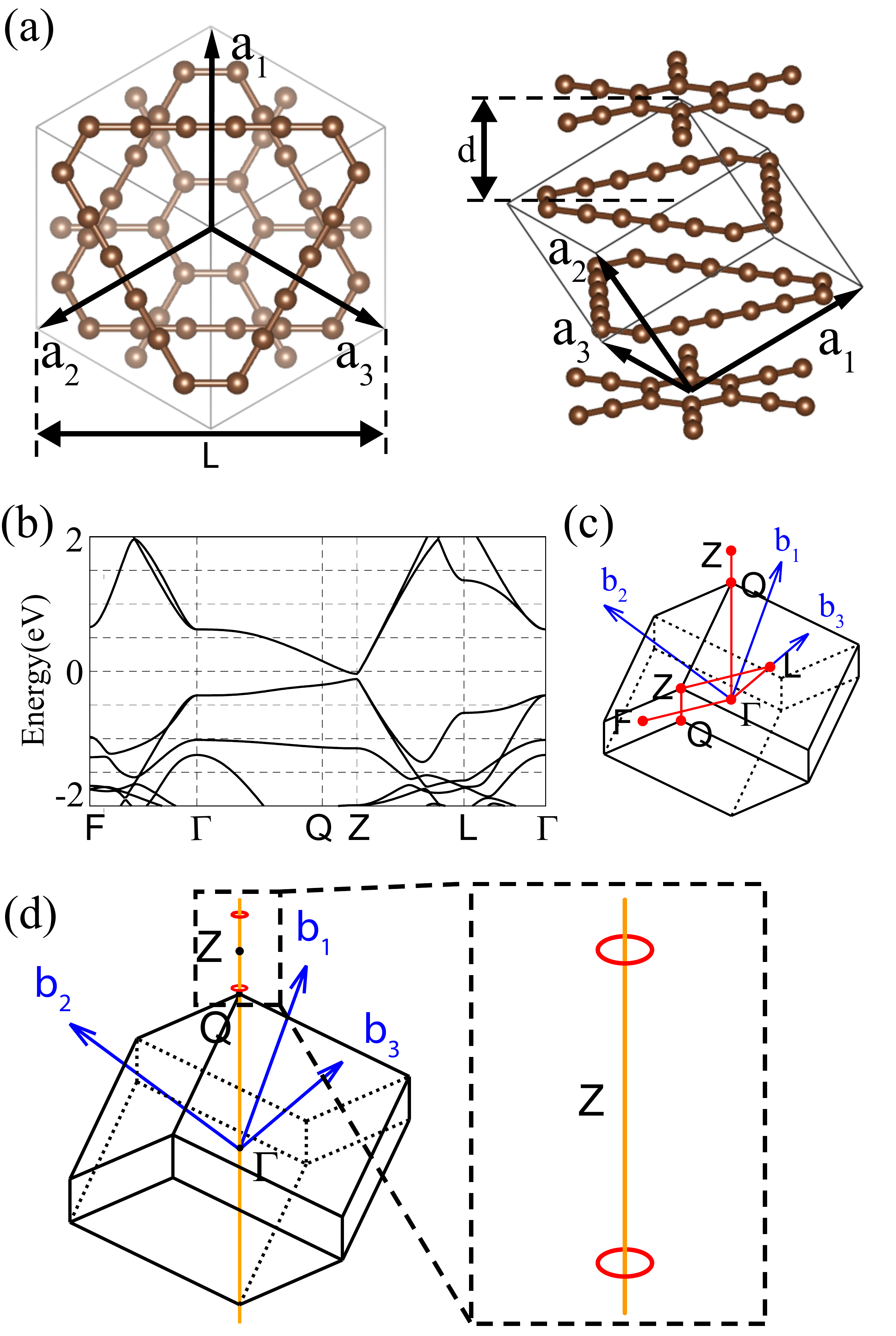} \\
\caption{(a) Atomic structure of ABC-stacked graphdiyne. Both the top (left panel) and orthographic (right panel) views are illustrated. The primitive vectors are indicated by the solid (black) arrows. The solid box represents a primitive unit cell.
 (b) Electronic energy band structure of ABC-stacked graphdiyne, plotted along the high-symmetry lines of the BZ presented in (c).
 (d) Illustration of nodal lines in momentum space. Red circles illustrate the $Z_2$NLs formed between the $N$th and $(N+1)$th bands. Orange line illustrates the second nodal line that is formed between the $N$th and $(N-1)$th bands. The secondary (orange) nodal line threads the $Z_2$NLs, exhibiting the proposed linking structure. The blue arrows indicate the reciprocal lattice vectors.
 }
\label{fig:material}
\end{figure}

In the main text, we predicted based on first-principles calculations that ABC-stacked graphdiyne realizes the $Z_2$NL with the proposed linking structure. ABC graphdiyne is a ABC stack of 2D graphdiyne layers. Graphdiyne is a planar sp2-sp carbon network of benzene rings connected by ethynyl chains [See Fig.\,\ref{fig:material}(a)]. Here we present our results in more detail.

\subsection{ABC-Stacked Graphdiyne}

We first show that 3D graphdiyne hosts nodal lines in momentum space. In good agreement with Nomura {\it et. al.}'s results, we were able to find the nodal lines occurring off the high-symmetry $Z$ point of the BZ. While the electronic band structure of ABC-stacked graphdiyne plotted along the high-symmetry $\mathbf{k}$ points of the Brillouin zone (BZ) displays well-defined band gap throughout the high-symmetry lines as shown in Fig.~\ref{fig:material}(b), the valence ($N$th) and conduction ($(N+1)$th) bands linearly touch each other along a pair of closed nodal lines colored by red in Figs.~\ref{fig:material}(d) and (e). (Here, $N$ is the filling - the number of electrons per unit cell.) In addition to these nodal lines, we also find that the occupied $(N-1)$th and $(N-2)$th bands form additional nodal line [colored by orange in Figs.~\ref{fig:material}(d) and (e)], which pierces the first nodal lines, realizing the proposed linking structure.


\begin{figure}[ht!]
\centering
\includegraphics[width=8.5cm]{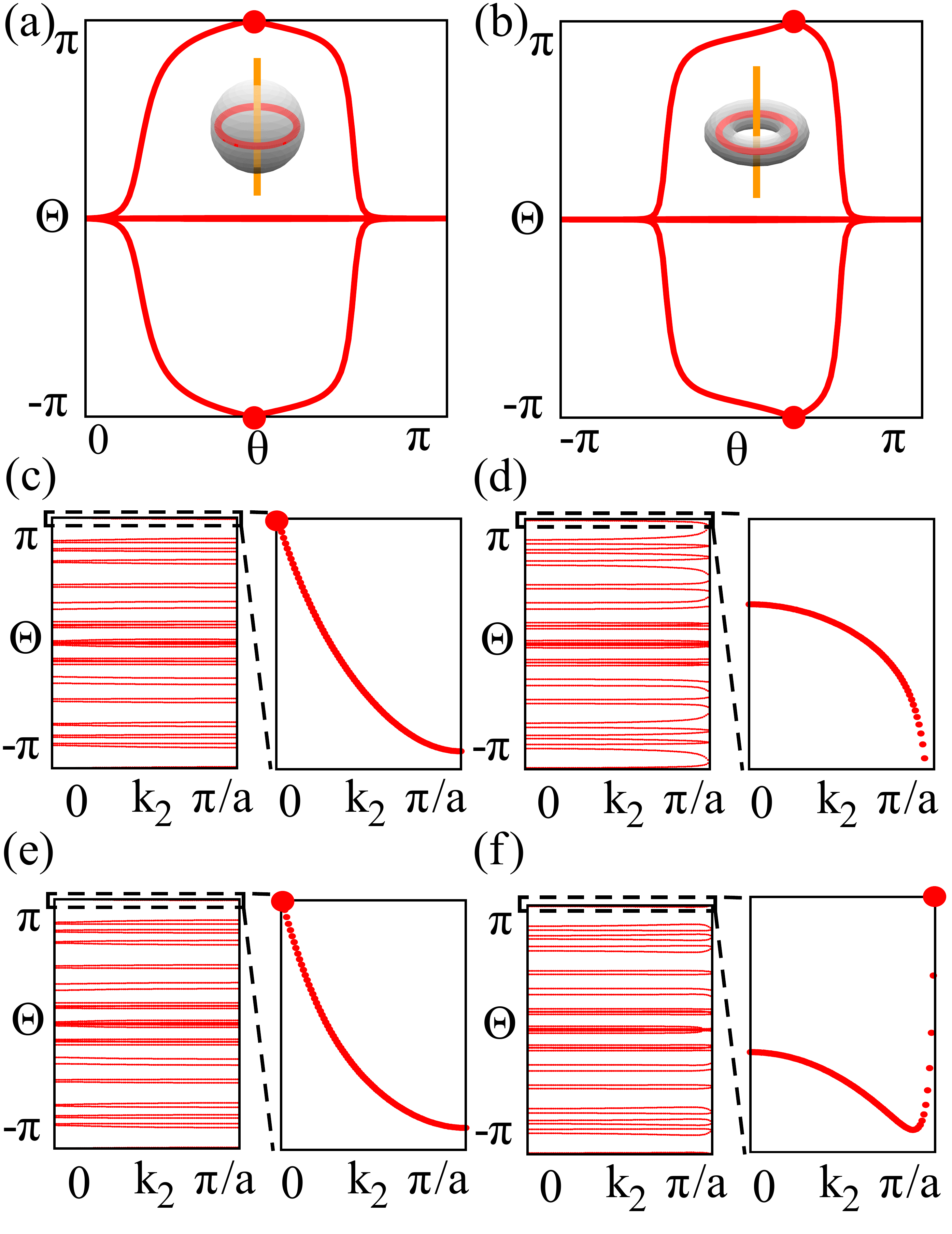} \\
\caption{
 Wilson loop spectra evaluated on (a) a sphere, (b) a torus. The insets of (a) and (b) illustrate the sphere the torus on which the Wilson loop eigenvalues are calculated. Both spectra feature an odd (=1) number of linear crossing at $\Theta = \pi$, which pictorially demonstrate the presence of non-trivial $Z_2$ monopole charge carried by the enclosed $Z_2$NL. Wilson loop operators are calculated along the $k_1$ direction on (c,e) the $k_3 = 0$ and (d,f) the $k_3 = \pi$ planes. Here ${\bf k}=k_1{\bf b}_1+k_2{\bf b}_2+k_3{\bf b}_3$. 
(c,d) Without external strain.
The insets of (c) and (d) magnify the regions enclosed by the black dashed boxes. The Wilson loops on the $k_3 = 0$ plane shows the linear crossing, leading to $\omega_2 = 1$. The inset of (d) shows a gap between the Wilson loop and the $\Theta = \pi$ line, indicating trivial $Z_2$ monopole charge with $\omega_2 = 0$. 
(e,f) Under the 6 \% of tensile strain applied along the perpendicular direction to the layers ($z$-direction). Both exhibit $\omega_2 = 1$. The insets show a magnified view of the Wilson spectra near the $\Theta = \pi$ line, which shows a linear crossing of the Wilson loop on the $\Theta = \pi$ line.
 }
\label{fig:material2}
\end{figure}

The linking geometry of the nodal lines is an indicative of the non-trivial $Z_2$ monopole charge carried by the first nodal line. We confirm this by conducting (1) parity analysis and (2) Wilson loop calculations. First, as the parity eigenvalues of the occupied Bloch states at the eight time-reversal invariant points $(\Gamma,3L,3F,Z)$ allow for the identification of a $Z_2$NL in momentum space via Eq.~(\ref{monopole_inversion}), 
we calculate the parity eigenvalues and find that $(-1)^{N^-_\mathrm {occ}(\Gamma)/2}= -1$, $(-1)^{N^-_\mathrm {occ}(F)/2}= 1$, $(-1)^{N^-_\mathrm {occ}(L)/2}= 1$, and $(-1)^{N^-_\mathrm {occ}(Z)/2}= 1$. This results in $N_{mp} = 1$, indicating that an odd number of $Z_2$NLs should occur in half the BZ with the other partner occurring in the other half of the BZ. The nontrivial $Z_2$ monopole charge is further confirmed by our Wilson loop spectrum calculated both on the a sphere and on  a torus enclosing the nodal line. The results are presented in Fig.\,\ref{fig:material2}(b). While the Wilson loop eigenvalues are pinned to 0 at the north and south poles of the sphere at $\theta = 0$ and $\pi$, one of the Wilson loop eigenvalues evolves to $\pi$ as $\theta$ evolves, while its time-reversal partner evolves to $-\pi$ as the polar angle $\theta$ varies from $0$ to $\pi$. This results in a single linear crossing occurring at $\pi$, which  pictorially demonstrate the nontrivial $Z_2$ monopole charge $\omega_2 =1$. The Wilson loop spectrum evaluated on the torus, which is calculated along the toroidal direction $\phi$ for a  fixed poloidal angle $\theta$, also demonstrates the nontrivial $Z_2$ charge, exhibiting a linear band crossing at the Wilson loop eigenvalue $\pm\pi$.

Having demonstrated the $Z_2$ NLSM in 3D graphdiyne, we now show that strain induces topological phase transition from the $Z_2$ NLSM to a 3D weak SWI. The pair of the $Z_2$NLs appearing near the $Z$ point fuse together and annihilate at the $\sim$ 3 \% of tensile strain, when we apply strain along the out-of-plane ($z$) direction to ABC-stacked graphdiyne with the rest of the lattice parameters fixed at the values obtained without strain. This process changes the parity eigenvalues of the occupied Bloch states at $Z$, resulting in $(-1)^{N^-_\mathrm {occ}(Z)/2}= -1$ while keeping the other parity eigenvalues. This leads to $N_{mp}= 0$, thus being consistent with the absence of the $Z_2$NL after the annihilation. We apply 5\,\% of strain along the $z$-direction and calculate the Wilson loop eigenvalue spectra on $k_3 = 0$ ($\Gamma - F - L - F$) and $k_3 = \pi$ ($F - L - L - Z$) planes. The results are presented in Fig.\,\ref{fig:material2}(e) and (f), which show that each plane hosts $\omega_2 = 1$, which confirms the 3D WSI hosted in 3D graphdiyne under the strain.

\subsection{Computational Methods}
 Our first-principles calculations are performed based on density functional theory (DFT). The Perdew--Burke--Ernzerhof type generalized gradient approximation \cite{Perdew96p3865} is used as implemented in \textsc{QUANTUM ESPRESSO} package \cite{Giannozzi09p395502}. Norm--conserving, optimized, designed nonlocal pseudopotentials for C, Ti, and B atoms are generated using \textsc{OPIUM} \cite{Rappe90p1227}. A plane--wave basis is used to represent the wave functions with the energy cutoff of 680 eV. The atomic structure is fully relaxed including cell-variation to a force tolerance of 0.005 eV/\AA\/. k-points are sampled in the 8$\times$8$\times$8 grid in the Monkhorst-Pack scheme \cite{Monkhorst76p5188}. For ABC graphdiyne, the lattice parameters are calculated as $L = 9.45$\ \AA, $d = 3.17$\ \AA, $b_1 = 1.43$\ \AA, $b_2 = 1.39$\ \AA, $b_3 = 1.23$\ \AA, $b_4 = 1.34$\ \AA. In order to test the effect of strain, we manually applied compressive strain along the out-of-plane direction from the fully relaxed atomic structure, with the other lattice parameters fixed at those of the fully relaxed structure.

\end{document}